\documentclass[a4paper,12pt]{article}
\pdfoutput=1
\usepackage[utf8]{inputenc}
 \usepackage{graphicx}
\usepackage{cancel}
\usepackage{amsmath, amssymb, amsfonts}
\usepackage{bbold}
\usepackage{latexsym}
\usepackage{subfigure}
\usepackage{authblk}
\usepackage{ dsfont }
\usepackage{ bbold }
\usepackage[font=small, format=plain, labelfont=bf,up, textfont=up]{caption}
\usepackage[top=3cm, bottom=3cm, left=2.5cm, right=2.5cm]{geometry}
\usepackage{setspace}
    \usepackage{sectsty}
    \sectionfont{\LARGE}
    \subsectionfont{\Large}
    \subsubsectionfont{\large}
\makeatletter
\def\maketag@@@#1{\hbox{\m@th\normalfont\normalsize#1}}%correct label size for smaller font size in equation
\makeatother

\usepackage{cite}

\usepackage[usenames,dvipsnames]{xcolor}

\usepackage{psfrag}
\usepackage{color, xcolor}
\usepackage{shadow}
\usepackage{feynarts}
\usepackage{fixltx2e}
\usepackage{mathrsfs}
\usepackage{fixltx2e}
\usepackage{tikz}
\usetikzlibrary{arrows,shapes}
\usetikzlibrary{trees}
\usetikzlibrary{matrix,arrows} 				% For commutative diagram
\usetikzlibrary{positioning}				% For "above of=" commands
\usetikzlibrary{calc,through}				% For coordinates
\usetikzlibrary{decorations.pathreplacing}  % For curly braces
\usetikzlibrary{patterns}
\usepackage{pgffor}							% For repeating patterns

\usetikzlibrary{decorations.pathmorphing}	% For Feynman Diagrams
\usetikzlibrary{decorations.markings}
\tikzset{
    vector/.style={decorate, decoration={snake}, draw},
	provector/.style={decorate, decoration={snake,amplitude=2.5pt}, draw},
	antivector/.style={decorate, decoration={snake,amplitude=-2.5pt}, draw},
    fermion/.style={draw=black, postaction={decorate},
        decoration={markings,mark=at position .55 with {\arrow[draw=black]{>}}}},
    fermionbar/.style={draw=black, postaction={decorate},
        decoration={markings,mark=at position .55 with {\arrow[draw=black]{<}}}},
    fermionnoarrow/.style={draw=black},
    gluon/.style={decorate, draw=black,
        decoration={coil,amplitude=4pt, segment length=5pt}},
    scalar/.style={dashed,draw=black, postaction={decorate},
        decoration={markings,mark=at position .55 with {\arrow[draw=black]{>}}}},
    scalarbar/.style={dashed,draw=black, postaction={decorate},
        decoration={markings,mark=at position .55 with {\arrow[draw=black]{<}}}},
    scalarnoarrow/.style={dashed,draw=black},
    electron/.style={draw=black, postaction={decorate},
        decoration={markings,mark=at position .55 with {\arrow[draw=black]{>}}}},
	bigvector/.style={decorate, decoration={snake,amplitude=4pt}, draw},
}

\tikzstyle{block} = [draw, rectangle, 
    minimum height=3em, minimum width=6em]
    
\newcommand{\ca}{c_{\alpha}}
\newcommand{\sa}{s_{\alpha}}
\newcommand{\cb}{c_{\beta}}
\newcommand{\sinb}{s_{\beta}}
\newcommand{\sw}{s_{W}}

\newcommand{\tb}{t_{\beta}}

\newcommand{\sib}{s_{\beta}}
\newcommand{\sibn}{s_{\beta_n}}
\newcommand{\sibc}{s_{\beta_c}}
\newcommand{\cbn}{c_{\beta_n}}
\newcommand{\cbc}{c_{\beta_c}}

\newcommand{\ci}{\widetilde{\chi}_i^{0}}

\newcommand{\bpm}{\begin{pmatrix}}
\newcommand{\epm}{\end{pmatrix}}
\newcommand{\mhp}{M_{H^{\pm}}}

\newcommand{\psq}{p^{2}}
\newcommand{\ps}{(p^{2})}
\newcommand{\mmat}{\textbf{M}}
\newcommand{\gmat}{\hat{\boldsymbol{\Gamma}}}
\newcommand{\dmat}{\boldsymbol{\Delta}}
\newcommand{\gh}{\hat{\Gamma}_}

\newcommand{\seff}{\hat{\Sigma}^{\rm{eff}}_{ii}}
\newcommand{\dseff}{\hat{\Sigma}^{\rm{eff'}}_{ii}}
\newcommand{\ddseff}{\hat{\Sigma}^{\rm{eff''}}_{ii}}

\newcommand{\dsef}{\hat{\Sigma}^{\rm{eff'}}}
\newcommand{\dsig}{\hat{\Sigma}^{'}}
\newcommand{\ddsig}{\hat{\Sigma}^{''}}
\newcommand{\Si}{\hat{\Sigma}_{ii}}
\newcommand{\Sj}{\hat{\Sigma}_{jj}}
\newcommand{\Sk}{\hat{\Sigma}_{kk}}
\newcommand{\Sij}{\hat{\Sigma}_{ij}}
\newcommand{\Sik}{\hat{\Sigma}_{ik}}

\newcommand{\Sjk}{\hat{\Sigma}_{jk}}
\newcommand{\Ski}{\hat{\Sigma}_{ki}}

\newcommand{\sig}{\hat{\Sigma}}
\newcommand{\Mm}{\mathcal{M}^{2}}
\newcommand{\Zb}{\hat{\textbf{Z}}}
\newcommand{\Zz}{\hat{Z}}
\newcommand{\Ub}{\textbf{U}}
\newcommand{\BW}{\Delta^{\textrm{BW}}}

\newcommand{\CP}{\mathcal{CP}}
\newcommand{\mhmod}{M_h^{\rm{mod+}}}
\newcommand{\mhmax}{M_h^{\rm{max}}}
\newcommand{\gev}{\,\textrm{GeV}}
\newcommand{\pat}{\phi_{A_t}}

\newcommand{\Zbf}{\hat{\textbf{Z}}}

\newcommand{\cp}{\mathcal{CP}}
\newcommand{\order}{\mathcal{O}}

\usepackage{titling}
\usepackage[colorlinks=true,citecolor=OliveGreen,urlcolor=OliveGreen,linktocpage=true,
linkcolor=blue]{hyperref}
\hypersetup{pageanchor=false}

\title{\hfill\small{\texttt{DESY 16-182}}\\ \vspace*{0.5cm}
\LARGE{
\textbf{Breit--Wigner approximation for propagators of mixed unstable states}}}

\author{
\textsc{Elina Fuchs}$^{a,}$\textsuperscript{1},~~
\textsc{Georg Weiglein}$^{b,}$\textsuperscript{2}
\vspace{0.6cm}\\
\textit{(a) Department of Particle Physics and Astrophysics, }\\
 \textit{Weizmann Institute of Science, Rehovot 76100, Israel}\\
\textit{(b)  DESY, Deutsches Elektronen-Synchrotron, Notkestr. 85,}\\ 
\textit{ 
D-22607 Hamburg, Germany}
}
\footnotetext[1]{elina.fuchs@weizmann.ac.il}
\footnotetext[2]{georg.weiglein@desy.de}
\date{}

\begin{document}
\maketitle
\thispagestyle{empty}
\begin{abstract}
For systems of unstable particles that mix with each other, an approximation
of the fully mo\-men\-tum-\,dependent propagator matrix is presented in terms of 
a sum of simple Breit--Wigner propagators that are multiplied with 
finite on-shell wave function normalisation factors. The latter are evaluated at
the complex poles of the propagators.
The pole structure of general propagator matrices is carefully analysed, and
it is demonstrated that in the proposed approximation 
imaginary parts arising from absorptive parts of loop integrals are properly
taken into account.
Applying the formalism to the neutral MSSM Higgs sector with complex 
parameters, very good numerical agreement is found
between cross sections based on the full propagators and the corresponding 
cross sections based on the described approximation.
The proposed approach does not only technically simplify the treatment of
propagators with non-vanishing off-diagonal contributions, it is
shown that it can also
facilitate an improved theoretical prediction of the considered observables
via a more precise implementation of the total widths of the involved
particles. It is also well-suited for the incorporation of interference
effects arising from overlapping resonances.

\end{abstract}
\newpage
\thispagestyle{empty}
\tableofcontents
\thispagestyle{empty}
\newpage

%%%%%%%%%%%%%%%%%%%%%%%%%%%%%%%%%%%%%%%%%%%%%%%%%%%%%%%%%%%%%%%%%%%%%%%%%%%%%%%%
%%%%%%%%%%%%%%%%
%%%%%%%%%%%%%%%%%%%%%%%%%%%%%%%%%%%%%%%%%%%%%%%%%%%%%%%%%%%%%%%%%%%%%%%%%%%%%%%%
%%%%%%%%%%%%%%%%
%%%%%%%%%%%%%%%%%%%%%%%%%%%%%%%%%%%%%%%%%%%%%%%%%%%%%%%%%%%%%%%%%%%%%%%%%%%%%%%%
%%%%%%%%%%%%%%%%

\section{Introduction}
\label{chap:intro}
\setcounter{page}{1}
As a consequence of electroweak symmetry breaking, particles with the
same quantum numbers of the electric charge and colour can mix with each
other. This is reflected in their propagators contributing to production
or decay processes in particle physics. Beyond lowest order those
propagators receive contributions from all possible mixings with other
particles, such that the propagators of the system of particles that can
mix with each other are of matrix-type. Since the involved particles are
in general unstable, a proper treatment of imaginary parts is necessary.
The mass of an unstable particle is
determined from the real part of the (gauge-invariant) complex
pole~\cite{Willenbrock:1991hu,Sirlin:1991fd,Stuart:1991xk,Veltman:1992tm},
while the imaginary part yields the total decay width of the particle. 
The eigenstates associated with the masses obtained from the complex
poles of the propagator matrix in general differ from the interaction
eigenstates that contribute at lowest order. 

As an example, in the Standard Model (SM) the neutral interaction
eigenstates arising from the ${\rm SU}(2)_{\rm I}$ and 
${\rm U}(1)_{\rm Y}$ gauge groups mix with each other to form the mass
eigenstates 
photon, $\gamma$, and $Z$. Higher-order contributions lead to a mixing
between these states,
giving rise to a $\gamma-Z$ propagator matrix. 
In a renormalisable gauge, the (would-be) Goldstone boson as an unphysical 
degree of freedom
does not vanish, and mixing contributions with this unphysical scalar
particle need to be taken into account as well.
In the quark sector of the SM the relation between the interaction and the
mass eigenstates is encoded in terms of the CKM matrix~\cite{Cabibbo:1963yz,Kobayashi:1973fv}. 
The additional states in models of physics beyond the SM (BSM) usually
follow the same pattern, i.e.\ interaction eigenstates mix with each other
to form mass eigenstates, and the propagators of the latter are of
matrix-type as a consequence of mixing contributions at higher orders (in
some cases it can also be convenient
to derive loop-corrected mass eigenstates directly from 
lowest-order interaction eigenstates, avoiding the introduction of
lowest-order mass eigenstates). 
In particular,
extended Higgs sectors usually contain several neutral Higgs states that can
mix with each other. If $\cp$-conservation is assumed, 
$\cp$-even and $\cp$-odd states can only mix among themselves. In the
general case where the possibility of $\cp$-violating interactions affecting
the Higgs sector is taken into account, non-vanishing mixing contributions
can occur between all neutral Higgs bosons of the model. 
Examples are general two-Higgs-doublet models,
the Higgs sector of the minimal supersymmetric extension of the
SM (MSSM), as well as the Higgs sectors of non-minimal supersymmetric models.
Similarly, propagator mixing between
vector resonances can occur for instance for the Kaluza--Klein excitations 
of the $Z$-boson and the photon ($Z'$ and $A'$, respectively), which  
can have similar masses and have a large mixing with each 
other\,\cite{Cacciapaglia:2009ic}. 
New, possibly nearly degenerate vector bosons can also 
arise in extended gauge symmetries and
theories with a strongly coupled sector, e.g.\ composite Higgs models. 
For the case of the mixing
of two $Z'$-like vectors, see e.g.\ Ref.\,\cite{deBlas:2012qp}.
An example involving mixing with a new fermion is the mixing
between the top quark $t$
and a top partner $t'$\,\cite{Pilaftsis:1997dr}.

In order to properly treat external particles in physical processes, the
correct on-shell properties of in- and outgoing particles have to be
ensured. While the S-matrix theory is formulated only for stable in- and
outgoing states~\cite{Eden:AnaSmatrix},
in practical applications one often needs to deal with the case of unstable
external particles. The proper normalisation of external particles requires in
particular that all external particles are on their mass shell, that their
residues are equal to one, and that the
mixing contributions with all other particles vanish on-shell at the
considered order. This requirement holds independently from the specific 
renormalisation scheme that has been adopted for a certain calculation. In
case the renormalisation conditions of the chosen scheme do not impose the
proper normalisation of external particles, the correct on-shell properties
need to be established via 
UV-finite wave function normalisation factors (``Z-factors'').
In the case of unstable particles, finite-width effects
have to be taken into account, and more generally imaginary parts arising from
the absorptive parts of loop integrals need to be properly treated. A
further complication arises if $\cp$-violating effects are taken into
account, since in this case the presence of complex parameters yields an
additional source of imaginary parts. In the SM the different sources of
imaginary parts affect for instance the renormalisation of the CKM matrix
from the two-loop level onwards, see e.g.\ the discussion in
Ref.~\cite{Espriu:2002xv} and references therein. 
In the MSSM with complex parameters the corresponding
effects arising from absorptive parts of loop integrals and complex model
parameters enter predictions for physical observables in the chargino and
neutralino sector already at the one-loop level, see
Refs.~\cite{Fowler:2009ay,Fowler:2010eba,Bharucha:2012nx}.

In the present paper we demonstrate that 
on-shell wave function normalisation factors, evaluated at the
complex poles, are well suited for approximating
the full mixing propagators also off-shell. In order to achieve this, the
elements of the propagator matrix are expressed in terms of simple
Breit--Wigner propagators that are multiplied with wave function
normalisation factors.
We show for various examples that this approximation works very well in
practice. 
In this context we discuss in detail the uncertainties that are associated
with the neglected terms in the approximation and provide estimates of their
impact, which we find to be in good agreement with the 
observed numerical deviations between the full result and the approximation.
We furthermore demonstrate that
besides very significant technical simplifications 
our approach also
has conceptual advantages.

Concerning technical simplifications, the evaluation of the 
propagator matrix for a given
value of the external momentum squared requires the full momentum dependence
of all contributing diagonal and off-diagonal self-energies. In our
approximation based on wave function normalisation factors that are evaluated 
at the complex poles, on the other hand, the only momentum dependence that
needs to be taken into account is the one of the simple Breit--Wigner propagators. 
We emphasize that our approach goes significantly beyond a simple
``effective coupling''-type approximation where all external momenta at the
loop level are neglected. While the latter approximation
would disregard the important absorptive parts of the 
loop functions, those contributions are taken account in our approach through
the evaluation of the Z-factors at the complex poles.
We will demonstrate in particular that 
expanding the full propagators around all of their complex poles indeed
results in the sum of Breit--Wigner propagators in combination with on-shell 
Z-factors.
Approximating the full propagator matrix in this simple and convenient way
has several conceptual advantages. In particular,
the formulation in terms of Breit--Wigner propagators and 
Z-factors facilitates the implementation of a more precise total width than
the width obtained from the imaginary part of the complex pole 
with self-energies evaluated at the same order.
Furthermore, for states with masses that are nearly degenerate (i.e.\ the
difference between the masses of the two states is of the same order as the
sum of the total widths of the two states) 
the mixing becomes resonantly enhanced and the resonances overlap, see 
e.g.\ Refs.~\cite{Pilaftsis:1997dr,Frank:2006yh,Cacciapaglia:2009ic,
deBlas:2012qp,Fuchs:2014ola,Gomez-Bock:2015sia}.
In such a case a single-pole approximation is not 
applicable~\cite{Cacciapaglia:2009ic,deBlas:2012qp,Fuchs:2014ola}. 
Instead, the contributions from the relevant
complex poles of the propagators have to be taken into account.
The approximation of the propagator matrix in terms of 
on-shell wave function normalisation factors evaluated at the
complex poles and Breit--Wigner propagators allows the implementation of
interference contributions using the prescription of Ref.~\cite{Fuchs:2014ola}. 

We perform in this paper 
a careful treatment of imaginary parts and
analyse the pole structure of matrix-type propagators in detail.
While in the no-mixing case every element of an $(n \times n)$ propagator matrix has a
single pole, we will demonstrate that in general every single element of 
an $(n \times n)$ propagator matrix (diagonal and off-diagonal) by
itself has $n$ poles. This fact makes the association of the
(loop-corrected) mass eigenstates with the propagator poles non-trivial. 
We will show that in principle all permutations of the $n$ propagator
poles with the $n$ mass eigenstates are physically equivalent. For practical
purposes, however, it is important to make a choice that leads to numerically stable
and well-behaved results. 

To be specific, in the following we use the language of the Higgs sector of 
the MSSM with complex parameters. In this case $\cp$-violating effects entering at 
the loop level give rise to a mixing between the lowest-order mass eigenstates
$h, H, A$, which are $\cp$-eigenstates ($h, H$ are $\cp$-even states, while
$A$ is a $\cp$-odd state). The $(3 \times 3)$ mixing between the neutral
Higgs states is the smallest type of propagator mixing that exhibits the
qualitative features of general $(n \times n)$ propagator matrices.
Furthermore, the MSSM example is well-suited to emphasise the distinction
between interaction eigenstates (in this case the component field of the two
Higgs doublets, e.g.\ $\phi_1$ and $\phi_2$), lowest-order mass eigenstates
($h, H, A$) and loop-corrected mass eigenstates, which in the MSSM case are
called $h_1, h_2, h_3$. The MSSM case also illustrates the possibility of
mixing with other spin states ($\gamma, Z$) and with unphysical degrees of
freedom ($G$). An important feature of models like the MSSM is the fact
that the masses are {\em predicted\/} from the input parameters of the
model, which makes the need for an appropriate assignment of mass
eigenstates with propagator poles particularly apparent. 
The generalisation of our results to other models, cases with 
$(n \times n)$ mixing for $n > 3$ and other spin states should be straightforward.

The paper is organised as follows.
After reviewing aspects of higher-order contributions 
and mixing effects in a system of three 
scalar states in Section\,\ref{sect:LoopMix}, we derive the pole 
structure of the propagator matrix and resulting relations between the 
wave function normalisation factors
in Section\,\ref{sect:poles}. An analytical derivation of the 
multi-Breit--Wigner approximation of the full propagators 
including an uncertainty estimate
is given in 
Section\,\ref{sect:fullBWZ}. We perform a detailed numerical comparison of the 
two approaches in Section\,\ref{sect:NumBWfull} before we conclude in 
Section\,\ref{sect:conclusions}.

%%%%%%%%%%%%%%%%%%%%%%%%%%%%%%%%%%%%%%%%%%%%%%%%%%%%%%%%%%%%%%%%%%%%%%%%%%%%%%%%%
\section{Loop-level mixing of scalar propagators}
\label{sect:LoopMix}
In a system of $n$ particles which have the same conserved quantum numbers
in general a non-trivial mixing between the different states will occur.
While at lowest order the mass matrix given in terms of the interaction
eigenstates can be diagonalised in order to obtain the mass eigenstates, at
the loop level momentum-dependent self-energy corrections enter. These
contributions give
rise to a momentum-dependent $(n \times n)$ propagator matrix which is in general
non-diagonal. The physical masses that are associated with the
loop-corrected propagator matrix need to be determined from the (complex)
poles of the propagator matrix. 

While the case of $(2\times 2)$ mixing provides the minimal setup for studying
such a propagator structure, the issue of how to associate the mass
eigenstates with the $n$ poles of the propagator matrix becomes fully
apparent only from the $(3\times 3)$ case onwards.
The case of $(3\times 3)$ mixing captures all relevant aspects of the pole 
structure and permutation of the states so that the conclusions can be 
generalised for larger mixing systems. 

For illustration and definiteness we use in this paper the language and the
formalism of the neutral Higgs sector of the MSSM with $\cp$-violating mixing, 
i.e.\ for the three lowest-order mass eigenstates $i,j,k = h,H,A$ and their mixing 
into the loop-corrected 
mass eigenstates $h_a,~a=1,2,3$. Since the analytic discussions in 
Chapters~\ref{sect:LoopMix}-\ref{sect:fullBWZ} do not rely on any 
model-dependent relations of the MSSM, the results can be transferred also to 
scalar sectors of other models by replacing $h,H,A$ by a different set of 
$i,j,k$, and analogously $h_a$ by a general $X_a$, where the generalisation
to the case of mixing among more than three particles 
should be straightforward.
Upon incorporation of the appropriate spin structure, the results can also
be generalised to the case of
mixing propagators of vector bosons or 
fermions.

The Higgs propagators in the MSSM in fact receive contributions from the
mixing with the longitudinal components of the neutral gauge bosons $Z$ and
$\gamma$. While the $Z$-boson propagator contributes to the diagonal
elements of the Higgs propagator matrix at the two-loop level, the photon
enters in those elements
only at the four-loop level. Besides the mixing with the longitudinal
component of the $Z$ boson, also the mixing with the unphysical (would-be)
Goldstone boson $G$ needs to be taken into account. Instead of considering a 
$(6\times6)$ propagator matrix of the states
$\left\lbrace h,H,A,G,Z,\gamma\right\rbrace$, we focus on the full mixing
contributions of the physical Higgs fields, i.e.\ we treat the case of a $(3
\times 3)$ propagator matrix (which would generalise to $(n \times n)$ for
the case of $n$ physical scalar fields). For the $(3 \times 3)$
propagator matrix of physical Higgs fields, higher-order effects from the
inversion of the matrix of the irreducible 2-point vertex functions are
taken into account. 
In contrast, we treat the mixing contributions with the gauge bosons
and the (would-be) Goldstone boson in a strict perturbative expansion up to
the desired order.
It should be noted that Higgs--$G/Z$ mixing contributions 
already appear at the one-loop level in processes with external Higgs bosons, 
see e.g.\ 
Refs.~\cite{Williams:2007dc,Fowler:2009ay,Williams:2011bu,Bharucha:2012nx}, 
and therefore need to be included in order to obtain a complete one-loop
result for processes of this kind. For contributions to the propagator
matrix with an incoming and outgoing Higgs boson the mixing contributions
with $G$ and $Z$ enter from the order of $(\sig_{i,G/Z})^{2}$ onwards, which
is of subleading two-loop order. For our numerical analysis below, see 
Sect.\,\ref{sect:NumBWfull}, we will use use the program 
\texttt{FeynHiggs}\,\cite{Heinemeyer:1998np, Heinemeyer:1998yj,
Degrassi:2002fi, Heinemeyer:2007aq}, where dominant two-loop corrections and
leading higher-order contributions are incorporated in the irreducible
self-energies. Since sub-leading two-loop contributions that are of the same
order as the mixing contributions with $G$ and $Z$\,\cite{Frank:2006yh} 
are neglected in this analysis, we include 
no further corrections beyond the $(3 \times 3)$ 
propagator matrix of the physical Higgs fields in this case.

In this section we introduce the relevant quantities and 
fix our notation for the propagator matrix and the wave
function normalisation factors, see 
Refs.\,\cite{Dabelstein:1994hb,Dabelstein:1995js,Heinemeyer:2000fa,
Heinemeyer:2001iy,Hahn:2002gm,Frank:2006yh,Hahn:2006np,Williams:2007dc,
Fowler:2010eba,Williams:2011bu}.
In Sect.\,\ref{sect:poles} we will analyse the pole structure 
of propagator matrices of unstable particles.

%%%%%%%%%%%%%%%%%%%%%%%%%%%%%%%%%%%%%%%%%%%%%%%%%%%%%%%%%%%%%%%%%%%%%%%%%%%%%

\subsection{Propagator matrix and the effective self-energy}

As explained above, we use the case of the MSSM with complex parameters in
order to illustrate propagator mixing between three physical scalar fields. 
If $\CP$ were assumed to be conserved in the MSSM, the $\CP$-violating 
self-energies would vanish, 
$\sig_{hA}=\sig_{HA}=0$, so that only the two $\CP$-even states $h$ and $H$
would mix with each other. Our treatment corresponds to the case where 
non-zero phases from complex parameters are taken
into account. Hence all renormalised self-energies $\Sij\ps$ of the Higgs 
bosons $i,j = h,H,A$ are in general non-vanishing, so that
the matrix $\mmat$ of mass 
squares consists of the tree-level masses $m_i^{2}$ on the diagonal and
renormalised self-energies on the diagonal and 
off-diagonal entries. Expressed in terms of the lowest-order mass
eigenstates $h, H, A$, which are also $\cp$ eigenstates, the matrix takes
the form
\begin{equation}
 \mmat(\psq) = \bpm
  m_h^{2}-\hat{\Sigma}_{hh}(\psq) & -\hat{\Sigma}_{hH}(\psq) &-\hat{\Sigma}_{hA}(\psq)\\
  -\hat{\Sigma}_{Hh}(\psq) & m_H^{2}-\hat{\Sigma}_{HH}(\psq) &-\hat{\Sigma}_{HA}(\psq)\\
  -\hat{\Sigma}_{Ah}(\psq) & -\hat{\Sigma}_{AH}(\psq) &m_A^{2}-\hat{\Sigma}_{AA}(\psq)
\epm. \label{eq:HiggsMassMatrix}
\end{equation}
The renormalised irreducible 2-point vertex functions
\begin{equation}
 \hat{\Gamma}_{ij}\ps = i\left[(\psq-m_i^{2})\delta_{ij}+\sig_{ij}\ps\right] \label{eq:IrepVert}
\end{equation}
can be collected in the $3\times3$ matrix $\gmat_{hHA}$ in terms of $\mmat$ as
\begin{equation}
 \gmat_{hHA}(\psq) = i\left[\psq \textbf{1} -\mmat(\psq) \right] \label{eq:Gamma}.
\end{equation}
Finally, the propagator matrix $\dmat_{hHA}$ equals, up to the sign, the inverse of $\gmat_{hHA}$,
\begin{equation}
 \dmat_{hHA}(\psq) = - \left[\gmat_{hHA}(\psq) \right]^{-1}. \label{eq:DeltaGamInv}
\end{equation}
Accordingly, the matrix inversion yields the individual propagators 
$\Delta_{ij}\ps$ as the the $(ij)$ elements of the $3\times3$ matrix 
$\dmat_{hHA}(\psq)$,
\begin{equation}
 \dmat_{hHA} = \bpm \Delta_{hh}	&\Delta_{hH}	&\Delta_{hA}\\
		    \Delta_{Hh}	&\Delta_{HH}	&\Delta_{HA}\\
		    \Delta_{Ah}	&\Delta_{AH}	&\Delta_{AA} 
 \epm \label{eq:DeltaMat}.
\end{equation}
The off-diagonal entries (for $i\neq j$) result in:
\begin{equation}
 \Delta_{ij}(\psq) = \frac{\gh{ij}\gh{kk}- \gh{jk}\gh{ki}}{\gh{ii}\gh{jj}\gh{kk} +2\gh{ij}\gh{jk}\gh{ki}-\gh{ii}\gh{jk}^{2} -\gh{jj}\gh{ki}^{2}-\gh{kk}\gh{ij}^{2}} \label{eq:Dij}.
\end{equation}
All 2-point vertex functions $\hat{\Gamma}\ps$ depend on $\psq$ via 
Eq.\,(\ref{eq:IrepVert}). Here we do not write the $\psq$-dependence explicitly 
for the purpose of a simpler notation, but the full $\psq$-dependence is implied 
also below.
The solutions of the diagonal propagators, $\Delta_{ii}$, can be 
expressed in the following compact way: 
\begin{align}
 \Delta_{ii}\ps &= \frac{\gh{jj}\gh{kk}-\gh{jk}^{2}}{-\gh{ii}\gh{jj}\gh{kk} + \gh{ii}\gh{jk}^{2} - 2\gh{ij}\gh{jk}\gh{ki} + \gh{jj}\gh{ki}^{2} +\gh{kk}\gh{ij}^{2}}\label{eq:Dii}\\
&= \frac{i}{\psq -m_i^{2} + \seff\ps},\label{eq:Diieff}
\end{align}
where the effective self-energy is introduced,
\begin{align}
   \seff(p^{2}) &= \hat{\Sigma}_{ii}(p^{2}) - i \frac{2\gh{ij}(p^{2})\gh{jk}(p^{2})\gh{ki}(p^{2}) - \gh{ki}^{2}(\psq)\gh{jj}(\psq)-\gh{ij}^{2}(\psq)\gh{kk}(\psq)}{\gh{jj}(\psq)\gh{kk}(\psq)-\gh{jk}^{2}(\psq)} \label{eq:seff}.
\end{align}
It contains the diagonal self-energy, $\Si$ (which exists already at 1-loop 
order), and the mixing 2-point functions (whose products only contribute 
to $\seff$ from 2-loop order onwards). Hence, replacing the pure self-energy 
$\hat{\Sigma}_{ii}$ by the effective one, $\seff$, includes also the $3\times 3$ 
mixing contributions to the diagonal propagator in Eq.\,(\ref{eq:Diieff}) while 
preserving formally the structure of the propagator as in the unmixed case. In 
the limit of no mixing, the second term in Eq.\,(\ref{eq:seff}) vanishes.

The composition of $\seff$ in terms of the unmixed and the mixing contributions 
in Eq.\,(\ref{eq:seff}) can be 
written in the alternative way~\cite{KarinaPhD,Fowler:2010eba,Fuchs:2015jwa}
\begin{align}
 \seff\ps &= \Si\ps + \frac{\Delta_{ij}\ps}{\Delta_{ii}\ps}\Sij\ps + \frac{\Delta_{ik}\ps}{\Delta_{ii}\ps}\Sik\ps \label{eq:seffratio}
\end{align}
for $i,j,k$ all different. Eq.\,(\ref{eq:seffratio}) represents the sum of
the diagonal and off-diagonal self-energies involving a Higgs boson $i$
where the off-diagonal contributions are weighted by the ratios of the
respective off-diagonal and the diagonal propagators. 
This expression follows from Eq.\,(\ref{eq:seff}) with the help of the
equality
\begin{align}
 \frac{\Delta_{ij}}{\Delta_{ii}} &= -\frac{\gh{ij}\gh{kk}-\gh{jk}\gh{ki}}{\gh{jj}\gh{kk}-\gh{jk}^{2}}
 = -\frac{\Sij\,(D_k+\Sk)-\Sjk \Ski}{(D_j+\Sj)\,(D_k+\Sk)-\Sjk^{2}}
 ,\label{eq:ratioijii}
\end{align}
where the shorthand
\begin{align}
 D_i\ps &= \psq - m_i^{2} \label{eq:Di}
\end{align}
has been used,
analogously for $j\leftrightarrow k$, and $\sig_{ij}=-i\gh{ij}$ from
Eq.\,(\ref{eq:IrepVert}) with $i\neq j$ (for the special case of
$(2\times 2)$ mixing, this relation can be directly read off from 
Eqs.\,(\ref{eq:Dii2x2}-\ref{eq:seff2x2}) given below).

\subsection{On-shell wave function normalisation factors}\label{sect:onshellZ}

\subsubsection{Complex poles}

The complex poles $\Mm_{a}$, where $a = 1, 2, 3$, are determined as
solutions of the equations
\begin{align}
\psq -m_i^{2} + \seff\ps = 0 ,\label{eq:complexpole}
\end{align}
with $i = h, H, A$, see Sect.~\ref{sect:poles} below. 
While the propagators have poles for $\psq=\Mm_a$, it is
interesting to note that the 
ratios of off-diagonal and diagonal propagators 
\begin{equation}
 R_{ij}^{(a)} := \frac{\Delta_{ij}\ps}{\Delta_{ii}\ps}\bigg\rvert_{\psq=\Mm_a},\label{eq:Rija}
\end{equation}
stay finite at the complex poles $\Mm_a,\,~a=1,2,3$.

\subsubsection{\texorpdfstring{$\Zb$}{Z}-matrix for on-shell properties of 
external particles}\label{sect:Zmatrix}
Higgs bosons appearing as \textit{external} particles in a process need the 
appropriate on-shell properties for a correct normalisation of the S-matrix. 
In the hybrid on-shell/$\overline{\rm{DR}}$ renormalisation scheme 
\cite{Frank:2006yh}, the masses are renormalised on-shell, but the 
$\overline{\textrm{DR}}$ renormalisation conditions for the fields and 
$\tan\beta$ \cite{Frank:2006yh},
\begin{eqnarray}
 \delta Z_{\mathcal{H}_1}^{\overline{\rm{DR}}} &=& -\text{Re}\left[\Sigma^{'(\rm{div})}_{HH}(m_{H}^{2})\right]_{\alpha=0}\label{eq:dZH1},\\
 \delta Z_{\mathcal{H}_2}^{\overline{\rm{DR}}} &=& -\text{Re}\left[\Sigma^{'(\rm{div})}_{hh}(m_{h}^{2})\right]_{\alpha=0}\label{eq:dZH2},\\
 \delta \tan\beta^{\overline{\rm{DR}}} &=& \frac{1}{2}(\delta Z_{\mathcal{H}_2}^{\overline{\rm{DR}}} - \delta Z_{\mathcal{H}_1}^{\overline{\rm{DR}}})\label{eq:dtanb},
\end{eqnarray}
where $\alpha$ is the mixing angle between $h$ and $H$, see
Eq.\,(\ref{eq:ncHmix}) below,
do not ensure proper on-shell properties of the Higgs bosons. In fact, the 
loop-corrected mass eigenstates $h_1,h_2,h_3$, which occur as external, 
on-shell, particles e.g.\ in decay processes, are a mixture of the lowest order 
states $h,H,A$. Thus, finite wave function normalisation factors need to be 
introduced in order to guarantee that the mixing vanishes on-shell and 
that the propagators of the external particles have unit residue. 

These so-called $Z$-factors for a neutral Higgs boson $i=h,H,A$ on an external 
line are obtained from the residue of the propagators at the complex pole 
$\Mm_a,\,a=1,2,3$,\,\cite{Dabelstein:1995js,Heinemeyer:2001iy}
\begin{align}
 \Zz_i^{a} &:= \textrm{Res}_{\Mm_a}\left\lbrace\Delta_{ii}\ps\right\rbrace   \label{eq:ZhaRes}.
 \end{align}
Expanding $\seff(\psq)$ around the complex pole $\psq=\Mm_a$, 
one obtains the diagonal propagator
\begin{align}
  \Delta_{ii}(p^{2}) &=\frac{i}{p^{2}-m_i^{2}+\seff(p^{2})}
=\frac{i}{p^{2}-\Mm_a}\cdot
\frac{1}{1+\dseff(\Mm_a)+ \mathcal{O}(\psq - \Mm_a)}
\label{eq:DiiApprdSeff}\,.
\end{align}
Performing the limit $\psq\rightarrow \Mm_a$ yields the residue of Eq.\,(\ref{eq:ZhaRes}),
 \begin{align}
 \Zz_i^{a}
 &= \frac{1}{\frac{\partial}{\partial \psq}\frac{i}{\Delta_{ii}\ps}}\bigg\rvert_{\psq=\Mm_a}
 =\frac{1}{1+\dseff(\Mm_a)} \label{eq:Zha}\,.
\end{align}
Considering a diagram with the Higgs boson $i$ on an external line, whose
propagator has three poles (see Sect.~\ref{sect:poles}),
there are three possibilities which residue to compute. If the amputated
Green's function is evaluated at $\Mm_a$, the external $i$-line has to be
multiplied by $\sqrt{\Zz_i^{a}}$ for the correct S-matrix
normalisation\footnote{In order to avoid sign ambiguities, taking the square
root of a $Z$-factor, which in general has two solutions in the complex
plane, refers here and in the following always to the principal square root, i.e.\,the solution with a non-negative real part.}. So the resulting mass eigenstate as an outgoing particle is $h_a$.
Alternatively, if the Green's function is evaluated
at $\Mm_b$, it has to be normalised by 
\begin{align}
\sqrt{\Zz_i^{b}}=\frac{1}{\sqrt{1+\dsef_{ii}(\Mm_b)}}
\end{align}
to achieve the correct S-matrix element. For this choice, the external mass eigenstate is $h_b$.
Moreover, the wave function normalisation factor for $i-j$ mixing on an
external on-shell line at $\Mm_a$ is composed of the overall normalisation
factor $\sqrt{\Zz_i^{a}}$ times the on-shell transition ratio 
\begin{equation}
 \Zz_{ij}^{a} \equiv R_{ij}^{(a)} = \frac{\Delta_{ij}\ps}{\Delta_{ii}\ps}\bigg\rvert_{\psq=\Mm_a} \label{eq:transij}\,.
\end{equation}
Since $\Delta_{ii}$ and $\Delta_{ij}$ have -- in the case of $3\times 3$
mixing -- 3 complex poles
(see Sect.~\ref{sect:poles}),
any of them can be chosen for the evaluation of
$\Zz_i$, $\Zz_{ij}$ and $\Zz_{ik}$; in the considered example 
this is 
$\Mm_a$. Correspondingly, $\Zz_j, \Zz_{ji}$ and $\Zz_{jk}$ will be evaluated
at $\Mm_b$ and $\Zz_k, \Zz_{ki}$ and $\Zz_{kj}$ at $\Mm_c$ where $a,b,c$ are
a permutation of 1,2,3, and $i,j,k$ a permutation of
$h,H,A$\,\cite{Williams:2007dc,Fuchs:2015jwa}. For $2\times 2$ mixing, 
of course only
the two indices involved in the mixing can be permuted.

All choices allowed by the mixing structure are generally possible because
of the pole structure of each propagator. However, they might not be
equally numerically stable. If the loop-corrected 
mass eigenstate $h_a$ contains only a
small admixture of the lowest-order state
$i$, the propagator still has
a pole at $\Mm_a$, but the contribution of $i$ to $h_a$ is suppressed at
$\psq \neq \Mm_a$.

In order to be definite, it is in any case necessary to define at which pole to evaluate which normalisation and mixing $\Zz$-factor. This choice corresponds to fixing an assignment between the indices $i,j,k$ of the lowest-order states and the indices $a,b,c$ of the higher-order mixed states and then using it consistently. The assignment $(i,a), (j,b), (k,c)$, which we label as scheme $I$, prescribes to evaluate $\Zz_i$, $\Zz_{ij}$ and $\Zz_{ik}$ at $\Mm_a$. Once the indices have been assigned we can clear up the notation by writing
\begin{equation}
 \Zz_a\big\rvert_I := \Zz_{i}^{a},~~ \Zz_{aj}\big\rvert_I := \Zz_{ij}^{a},~~ \Zz_{bi}\big\rvert_I := \Zz_{ji}^{b}\,,\label{eq:Zidentification}
\end{equation}
and accordingly for the other indices such that the first index always
refers to a loop-corrected mass eigenstate ($a,b,c\in\{1,2,3\}$)\footnote{This index notation differs from the conventions in Refs.\,\cite{Frank:2006yh,Hahn:2006np,Williams:2007dc,Fowler:2010eba,Williams:2011bu}.}
and the second index to a lowest-order state ($i,j,k\in\{h,H,A\}$). 
Note that $\Zz_{ai}=\Zz_{bj}=\Zz_{ck}\equiv1$ in the index scheme $I$ defined above. 
Once the index scheme has been specified, one can leave out the subscript $I$. 
For the scheme-independence of physical results, see 
Sect.\,\ref{sect:schemeIndep}.

Furthermore, it is convenient\,\cite{Frank:2006yh} to arrange the products of the normalisation factors $\sqrt{\Zz_a}$ and transition ratios $\Zz_{aj}$ as
\begin{align}
\Zb_{aj} &= \sqrt{\Zz_a}\Zz_{aj}
\label{eq:ZiZij}
\end{align}
(note the difference between $\Zz_{aj}$ and $\Zb_{aj}$) 
into a non-unitary matrix:
\begin{align}
 \hat{\textbf{Z}} &= \bpm
\sqrt{\hat{Z}_1}\Zz_{1h} & \sqrt{\hat{Z}_1}\hat{Z}_{1H} &\sqrt{\hat{Z}_1}\hat{Z}_{1A}\\
\sqrt{\hat{Z}_2}\hat{Z}_{2h} & \sqrt{\hat{Z}_2}\Zz_{2H} &\sqrt{\hat{Z}_2}\hat{Z}_{2A}\\
\sqrt{\hat{Z}_3}\hat{Z}_{3h} & \sqrt{\hat{Z}_3}\hat{Z}_{3H} &\sqrt{\hat{Z}_3}\Zz_{3A}
\epm\label{eq:Zmatrix}.
\end{align}
The $\Zb$-matrix defined in Eq.\,(\ref{eq:Zmatrix}) fulfils the on-shell
conditions
(unit residue and vanishing mixing), which can be 
written in the following compact form\,\cite{Williams:2007dc,KarinaPhD,Fowler:2010eba,Williams:2011bu}:
\begin{align}
 \lim_{\psq\rightarrow \Mm_a} -\frac{i}{\psq-\Mm_a}\left(\Zb\cdot 
\gmat_{hHA}\cdot \Zb^{T} \right)_{hh} ~&= 1
\label{eq:Reshha},\\
 \lim_{\psq\rightarrow \Mm_b} -\frac{i}{\psq-\Mm_b}\left(\Zb\cdot 
\gmat_{hHA}\cdot 
\Zb^{T} \right)_{HH} &= 1
\label{eq:ResHHb},\\
 \lim_{\psq\rightarrow \Mm_c} -\frac{i}{\psq-\Mm_c}\left(\Zb\cdot 
\gmat_{hHA}\cdot \Zb^{T} \right)_{AA} &= 1
\label{eq:ResAAc},
\end{align}
with vanishing off-diagonal entries. 
It is equally possible to begin with these equations 
(\ref{eq:Reshha}-\ref{eq:ResAAc}), i.e.\ to require unit residues and to demand 
vanishing mixing on-shell, to derive the elements of the $\Zb$-matrix 
whose solutions are given in Eqs.\,(\ref{eq:Zha}) and (\ref{eq:transij}).

In Ref.\,\cite{Williams:2007dc} the evaluation at the full complex poles and the 
inclusion of imaginary parts were introduced
(see Eq.\,(\ref{eq:SigijReIm}) below for the treatment of loop functions of
complex arguments). 
The evaluation at the complex poles 
leads to numerically more stable results than 
the evaluation at real $\psq=M_{h_a}^{2}$,
as well as to the physically equivalent choices of index assignments. 
The calculation of the $\Zb$-factors of MSSM Higgs bosons can be performed with the program \texttt{FeynHiggs}.

\subsection{Use of \texorpdfstring{$\Zb$}{Z}-factors for external states}\label{sect:ExtH}
The $\Zb$-factors have been introduced for the correct normalisation of
matrix elements with external (on-shell)
Higgs bosons $h_a,~\psq=\Mm_a$. It should be noted that $\Zb$ does not
provide a unitary transformation between the basis of lowest-order states 
and the basis of loop-corrected mass eigenstates. 
The fact that $\Zb$ is a non-unitary matrix is related to the imaginary
parts appearing in the propagators of unstable particles.
Using the $\Zb$-matrix, one can express the one-particle irreducible (1PI)
vertex functions $\gh {h_a}$ involving a loop-corrected
mass eigenstate $h_1,h_2,h_3$ as an external particle as a linear
combination of the 
%well defined 
1PI vertex functions of the lowest-order states, $\gh i$:
\begin{align}
 \gh{h_a}&=\Zb_{ah}\gh h+\Zb_{aH}\gh H+\Zb_{aA}\gh A + ...\label{eq:Zextsum}\\
 &=\sqrt{\Zz_a}\,\left(\Zz_{ah}\gh h+\Zz_{aH}\gh H+\Zz_{aA}\gh A\right) + \dots \label{eq:Vertexha},
\end{align}
where the ellipsis refers to additional terms arising from the mixing with Goldstone and vector bosons, which are not described by the $\Zb$-matrix.
Thus, the overall normalisation factor $\sqrt{\Zz_a}$ accounts for the 
%unstable 
particle $h_a$ appearing at an external line. In addition, the factors
$\Zz_{ai}$ given in Eqs.\,(\ref{eq:transij}) and (\ref{eq:Zidentification}) as
ratios of propagators at $\psq=\Mm_a$ describe the transition between the
states $h_a$ and $i$. The transition factor $\Zz_{ai}$ occurs in a diagram
where $h_a$ is the external particle, but $i$ directly couples to the
vertex. All possibilities for $i=h,H,A$ need to be included for each $h_a$,
hence the sum arises. This is depicted in Fig.\,\ref{fig:Zext} (cf. also
Refs.\,\cite{Hahn:2007fq,KarinaPhD}).
\begin{figure}[ht!]
\begin{tikzpicture}[line width=0.8 pt, scale=0.6]
	\node[color=black] at (2.5,-0.6) {$\psq=\Mm_a$};
	\draw[color=blue,dashed] (1,0)--(4,0);	
	\node[color=blue] at (2.5,0.6) {$\boldsymbol{h_a}$};
	\fill[black!30!white] (4.3,0) circle (0.3);
	\draw (5.6,1)--(4.6,0.1);
	\draw (5.6,-1)--(4.6,-0.1);
	\node at (5.5,0) {$\hat\Gamma_{{\color{blue} h_a}}$};
	\node at (7.2,0) {= $\sqrt{\Zz_{{\color{blue} a}}}\Bigg($};
\begin{scope}[shift={(7.5,0)}]
	\node[color=black] at (2.5,-0.8) {$\Zz_{{\color{blue}a}{\color{OliveGreen}h } }$};
	\draw[color=blue,dashed] (1,0)--(2.2,0);	
	\draw[color=OliveGreen,dashed] (2.8,0)--(4,0);	
	\node[color=blue] at (1.6,0.6) {$\boldsymbol{h_a}$};
	\node[color=OliveGreen] at (3.4,0.6) {$\boldsymbol{h}$};
	\fill[black!30!white] (2.5,0) circle (0.3);
	\draw (5,1)--(4,0)--(5,-1);
	\node at (5,0) {$\hat\Gamma_{{\color{OliveGreen} h}}$};
	\node at (6,0) {+};	 
\end{scope}
\begin{scope}[shift={(13,0)}]
	\node[color=black] at (2.5,-0.8) {$\Zz_{{\color{blue}a}{\color{OliveGreen}H } }$};
	\draw[color=blue,dashed] (1,0)--(2.2,0);	
	\draw[color=OliveGreen,dashed] (2.8,0)--(4,0);	
	\node[color=blue] at (1.6,0.6) {$\boldsymbol{h_a}$};
	\node[color=OliveGreen] at (3.4,0.6) {$\boldsymbol{H}$};
	\fill[black!30!white] (2.5,0) circle (0.3);
	\draw (5,1)--(4,0)--(5,-1);
	\node at (5,0) {$\hat\Gamma_{{\color{OliveGreen} H}}$};
	\node at (6,0) {+};	 
\end{scope}
\begin{scope}[shift={(18.5,0)}]
	\node[color=black] at (2.5,-0.8) {$\Zz_{{\color{blue}a}{\color{OliveGreen}A } }$};
	\draw[color=blue,dashed] (1,0)--(2.2,0);	
	\draw[color=OliveGreen,dashed] (2.8,0)--(4,0);	
	\node[color=blue] at (1.6,0.6) {$\boldsymbol{h_a}$};
	\node[color=OliveGreen] at (3.4,0.6) {$\boldsymbol{A}$};
	\fill[black!30!white] (2.5,0) circle (0.3);
	\draw (5,1)--(4,0)--(5,-1);
	\node at (5,0) {$\hat\Gamma_{{\color{OliveGreen} A}}$};
\end{scope}
	\node at (25,0) {$\Bigg)_{\psq=\Mm_a}$};
	\node at (26.4,0) {$+\dots$};
\end{tikzpicture}
	\caption[$\Zb$-factors for external Higgs bosons.]{$\Zb$-factors for
external Higgs bosons: The vertex function $\hat\Gamma_{h_a}$ is constructed
from vertex functions $\hat\Gamma_i,\,i=h,H,A$, the transition factors
$\Zz_{ai}$ and the overall normalisation factor $\sqrt{\Zz_a}$. The ellipsis
refers to contributions from mixing with
Goldstone and gauge bosons.}
	\label{fig:Zext}
\end{figure}
Conveniently, Eq.\,(\ref{eq:Vertexha}) can be written in matrix form for all $h_1,h_2,h_3$ as
\begin{equation}
 \bpm \gh {h_1}\\\gh {h_2}\\\gh {h_3}\epm = \Zb\cdot \bpm \gh {h}\\\gh {H}\\\gh {A}\epm +\dots\label{eq:vertZ}~.
\end{equation}
In this way, propagator corrections at external legs are effectively absorbed into the vertices of neutral Higgs bosons. 
If $\Zb$-factors are applied to supplement the Born result, only non-Higgs propagator type corrections (such as mixing with the Goldstone and $Z$-bosons) as well as vertex, box and real corrections need to be calculated individually.

As it is the case for the usual (UV-divergent) on-shell field renormalisation 
constants, the finite wave function normalisation factors are in general
gauge-dependent quantities. While the gauge-parameter dependence drops out at a
fixed order in perturbation theory, the inclusion of higher-order
contributions into the $\Zb$-factors described above can give rise to a
residual gauge-parameter dependence, which is formally of sub-leading higher
order.
This is caused by the fact that the 
$\Zb$-factors contain not only gauge-parameter independent leading
higher-order contributions but also sub-leading higher-order contributions
that are in general gauge-parameter dependent. Those 
sub-leading higher-order contributions would cancel with corresponding
sub-leading higher-order contributions from other Green functions such as
vertex and box contributions.

\subsection{Effective couplings}
\label{sect:Ueff}
Since the $\Zb$-matrix is not unitary, it does not represent a unitary transformation between the $\{h,H,A\}$ and the $\{h_1,h_2,h_3\}$ basis. 
However, it is not necessary to diagonalise the mass matrix for the determination of the poles of the propagators. Hence there is a priori no need to introduce a unitary transformation. 
Though, if a unitary matrix $\Ub$ is desired for the definition of effective couplings, an approximation of the momentum dependence of $\Zb$ is required. 
There is no unique prescription of how to achieve a unitary mixing matrix as
an approximation of the  $\Zb$-matrix, but a possible choice is the $\psq=0$
approximation\,\cite{Frank:2006yh,Hahn:2007fq}. As in the effective
potential approach, the external momentum $\psq$ is set to zero in the
renormalised self-energies $\Sij(\psq)\rightarrow \Sij(0)$ so that they
become real. Then $\Ub$ diagonalises the real matrix $\mmat(0)$.
$\Ub$ can be chosen real and it transforms the $\CP$-eigenstates into the mass eigenstates,
\begin{align}
 \bpm h_1\\ h_2\\ h_3 \epm
 = \Ub \bpm h\\ H\\ A \epm ,~~~~
 \Ub = \bpm \Ub_{1h}& \Ub_{1H}& \Ub_{1A}\\
	    \Ub_{2h}& \Ub_{2H}& \Ub_{2A}\\
	    \Ub_{3h}& \Ub_{3H}& \Ub_{3A}
	\epm
 \label{eq:Urotate},
\end{align}
so that $U_{aA}^{2}$ quantifies the admixture of a $\CP$-odd component inside $h_a$\,\cite{Frank:2006yh}. 
The elements of $\Ub$ can then be used to introduce effective couplings of the loop-corrected states $h_a$ to any other particles $X$ in terms of the couplings of the unmixed states $i$ by the relation
\begin{equation}
 C_{h_a X}^{U} = \sum_{i=h,H,A} \Ub_{ai} C_{i X} \label{eq:EffCoup}.
\end{equation}
absorbing some higher-order corrections, but neglecting imaginary parts and the 
full momentum dependence of the self-energies. Hence, the application of $\Ub$ 
resembles the use of $\Zb$-factors in Eq.\,(\ref{eq:Zextsum}) for the purpose 
of implementing partial higher-order effects into an improved Born result. 
Yet, the rotation matrix $\Ub$ introduced for effective couplings as a unitary 
approximation is conceptually different from the $\Zb$-matrix arising from 
propagator corrections and introduced for the correct normalisation of the 
$S$-matrix. 

\subsection{Technical treatment of imaginary parts}

In order to account for complex momenta and imaginary parts of 
self-energies, we employ an expansion of
the self-energies around the real part of the complex 
momentum,
\begin{align}
 \psq &\equiv \psq_r + i \psq_i \label{eq:psqReIm},\\
 \Sij(\psq) &\simeq \Sij(\psq_r) + i \psq_i \dsig_{ij}(\psq_r)\label{eq:SigijReIm},
\end{align}
where $\dsig_{ij}\ps\equiv \frac{d\sig_{ij}\ps}{d\psq}$.
This expansion enables the calculation of self-energies at complex momentum in terms of self-energies evaluated at real momentum (e.g. by \texttt{FeynHiggs}).
For the inclusion of all products of real and imaginary parts, we do not
expand the effective self-energy from Eq.\,(\ref{eq:seff}) directly
according to Eq.\,(\ref{eq:psqReIm}). Instead, in the same way as in
Refs.\,\cite{KarinaPhD,Fowler:2010eba,Fuchs:2015jwa}, we expand all
$\hat{\Gamma}_{ij}\ps$ 
in Eq.\,(\ref{eq:seff})
individually before combining them into $\seff$.

\section{Pole structure of propagator matrices}
\label{sect:poles}

In this section, we will first discuss in detail the solutions of the 
poles of the propagator matrix depending on the mixing. Subsequently we will 
prove the equivalence of different assignments between the lowest-order
states and the loop-corrected
mass eigenstates for the physical $\Zb$-matrix.

\subsection{Determination of the poles}

Imaginary parts of the self-energies lead to propagator poles in the 
complex momentum plane. The Higgs masses are determined from the real part of 
the complex poles $\Mm$ of the diagonal propagators. 
Equivalently, the complex poles are obtained as the 
zeros of the inverse diagonal propagators. 
%%%% basic linear algebra relations, no physics input needed: %%%%
Due to 
$\boldsymbol{A}^{-1}\propto \det\left[\boldsymbol{A}\right]^{-1}$,
$\det\left[\boldsymbol{A}^{-1}\right]=\det\left[\boldsymbol{A}\right]^{-1}$ and
$\det\left[-\boldsymbol{A}\right]=(-1)^n\,\det\left[\boldsymbol{A}\right]$ 
for any $n\times n$-matrix $\boldsymbol{A}$, finding the roots of 
$\dmat^{-1}\ps$ is equivalent to solving
%%%% application to the physical matrices Delta = -Gamma^{-1}  %%%
\begin{align}
-\frac{1}{\det\left[\dmat_{hHA}\ps\right]}
&=\det\left[\gmat_{hHA}\ps\right]   \stackrel{!}{=} 0,\label{eq:det0}
 \end{align}
because
of the relation between $\dmat_{hHA}$ and 
$\gmat_{hHA}$ from Eq.\,(\ref{eq:DeltaGamInv})\footnote{Explicitly, all roots 
of $\dmat_{hHA}\ps^{-1}$ are roots of $\det\left[\dmat_{hHA}\ps\right]^{-1} 
\equiv \det\left[-\gmat_{hHA}\ps^{-1}\right]^{-1} = 
-\det\left[\gmat_{hHA}\ps\right]$.}. 
Then the loop-corrected masses $M$ are obtained from the real parts of the complex poles and the total widths $\Gamma$ from the imaginary parts via
\begin{equation}
 \Mm = M^{2} - i M\Gamma. \label{eq:MGamma}
\end{equation}
In the following, the impact of higher-order and mixing contributions on the pole structure of the propagators will be discussed.
 
\paragraph{Lowest order}
At lowest order, the self-energy contributions in Eq.\,(\ref{eq:IrepVert}) are absent and the matrix $\gmat$ simply reads
\begin{align}
 \gmat^{(0)}_{hHA}\ps &= i\,\textrm{diag}\left\lbrace D_h\ps, D_H\ps,
D_A\ps\right\rbrace \label{eq:GamLowestOrder},
\end{align}
where the shorthand of Eq.\,(\ref{eq:Di}) has been used.
The solutions of Eq.\,(\ref{eq:det0}) are the three tree-level masses $m_i^{2}$.

\paragraph{Higher order without mixing}
Beyond tree-level, the self-energies are added at the available order. Restricting them to the unmixed case, $\sig_{ij}=0$ for $i\neq j$, leads to
\begin{align}
 \gmat^{(\textrm{no mix})}_{hHA}\ps &= i\,\textrm{diag}\left\lbrace D_h\ps + \sig_{hh}\ps, D_H\ps+\sig_{HH}\ps, D_A\ps+\sig_{AA}\ps\right\rbrace \label{eq:GamNoMix},
\end{align}
so that 
\begin{equation}
\det\left[\gmat^{(\textrm{no mix})}_{hHA}\ps\right]=\prod_{i=h,H,A}\left(D_i\ps+\sig_{ii}\ps\right)=0 \label{eq:DetUnmix}
\end{equation}
is achieved if $\psq$ fulfils the following on-shell relation
\begin{equation}
 \psq - m_i^{2} + \sig_{ii}\ps = 0 \label{eq:0unmixed}
\end{equation}
for any $i=h,H,A$. Thus, the full propagator matrix $\dmat$ has three poles
and each propagator $\Delta_{ii}\ps$ has exactly one pole $\psq=\Mm_i$ that
solves Eq.\,(\ref{eq:0unmixed}) in this case.

\paragraph{Higher order with \texorpdfstring{$2\times2$}{2x2} mixing}
If now the mixing between $h$ and $H$ is taken into account, 
corresponding to the $\cp$-conserving case,
the matrix $\gmat$ 
becomes block-diagonal with the $2\times 2$ matrix $\gmat_{hH}$ and the 2-point 
vertex function of $A$, which does not mix with the other states:
\begin{align}
 \gmat_{hHA}\ps &= \bpm \gmat_{hH}\ps 	&0\\
			0		&\hat{\Gamma}_A\ps 
 \epm, \label{eq:blockdiag}\\
 \det\left[\gmat_{hHA}\ps\right] &= \det\left[\gmat_{hH}\ps\right]\cdot\hat\Gamma_{A}\ps\,.
\end{align}
For a closer look at the relation between the roots of the determinant and the roots of the inverse propagator, we write down the propagators and the effective self-energy of the $\left\lbrace h,H\right\rbrace$ system explicitly. They follow from Eqs.(\ref{eq:Dij}), (\ref{eq:Dii}) and (\ref{eq:seff}) by setting $\sig_{hA}=\sig_{HA}=0$ or equivalently from the inversion of the $2\times 2$ submatrix $\gmat_{hH}$: 
\begin{align}
\Delta_{ii}(\psq) &= \frac{i\left[D_j\ps + \Sj(\psq)\right]}{\left[D_i\ps+\Si(\psq)\right] \left[D_j\ps+\Sj(\psq)\right]-\Sij^{2}(\psq)}
=\frac{i}{\psq-m_i^{2}+\seff(\psq)},\label{eq:Dii2x2}\\
 \Delta_{ij}(\psq) 
&= \frac{-i\Sij(\psq)}{\left[D_i\ps+\Si(\psq)\right]\,\left[D_j\ps+\Sj(\psq)\right]-\Sij^{2}(\psq)}\label{2x2eq:Dij},\\
\seff(\psq) &= \Si(\psq) -\frac{\Sij^{2}(\psq)}{D_j(\psq)+\Sj(\psq)}\label{eq:seff2x2}.
\end{align}
Comparing the inverse diagonal propagators with the determinant of the submatrix $\gmat_{hH}$, we find for $i,j\in\left\lbrace h,H \right\rbrace,~i\neq j,$
\begin{align}
 \frac{1}{\Delta_{ii}\ps} = \frac{i}{D_j\ps+\sig_{jj}\ps}\,\det\left[\gmat_{hH}\ps\right]. \label{eq:InvPropDet2x2}
\end{align}
Eq.\,(\ref{eq:InvPropDet2x2}) reveals that both inverse diagonal propagators, $1/\Delta_{hh}$ and $1/\Delta_{HH}$, are proportional to the determinant of $\gmat_{hH}$, which has two zeros. 
As opposed to the unmixed case, both zeros of
$\det\left[\gmat_{hH}\ps\right]$ are poles of \textit{each} of the diagonal
propagators $\Delta_{hh},\Delta_{HH}$. The lowest-order states
$h$ and $H$ are mixed into the loop-corrected 
mass eigenstates $h_1$ and $h_2$
with the loop-corrected masses $M_{h_1}, M_{h_2}$. The corresponding poles
$\psq=\Mm_{h_1}, \Mm_{h_2}$ solve
\begin{equation}
 \psq - m_i^{2} + \hat{\Sigma}^{\textrm{eff}}_{ii}(\psq) = 0 \label{eq:0mixed}
\end{equation}
for $\psq=\Mm_{h_a}$ in any combination of $i=h,H$ and $a=1,2$, where the 
effective self-energy is given in Eq.\,(\ref{eq:seff2x2}). In the $2\times 2$ 
mixing system, it is convenient to choose $M_{h_1}\leq M_{h_2}$. As for the 
nomenclature in the $2\times 2$ case, the lighter mass eigenstate $h_1$ is often 
denoted as $h$ and the heavier one as $H$ because both are $\CP$-even states. 
It should be noted that both roots of $\gmat_{hH}$, $\Mm_{h_1}$ and $\Mm_{h_2}$, are also 
complex poles of the off-diagonal propagators 
$\Delta_{hH}\ps\equiv\Delta_{Hh}\ps$ due to
\begin{align}
  \frac{1}{\Delta_{ij}\ps} = \frac{-i}{\sig_{ij}\ps}\,\det\left[\gmat_{hH}\ps\right]. \label{eq:InvPropijDet2x2}
\end{align}
Since in this case $A$ does not mix with $h$ and $H$, the third pole $\Mm_A$ 
solely solves
\begin{equation}
 \Mm_{A} - m_A^{2} + \sig_{AA}(\Mm_{A}) = 0 \label{eq:solveA},
\end{equation}
but no other combination of $A$ and $h_a$ satisfies the on-shell condition. 
$M_A$ is the loop-corrected mass of the (mass and interaction) eigenstate 
$A$. 

\paragraph{Higher order with \texorpdfstring{$3\times3$}{3x3} mixing}
Now we turn to the 
case where complex MSSM parameters lead to $\CP$-violating self-energies $\sig_{hA}, \sig_{HA}$. Thus, all three neutral Higgs 
lowest-order and $\CP$-eigenstates $h,H,A$ mix into the loop-corrected mass eigenstates $h_1, h_2, h_3$, which have no longer well-defined $\CP$ quantum numbers, but are admixtures of $\CP$-even and $\CP$-odd components. In this framework, $\gmat_{hHA}$ is a full $3\times 3$ matrix with the determinant
\begin{align}
 \det[\gmat_{hHA}] &= -i\left[(D_{h}+\sig_{hh})(D_{H}+\sig_{HH})(D_{A}+\sig_{AA})+2\sig_{hH}\sig_{HA}\sig{hA}\right.\nonumber\\
 &\left.~~~~~~~~-(D_{h}+\sig_{hh})\sig_{HA}^2-(D_{H}+\sig_{HH})\sig_{hA}^2-(D_{A}+\sig_{AA})\sig_{hH}^2 \right] \label{eq:Det3mix},
\end{align}
where we dropped the explicit $\psq$-dependence of each term for an ease of 
notation. Comparing Eq.\,(\ref{eq:Det3mix}) with the diagonal and off-diagonal 
propagators from Eqs.\,(\ref{eq:Dii}) and (\ref{eq:Dij}), respectively, we see 
that their inverse is proportional to the determinant of $\gmat_{hHA}$:
\begin{align}
 \frac{1}{\Delta_{ii}} &= \frac{\det\left[\gmat_{hHA}\right]}{(D_{j}+\sig_{jj})(D_{k}+\sig_{kk})-\sig_{jk}^{2}} \label{eq:InvPropDet},\\
 \frac{1}{\Delta_{ij}} &= \frac{\det\left[\gmat_{hHA}\right]}{\sig_{jk}\sig_{ki}-\sig_{ij}(D_k+\sig_{kk}) } \label{eq:InvPropijDet}\,,
\end{align}
where $i\neq j\neq k\neq i$ and without summation over the indices.
From Eq.\,(\ref{eq:InvPropDet}) we conclude that all three roots 
$\psq=\Mm_{h_a},\,a=1,2,3,$ of $\det\left[\gmat_{hHA}\ps\right]$ are complex 
poles of \textit{each} of the three diagonal propagators 
$\Delta_{ii},\,i=h,H,A$. This means that
\begin{equation}
   \Mm_{h_a} - m_i^{2}+\hat{\Sigma}^{\rm{eff}}_{ii}(\Mm_{h_a}) =0\label{eq:SolvePole}
\end{equation}
holds for any combination of $i$ and $a$ in the presence of $3\times 3$ mixing. 
Moreover, Eq.\,(\ref{eq:InvPropijDet}) implies that also the off-diagonal 
propagators have as many poles as the determinant has zeros, namely three in 
the case of $\CP$-violating mixing. 
In the unmixed case, $\seff = \Si$ and each propagator has exactly one pole so 
that there is a unique mapping between $i$ and $a$, see 
Eq.\,(\ref{eq:0unmixed}). On the other hand, for the general mixing case it is 
not unique how to relate the mass eigenstates to the interaction eigenstates. An 
assignment will be needed for the definition of on-shell wave function 
normalisation factors in Sect.\,\ref{sect:onshellZ}.

\subsection{Index scheme independence of the \texorpdfstring{$\Zb$}{Z}-matrix}
\label{sect:schemeIndep}
As discussed above, the pole structure of the full propagators 
provides the freedom at which pole to evaluate which $\Zz$-factor, or
equivalently which loop-corrected 
mass eigenstate index ($a,b,c$) to assign to a lowest-order mass
eigenstate index ($i,j,k$). We denote two choices for such index schemes by
\begin{align}
 I &\leftrightarrow  (i,a), (j,b), (k,c)\label{eq:schemeI},\\
II &\leftrightarrow  (j,a), (i,b), (k,c)\label{eq:schemeII}.
\end{align}
This initial ambiguity, however, results in physically equivalent results. 
Using properties of ratios of diagonal and off-diagonal propagators and exploiting relations at a complex pole (for details, see Ref.\,\cite{Fuchs:2015jwa}),
we are able to show that
\begin{align}
 \frac{1+\dsef_{jj}(\Mm_a)}{1+\dsef_{ii}(\Mm_a)}
=\left(\frac{\Delta_{ji}\ps}{\Delta_{jj}\ps}\right)^{2}_{\psq=\Mm_a}.
\end{align}
This equality provides a transformation between scheme $I$ (where $i$ and $a$ are associated, hence $\Zz_{ai}\equiv1$) and scheme $II$ (where $j$ and $a$ are matched):
\begin{align}
 \Zb_{ai}|_{I} = \left(\sqrt{\Zz_a}\Zz_{ai}\right)_{I} &= \frac{1}{\sqrt{1+\dseff(\Mm_a)}}\nonumber\\
 &= \frac{1}{\sqrt{1+\dsef_{jj}(\Mm_a)}}\frac{\Delta_{ji}}{\Delta_{jj}}\bigg\rvert_{\psq=\Mm_a} 
 = \left(\sqrt{\Zz_a} \Zz_{ai} \right)_{II} = \Zb_{ai}\bigg\rvert_{II}\label{eq:TransformScheme}.
 \end{align}
While the values of $\Zz_a$ and $\Zz_{ai}$ depend on the choice of the index 
mapping, Eq.\,(\ref{eq:TransformScheme}) ensures that the elements of the 
$\Zb$-matrix, which appear in physical processes, are invariant under the 
choice of $a,b,c$ as a permutation of ${1,2,3}$. We also tested this relation 
numerically for various parameter points and always found agreement within the 
numerical precision.

% 
% %%%%%%%%%%%%%%%%%%%%%%%%%%%%%%%%%%%%%%%%%%%%%%%%%%%%%%%%%%%%%%%%%%%%%%%%%%%%%
\section{Breit--Wigner approximation of the full propagators}\label{sect:fullBWZ}
% %%%%%%%%%%%%%%%%%%%%%%%%%%%%%%%%%%%%%%%%%%%%%%%%%%%%%%%%%%%%%%%%%%%%%%%%%%%%% 
The propagator matrix depends on the momentum $\psq$ in a twofold way. On the one hand, the tree-level
propagator factors $D_i\ps=\psq-m_i^{2}$ give rise to an explicit $\psq$-dependence. 
On the other hand, the self-energies $\Sij\ps$ depend on the momentum as well, but away from thresholds their $\psq$ dependence is not particularly pronounced. In this chapter, we will develop an approximation of the full mixing propagators with the aim to maintain the leading momentum dependence, but to 
greatly simplify the mixing contributions by making use of the on-shell $\Zb$-factors.

\subsection{Unstable particles and the total decay width}
In the context of determining complex poles of propagators, we now briefly discuss resonances and unstable particles, see e.g.\
Refs.~\cite{Wackeroth:1996hz,Freitas:2002ja,Grunewald:2000ju,Argyres:1995ym,Beenakker:1996kn}. 
Stable particles 
are associated with a
real pole of the S-matrix, 
whereas the self-energies of unstable particles develop an imaginary part,
so that the pole of the propagator is located within the
complex momentum plane off the real axis. 
For a single pole, 
the scattering amplitude $\mathcal{A}$ can 
be schematically written near the complex pole $\Mm_a$ 
in a gauge-invariant way as 
\begin{equation}
 \mathcal{A}(s) = \frac{R}{s-\Mm_a}+F(s) \label{eq:gaugeinv},
\end{equation}
where $s$ is the squared centre-of-mass energy, $R$ denotes the residue,
while $F$ represents non-resonant contributions.
The mass $M_{h_a}$ of the
unstable particle $h_a$ 
is obtained from the real part of the complex pole
$\Mm_a=M_{h_a}^{2}-iM_{h_a}\Gamma_{h_a}$, 
while the imaginary part gives rise to the total width. Accordingly, the
expansion around the complex pole $\Mm_a$ leads to a Breit--Wigner propagator with a
constant width,
\begin{equation}
\BW_a\ps:=\frac{i}{\psq - \Mm_{a}} = \frac{i}{\psq - M_{h_a}^2 + i M_{h_a}\Gamma_{h_a}} . \label{eq:BWdef} 
\end{equation}
In the following, we will use a Breit--Wigner propagator of this form
to describe the contribution
of the unstable scalar $h_a$ with mass $M_{h_a}$ and total width $\Gamma_{h_a}$ in the
resonance region.

\subsection{Expansion of the full propagators around the complex poles}

Eqs.\,(\ref{eq:InvPropDet}) and (\ref{eq:InvPropijDet}) imply for $3\times
3$ mixing that each propagator $\Delta_{ii}, \Delta_{ij}$ has a pole at
$\Mm_1,\Mm_2$ and $\Mm_3$. Because of this structure, an expansion of the
full propagators near one single pole is not expected to yield a sufficient
approximation. Instead, we will perform an expansion of the full propagators 
around all of their complex poles. The final expression obtained from combining
the contributions from the different poles will constitute a main result of
the present paper.

\subsubsection{Expansion of the diagonal propagators}
We begin with an expansion of $\Delta_{ii}\ps$ in the vicinity of $\Mm_a$ as in Eq.\,(\ref{eq:DiiApprdSeff}), where the first factor equals the definition of the Breit--Wigner propagator of the state $h_a$, and the second factor corresponds to $\Zz_a$ in scheme $I$ where $i$ and $a$ are associated indices. On top of that, $\Zz_a\big\rvert_I=\Zb_{ai}^{2}$ 
as defined in Eq.\,(\ref{eq:ZiZij}), and the elements of the $\Zb$-matrix are independent of the index scheme (see Eq.\,(\ref{eq:TransformScheme})). Thus, the following scheme-independent approximation holds for $\psq\simeq \Mm_a$:
\begin{align}
 \Delta_{ii}\ps &\simeq \BW_a\ps\,\Zz_a\big\rvert_I = \BW_a\ps\,\Zb_{ai}^{2}
\label{eq:DeltaiiBWa}.
\end{align}
In this approach, the mixing contributions are summarised in the on-shell $Z$-factor evaluated at $\Mm_a$. In contrast, the leading momentum dependence is contained in the Breit--Wigner propagator parametrised by the loop-corrected mass $M_{h_a}$ and the total width $\Gamma_{h_a}$ from the complex pole.
In addition, $\Delta_{ii}(\psq)$ has a second pole at $\Mm_b$ because 
$\psq-m_i^{2}+\seff(p^{2})=0$
holds also at $\psq=\Mm_b$. Analogously, we can expand $\seff$ around $\Mm_b$ and obtain for the diagonal propagator
\begin{align}
  \Delta_{ii}(p^{2}) &=\frac{i}{p^{2}-m_i^{2}+\seff(p^{2})}\\
&\simeq \frac{i}{p^{2}-m_i^{2} + \seff(\Mm_b) +(p^{2}-\Mm_b)\cdot \dseff(\Mm_b)}\\
&=\frac{i}{(p^{2}-\Mm_b)\cdot\left[1+\dseff(\Mm_b)\right]}\label{eq:DeltaiiMb}.
\end{align}
Formally, $\frac{1}{1+\dseff(\Mm_b)}$ has the structure of the definition of a $\Zz$-factor from Eq.\,(\ref{eq:Zha}), but in the index scheme $II$ where $b$ is assigned to $i$, whereas Eq.\,(\ref{eq:DeltaiiBWa}) has been obtained in scheme $I$ with the $(i,a)$ assignment. Using the relation (\ref{eq:TransformScheme}), we can rewrite Eq.\,(\ref{eq:DeltaiiMb}) as
\begin{align}
 \Delta_{ii}(\psq) &\simeq \BW_b(\psq) \cdot \frac{1}{1+\dseff(\Mm_b)}\\
 &= \BW_b(\psq)\cdot \frac{1}{1+\dsef_{jj}(\Mm_b)} \left(\frac{\Delta_{ji}}{\Delta_{jj}}\right)^{2}_{\psq=\Mm_b} \label{eq:BWbinDeltaii}\\
 &= \BW_b(\psq)\cdot \left(\Zz_b\, \Zz_{bi}^{2}\right)\bigg\rvert_{I}\label{eq:BWbSchemeia}\\
 &= \BW_b(\psq)\cdot \Zb_{bi}^{2}%\bigg\rvert_{(i,a),(j,b)}
 \label{eq:BWbZbi},
\end{align}
where the $\Zz$-factors in Eq.\,(\ref{eq:BWbSchemeia}) are expressed in the same scheme as in Eq.\,(\ref{eq:DeltaiiBWa}).
Hence, in the vicinity of $\psq\simeq \Mm_b$, the diagonal propagator $\Delta_{ii}$ can be approximated by the Breit--Wigner propagator of $h_b$ weighted by the square of $\Zb_{bi}$ that ensures the coupling to the incoming fields as Higgs boson $i$, propagation as the mass eigenstate $h_b$ and the coupling to the outgoing fields again as Higgs boson $i$. In the same manner, $\Delta_{ii}$ can be expanded around the third complex pole, $\Mm_c$, yielding
\begin{align}
 \Delta_{ii}(\psq) &\simeq \frac{i}{(p^{2}-\Mm_c)\cdot\left[1+\dseff(\Mm_c)\right]}\label{eq:DeltaiiMc}\\
 &\simeq \BW_c(\psq)\cdot \frac{1}{1+\dsef_{kk}(\Mm_c)} \left(\frac{\Delta_{ki}}{\Delta_{kk}}\right)^{2}\bigg\rvert_{\psq=\Mm_c} \label{eq:BWcinDeltaii}\\
 &= \BW_c(\psq)\cdot \Zb_{ci}^{2}\label{eq:BWcZci}.
\end{align}
Thus, close to one of the complex poles (e.g.\ $\Mm_a$), the dominant contribution to the full propagator $\Delta_{ii}$ can be approximated by the corresponding Breit--Wigner propagator ($\BW_a$) multiplied by the square of the respective $\Zb$-factor, ($\Zb_{ai}^{2}$). However, close-by poles may cause overlapping resonances. In order to include this possibility and to extend the range of validity of the Breit--Wigner approximation to a more general case, we take the sum of all three Breit--Wigner contributions into account:
\begin{align}
 \Delta_{ii}(\psq)\simeq \BW_a(\psq) \,\Zb_{ai}^{2} + \BW_b(\psq) \,\Zb_{bi}^{2} + \BW_c(\psq) \,\Zb_{ci}^{2} 
 = \sum_{a=1}^{3}\BW_a(\psq)\,\Zb_{ai}^{2}
 \label{eq:iiBWsum}.
\end{align}

\subsubsection{Expansion of the off-diagonal propagators}
\label{sec:ExpansionOffD}
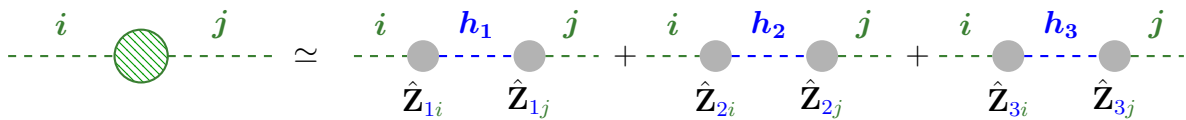
\begin{figure}[ht!]
 \begin{center}
 \begin{tikzpicture}[line width=0.8 pt, scale=0.7]
	\draw[color=OliveGreen,dashed] (0,0)--(2,0);	
	\node[color=OliveGreen] at (1,0.6) {$\boldsymbol{i}$};
	\draw[pattern=north west lines, pattern color=OliveGreen, draw=OliveGreen] (2.5,0) circle (0.5);
	\draw[color=OliveGreen,dashed] (3,0)--(5,0);	
	\node[color=OliveGreen] at (4,0.6) {$\boldsymbol{j}$};
	\node[color=black] at (5.6,0) {$\simeq$};
\begin{scope}[shift={(6.5,0)}]
	\draw[color=OliveGreen,dashed] (0,0)--(1,0);	
	\node[color=OliveGreen] at (0.5,0.6) {$\boldsymbol{i}$};
	\fill[black!30!white] (1.3,0) circle (0.3);
	\draw[color=blue,dashed] (1.6,0)--(3,0);
	\node[color=blue] at (2.3,0.6) {$\boldsymbol{h_1}$};
	\fill[black!30!white] (3.3,0) circle (0.3);
	\draw[color=OliveGreen,dashed] (3.6,0)--(4.6,0);	
	\node[color=OliveGreen] at (4.1,0.6) {$\boldsymbol{j}$};
	\node[color=black] at (5.1,0) {$+$};
	\node[color=black] at (1.3,-0.8) {$\Zb_{{\color{blue}1}{\color{OliveGreen}i } }$};
	\node[color=black] at (3.3,-0.8) {$\Zb_{{\color{blue}1}{\color{OliveGreen}j } }$};
\end{scope}
\begin{scope}[shift={(12,0)}]
	\draw[color=OliveGreen,dashed] (0,0)--(1,0);	
	\node[color=OliveGreen] at (0.5,0.6) {$\boldsymbol{i}$};
	\fill[black!30!white] (1.3,0) circle (0.3);
	\draw[color=blue,dashed] (1.6,0)--(3,0);
	\node[color=blue] at (2.3,0.6) {$\boldsymbol{h_2}$};
	\fill[black!30!white] (3.3,0) circle (0.3);
	\draw[color=OliveGreen,dashed] (3.6,0)--(4.6,0);	
	\node[color=OliveGreen] at (4.1,0.6) {$\boldsymbol{j}$};
	\node[color=black] at (5.1,0) {$+$};
	\node[color=black] at (1.3,-0.8) {$\Zb_{{\color{blue}2}{\color{OliveGreen}i } }$};
	\node[color=black] at (3.3,-0.8) {$\Zb_{{\color{blue}2}{\color{OliveGreen}j } }$};
\end{scope}
\begin{scope}[shift={(17.5,0)}]
	\draw[color=OliveGreen,dashed] (0,0)--(1,0);	
	\node[color=OliveGreen] at (0.5,0.6) {$\boldsymbol{i}$};
	\fill[black!30!white] (1.3,0) circle (0.3);
	\draw[color=blue,dashed] (1.6,0)--(3,0);
	\node[color=blue] at (2.3,0.6) {$\boldsymbol{h_3}$};
	\fill[black!30!white] (3.3,0) circle (0.3);
	\draw[color=OliveGreen,dashed] (3.6,0)--(4.6,0);	
	\node[color=OliveGreen] at (4.1,0.6) {$\boldsymbol{j}$};
	\node[color=black] at (1.3,-0.8) {$\Zb_{{\color{blue}3}{\color{OliveGreen}i } }$};
	\node[color=black] at (3.3,-0.8) {$\Zb_{{\color{blue}3}{\color{OliveGreen}j } }$};
\end{scope}
\end{tikzpicture}
\caption[Diagrammatic illustration of the approximation of the full mixing Higgs propagators in terms of the Breit--Wigner propagators with $\Zb$-factors.]{Diagrammatic illustration of the full mixing Higgs propagators compared to the Breit--Wigner propagators where the $\Zb$-factors encode the transition between the interaction and the mass eigenstates.}  
\label{fig:fullvsBWZdiagram}
 \end{center}
\end{figure}

We proceed similarly for the off-diagonal propagators, which also have three complex poles so that we can expand the propagators around them. 
Note that $\Zb_{ai}=\sqrt{\Zz_a}$ and $\Zb_{aj}=\sqrt{\Zz_a}\Zz_{aj}$ as defined in Eq.\,(\ref{eq:ZiZij}).
Starting at $\psq\simeq\Mm_a$, we express the $\Zz$-factors in scheme $I$,
\begin{align}
 \Delta_{ij}\ps &= \frac{\Delta_{ij}\ps}{\Delta_{ii}\ps} \Delta_{ii}\ps
 \simeq \Zz_{aj} \Zb_{ai}^{2}\,\BW_a\ps
 = \Zb_{aj}\Zb_{ai}\,\BW_a\ps \label{eq:DeltaijBWa},
\end{align}
Next, we approximate $\Delta_{ij}$ near $\psq=\Mm_b$:
\begin{align}
 \Delta_{ij}\ps &= \frac{\Delta_{ji}\ps}{\Delta_{jj}\ps} \Delta_{jj}\ps
 \simeq \Zz_{bi} \Zb_{bj}^{2}\,\BW_b\ps
 = \Zb_{bi}\Zb_{bj}\,\BW_b\ps \label{eq:DeltaijBWb}.
\end{align}
For $\psq\simeq\Mm_c$, we switch to a scheme where the indices $i$ and $c$ belong together. Thereby we can write
\begin{align}
 \Delta_{ij}\ps &= \frac{\Delta_{ij}\ps}{\Delta_{ii}\ps} \Delta_{ii}\ps
 \simeq \Zz_{cj} \Zb_{ci}^{2}\,\BW_c\ps
 = \Zb_{cj}\Zb_{ci}\,\BW_c\ps \label{eq:DeltaijBWc},
\end{align}
which is expressed
in terms of scheme-invariant $\Zb$-factors. Finally, we take the sum of Eqs.\,(\ref{eq:DeltaijBWa})-(\ref{eq:DeltaijBWc}) to obtain
\begin{align}
 \Delta_{ij}(\psq) &\simeq \sum_{a=1}^{3}\Zb_{ai}\,\BW_a(\psq)\,\Zb_{aj} \label{eq:ijBWsum}.
\end{align}
This sum is illustrated diagrammatically in Fig.\,\ref{fig:fullvsBWZdiagram}.

Eq.\,(\ref{eq:ijBWsum}) represents the central result of this section, covering 
also the diagonal propagators in the special case of $i=j$. It shows how the 
full propagator can be approximated by the contributions of the three resonance 
regions, expressed by the Breit--Wigner propagators $\Delta_a\ps,\,a=1,2,3$, 
reflecting the main momentum dependence. The mixing among the Higgs bosons is 
comprised in the $\Zb$-factors which are evaluated on-shell. Nonetheless, even a 
part of the momentum dependence of the self-energies is accounted for because 
the derivation of Eq.\,(\ref{eq:iiBWsum}) is based on a first-order expansion of 
the momentum-dependent effective self-energies. Furthermore, the $\Zb$-factors 
serve as transition factors between the
loop-corrected 
mass eigenstates $h_a$ and the 
lowest-order states
$i$ (although $\Zb$ is not a unitary matrix 
transforming the states into each other). 
Pictorially, $\Delta_{ij}$ is a 
propagator that begins on the state $i$ and ends on $j$ while mixing occurs in 
between. Also on the RHS of Fig.\,\ref{fig:fullvsBWZdiagram} and Eq.\,(\ref{eq:ijBWsum})
the propagator in the $h_a$-basis begins with $i$ and ends on $j$. Thus, the coupling to the rest of 
any diagram connected to the propagator is well-defined. 
In between, each of the $h_a$ can propagate, and the correct transition is ensured by $\Zb_{ai}$ and 
$\Zb_{aj}$. All three combinations are visualised in 
Fig.\,\ref{fig:fullvsBWZdiagram}.

If $\CP$ is conserved and only $h$ and $H$ mix, or if two states are nearly degenerate and their resonances widely separated from the remaining complex pole, the full $3\times 3$ mixing is (exactly or approximately) reduced to the $2\times 2$ mixing
case. Then the off-diagonal $\Zb$-factors involving the unmixed state vanish or become negligible so that some terms in Eq.\,(\ref{eq:ijBWsum}) approach zero.

Beyond that, if no mixing occurs among the neutral Higgs bosons, all off-diagonal full propagators as well as the off-diagonal $\Zb$-factors vanish and each diagonal full propagator consists of only a single Breit--Wigner term where the $\Zb$-factor is based on the diagonal self-energy instead of the effective self-energy. Thus, Eq.\,(\ref{eq:ijBWsum}) covers all special cases of the a priori $3\times 3$ mixing among the neutral Higgs bosons.

\subsubsection{Uncertainty estimate}
\label{sec:uncertainty}
As for the usual Breit--Wigner approximation in the case of a single pole, 
the approximation of Eq.\,(\ref{eq:ijBWsum}) contains all resonant
contributions for the case with mixing, while the non-resonant contributions
that are neglected in this approximation are suppressed by powers of 
$\Gamma_{h_a}/M_{h_a}$.
While a narrow width enhances the resonance contribution near its pole, it
also causes it to fall rapidly for momenta away from the pole and thereby
suppresses the non-resonant, far off-shell propagators, as well as their
possible interference with a resonant term. The approximation in the
off-shell regime is therefore of a comparable quality, determined by the
ratios $\Gamma_{h_a}/M_{h_a}$ for the different resonances, 
as the standard narrow-width approximation without
mixing.
Interference effects between different resonances are accounted for by
employing the approximation of Eq.\,(\ref{eq:ijBWsum}). As we will discuss
in more detail below, interference effects can be large for resonances that
are close in mass, while on the other hand interferences between widely
separated resonances are suppressed. The effects of the interference of
resonances with a non-resonant background continuum can be inferred from the
case without mixing.

While the on-shell approximation of Eq.\,(\ref{eq:ijBWsum}) is based on
the expansion around the complex pole, the description of the momentum 
dependence can be improved by incorporating 
higher powers of $(\psq- \Mm_a)$. It should be noted in this context that
apart from thresholds the self-energies generally depend only rather 
mildly on the incoming momentum.

In the following, we will assess the main sources of uncertainties that apply in the vicinity of the complex poles.
The approximation given in Eq.\,(\ref{eq:ijBWsum}) of the fully
momentum-dependent propagators in terms of Breit--Wigner propagators and
on-shell $\Zb$-factors introduces an uncertainty by neglecting higher orders
in the expansion of the effective self-energies and in the evaluation of the
self-energies depending on complex momenta in Eq.\,(\ref{eq:SigijReIm}).
Besides, the pole condition is numerically not exactly fulfilled,
which may represent
an additional source of uncertainty. 

\paragraph{Expansion of self-energies around the real part of a complex momentum}
As defined in Eq.\,(\ref{eq:SigijReIm}), we expand all self-energies as functions of a complex squared momentum $\psq$ around the real part of $\psq$ up to the first order in the imaginary part 
$p_i^2\equiv{\rm Im}\,\psq$.
Furthermore, higher powers of $p_i^2$ can enter,
\begin{align}
 \Sij(\psq) &\simeq \Sij(\psq_r) + i \psq_i\, \dsig_{ij}(\psq_r)
 + \frac{1}{2}\left(i \psq_i\right)^2\, \ddsig_{ij}(\psq_r)\,.
\end{align}
In order to expand the propagators in terms of the second-order contribution we define
\begin{align}
 \epsilon^{{\rm Im}}_{ij} := \frac{\frac{1}{2}\left(i \psq_i\right)^2\, \ddsig_{ij}(\psq_r)}{\Sij(\psq_r) + i \psq_i\, \dsig_{ij}(\psq_r)}\,,
\end{align}
and calculate $\Delta_{ii}$ 
to first order in $ \epsilon^{{\rm Im}}_{ij}$. 
One should keep in mind, though, that we also evaluate the full propagators using the above-mentioned expansion of self-energies around the real part of the momentum. 
Thus, the comparison between the full and the approximated propagators is not directly affected by omitting terms of $\mathcal{O}(\epsilon^{{\rm Im}}_{ij})$, but it represents a source of uncertainty in both methods.

\paragraph{Expansion of effective self-energies around a complex pole}
Regarding the effective self-energies,
the expansion of $\seff\ps$ around $\Mm_a$ in Eq.\,(\ref{eq:DiiApprdSeff}) includes the term of $\mathcal{O}\left(p^2-\Mm_a\right)$. Additional terms in the expansion up to 
$\mathcal{O}\left((p^2-\Mm_a)^2\right)$,
 \begin{align}
 p^{2}-m_i^{2}+\seff(p^{2})
\simeq\left(p^{2}-\Mm_a\right)\cdot
\left(1+\dseff(\Mm_a)+\frac{1}{2}(\psq - \Mm_a)\ddseff(\Mm_a)\right)
\label{eq:d2Seff}\,,
\end{align}
contribute if the term involving the 
second derivative of the effective self-energy at the complex pole,
$\ddseff(\Mm_a)$, is non-negligible.
We define
\begin{align}
 \epsilon^p_{a,i} &= \frac{\frac{1}{2}(\psq - \Mm_a)\ddseff(\Mm_a)}{1+\dseff(\Mm_a)}
 \simeq \frac{1}{2} (x-1) \Mm_a\, \ddseff(\Mm_a)\,,
 \label{eq:epsp}
\end{align}
where the approximation holds for ${\rm Re}\,\ddseff(\Mm_a),\,{\rm Im}\,\ddseff(\Mm_a) \ll 1$ and the variation of $\psq = x\Mm_a$ around $\Mm_a$. 
The impact of the neglected terms on the propagator $\Delta_{ii}$ reads at first order in $\epsilon^p_{a,i}$:
\begin{align}
 \Delta_{ii}\ps \simeq \frac{i}{(\psq -\Mm_a)\, (1+\dseff(\Mm_a))}\, \cdot (1-\epsilon^p_{a,i})\,.
\end{align}

The higher powers of $(\psq - \Mm_a)$ directly translate to an expansion in terms of $\Gamma_{h_a}/M_{h_a}$ via ${\rm Im}\,\Mm_a = -iM_{h_a} \Gamma_{h_a}$, analogously to the unmixed case.
Expanding the diagonal propagator $\Delta_{ii}\ps$ around the complex pole $\psq \sim \Mm_a$ yields
\begin{align}
 \Delta_{ii}\ps\simeq &\frac{i}{\psq - M_{h_a}^2 + iM_{h_a}\Gamma_{h_a}  }
  \cdot \left[ 1-\sum_{n=1}^{\infty} \frac{1}{n!} \left(\frac{\psq}{M_{h_a}^2 }-1 +
i\frac{\Gamma_{h_a} }{M_{h_a} }\right)^{n-1}\,M_{h_a}^{2(n-1)}\,
\hat\Sigma_{ii}^{\textrm{eff},(n)}\left(\Mm_a\right)
 \right]\,, \label{eq:expandGammaOverM}
\end{align}
where
$\hat\Sigma_{ii}^{\textrm{eff},(n)}\left(\Mm_a\right) \equiv \left.\frac{\partial^n \seff\ps }{\partial (\psq)^n}\right|_{\Mm_a}$
denotes the $n$th derivative of the effective self-energy evaluated at the complex pole $\Mm_a$. The expression $\frac{\psq}{M_{h_a}^2 }-1$ approaches zero at the resonance for $\psq \rightarrow M_{h_a}$, both in the Breit--Wigner propagator in front of the square bracket and in the correction term. Beyond that, the correction scales with powers of $\Gamma_{h_a}/M_{h_a}$ multiplied by derivatives of $\seff$. While the term of the first derivative giving rise to the diagonal $\Zbf$-factor $Z_{ai}^2$ is included in the approximation in Eq.\,(\ref{eq:iiBWsum}), the term linear in $\Gamma_{h_a}/M_{h_a}$ times the second derivative, and higher-order terms, are omitted and therefore represent a source of uncertainty. These higher terms exist at the real part of the resonance, $\psq=M_{h_a}^2$, but vanish at the complex pole, $\psq = \Mm_a$. In any case, they are not only suppressed by a narrow width compared to the mass, but also by the smallness of the higher derivatives at the pole. We confirmed numerically that already the second derivatives are suppressed by several orders compared to the first derivatives.

The expansion in Eq.\,(\ref{eq:expandGammaOverM}) indeed corresponds exactly
to the well-known unmixed case for the simplification of $M_{h_a}\rightarrow
M, \Gamma_{h_a} \rightarrow \Gamma$ and $\seff \rightarrow \Sigma$. In
contrast, the presence of mixing necessitates the expansion around all
complex poles. The mixing is accounted for by the effective self-energies,
but the structure of the dependence on $\Gamma_{h_a}/M_{h_a}$ remains in
analogy to the unmixed case. 

Even in the case of no mixing, the inclusion of the diagonal $\Zbf$-factor does not only represent a conceptual, but also a numerical improvement of the agreement between the full and the approximate propagator compared to the pure Breit--Wigner propagator.

\paragraph{Pole condition}
As an additional source of uncertainties, Eq.\,(\ref{eq:0mixed}) is numerically not exactly fulfilled. Instead,
\begin{equation}
 \Mm_a - m_i^{2} + \hat{\Sigma}^{\textrm{eff}}_{ii}(\Mm_a) = o_{a,i} \label{eq:PoleNumericalZero}
\end{equation}
sums up to a small number $o_{a,i}$.
However, evaluating the dimensionless ratio 
$\epsilon^o_{a,i}:= \frac{o_{a,i}}{m_i^2}$
for all combinations of $a$ and $i$ 
and its impact on the propagators,
this source of uncertainty is found to be negligible for the scenarios analysed in Sections~\ref{sect:prop3light} and \ref{sect:propMixextreme}.

\subsubsection{Amplitude with mixing based on full or Breit--Wigner propagators}

\begin{figure}[ht!]
 \begin{center}
 \begin{tikzpicture}[line width=0.9 pt, scale=0.65]
%  h-j
	\draw (-1,1)--(0,0)--(-1,-1);
	\draw[color=OliveGreen,dashed] (0,0)--(2,0);	
	\draw[pattern=north west lines, pattern color=OliveGreen, draw=OliveGreen] (2.5,0) circle (0.5);
	\draw[color=OliveGreen,dashed] (3,0)--(5,0);
	\draw (6,1)--(5,0)--(6,-1);
	\node[color=OliveGreen] at (1,0.6) {$\boldsymbol{h}$};
	\node[color=OliveGreen] at (4,0.6) {$\boldsymbol{h}$};
	\node[color=OliveGreen] at (2.5,-1) {$\boldsymbol{\Delta_{hh}}$};
\begin{scope}[shift={(8,0)}]
 	\draw (-1,1)--(0,0)--(-1,-1);
	\draw[color=OliveGreen,dashed] (0,0)--(2,0);	
	\draw[pattern=north west lines, pattern color=OliveGreen, draw=OliveGreen] (2.5,0) circle (0.5);
	\draw[color=OliveGreen,dashed] (3,0)--(5,0);
	\draw (6,1)--(5,0)--(6,-1);
	\node[color=OliveGreen] at (1,0.6) {$\boldsymbol{h}$};
	\node[color=OliveGreen] at (4,0.6) {$\boldsymbol{H}$};
	\node[color=OliveGreen] at (2.5,-1) {$\boldsymbol{\Delta_{hH}}$};
\end{scope}
\begin{scope}[shift={(16,0)}]
 	\draw (-1,1)--(0,0)--(-1,-1);
	\draw[color=OliveGreen,dashed] (0,0)--(2,0);	
	\draw[pattern=north west lines, pattern color=OliveGreen, draw=OliveGreen] (2.5,0) circle (0.5);
	\draw[color=OliveGreen,dashed] (3,0)--(5,0);
	\draw (6,1)--(5,0)--(6,-1);
	\node[color=OliveGreen] at (1,0.6) {$\boldsymbol{h}$};
	\node[color=OliveGreen] at (4,0.6) {$\boldsymbol{A}$};
	\node[color=OliveGreen] at (2.5,-1) {$\boldsymbol{\Delta_{hA}}$};
\end{scope}
% H-j
\begin{scope}[shift={(0,-3)}]
 	\draw (-1,1)--(0,0)--(-1,-1);
	\draw[color=OliveGreen,dashed] (0,0)--(2,0);	
	\draw[pattern=north west lines, pattern color=OliveGreen, draw=OliveGreen] (2.5,0) circle (0.5);
	\draw[color=OliveGreen,dashed] (3,0)--(5,0);
	\draw (6,1)--(5,0)--(6,-1);
	\node[color=OliveGreen] at (1,0.6) {$\boldsymbol{H}$};
	\node[color=OliveGreen] at (4,0.6) {$\boldsymbol{h}$};
	\node[color=OliveGreen] at (2.5,-1) {$\boldsymbol{\Delta_{Hh}}$};
\begin{scope}[shift={(8,0)}]
 	\draw (-1,1)--(0,0)--(-1,-1);
	\draw[color=OliveGreen,dashed] (0,0)--(2,0);	
	\draw[pattern=north west lines, pattern color=OliveGreen, draw=OliveGreen] (2.5,0) circle (0.5);
	\draw[color=OliveGreen,dashed] (3,0)--(5,0);
	\draw (6,1)--(5,0)--(6,-1);
	\node[color=OliveGreen] at (1,0.6) {$\boldsymbol{H}$};
	\node[color=OliveGreen] at (4,0.6) {$\boldsymbol{H}$};
	\node[color=OliveGreen] at (2.5,-1) {$\boldsymbol{\Delta_{HH}}$};
\end{scope}
\begin{scope}[shift={(16,0)}]
 	\draw (-1,1)--(0,0)--(-1,-1);
	\draw[color=OliveGreen,dashed] (0,0)--(2,0);	
	\draw[pattern=north west lines, pattern color=OliveGreen, draw=OliveGreen] (2.5,0) circle (0.5);
	\draw[color=OliveGreen,dashed] (3,0)--(5,0);
	\draw (6,1)--(5,0)--(6,-1);
	\node[color=OliveGreen] at (1,0.6) {$\boldsymbol{H}$};
	\node[color=OliveGreen] at (4,0.6) {$\boldsymbol{A}$};
	\node[color=OliveGreen] at (2.5,-1) {$\boldsymbol{\Delta_{HA}}$};
\end{scope}
\end{scope}
% A-j
\begin{scope}[shift={(0,-6)}]
	\draw (-1,1)--(0,0)--(-1,-1);
	\draw[color=OliveGreen,dashed] (0,0)--(2,0);	
	\draw[pattern=north west lines, pattern color=OliveGreen, draw=OliveGreen] (2.5,0) circle (0.5);
	\draw[color=OliveGreen,dashed] (3,0)--(5,0);
	\draw (6,1)--(5,0)--(6,-1);
	\node[color=OliveGreen] at (1,0.6) {$\boldsymbol{A}$};
	\node[color=OliveGreen] at (4,0.6) {$\boldsymbol{h}$};
	\node[color=OliveGreen] at (2.5,-1) {$\boldsymbol{\Delta_{Ah}}$};
\begin{scope}[shift={(8,0)}]
 	\draw (-1,1)--(0,0)--(-1,-1);
	\draw[color=OliveGreen,dashed] (0,0)--(2,0);	
	\draw[pattern=north west lines, pattern color=OliveGreen, draw=OliveGreen] (2.5,0) circle (0.5);
	\draw[color=OliveGreen,dashed] (3,0)--(5,0);
	\draw (6,1)--(5,0)--(6,-1);
	\node[color=OliveGreen] at (1,0.6) {$\boldsymbol{A}$};
	\node[color=OliveGreen] at (4,0.6) {$\boldsymbol{H}$};
	\node[color=OliveGreen] at (2.5,-1) {$\boldsymbol{\Delta_{AH}}$};
\end{scope}
\begin{scope}[shift={(16,0)}]
 	\draw (-1,1)--(0,0)--(-1,-1);
	\draw[color=OliveGreen,dashed] (0,0)--(2,0);	
	\draw[pattern=north west lines, pattern color=OliveGreen, draw=OliveGreen] (2.5,0) circle (0.5);
	\draw[color=OliveGreen,dashed] (3,0)--(5,0);
	\draw (6,1)--(5,0)--(6,-1);
	\node[color=OliveGreen] at (1,0.6) {$\boldsymbol{A}$};
	\node[color=OliveGreen] at (4,0.6) {$\boldsymbol{A}$};
	\node[color=OliveGreen] at (2.5,-1) {$\boldsymbol{\Delta_{AA}}$};
\end{scope}
\end{scope}
\end{tikzpicture}
\end{center}
\caption{Contributions from the full mixing propagators $\Delta_{ij}\ps$ for $i,j=h,H,A$ to a generic amplitude (cf. Ref.\,\cite{Fowler:2010eba}). If the $\Zb$-factor approach is applied, each of the 9 full propagators needs to be approximated by the sum of the three corresponding Breit--Wigner diagrams as shown in Fig.\,\ref{fig:fullvsBWZdiagram}.}
\label{fig:FullPropAmpl}
\end{figure}
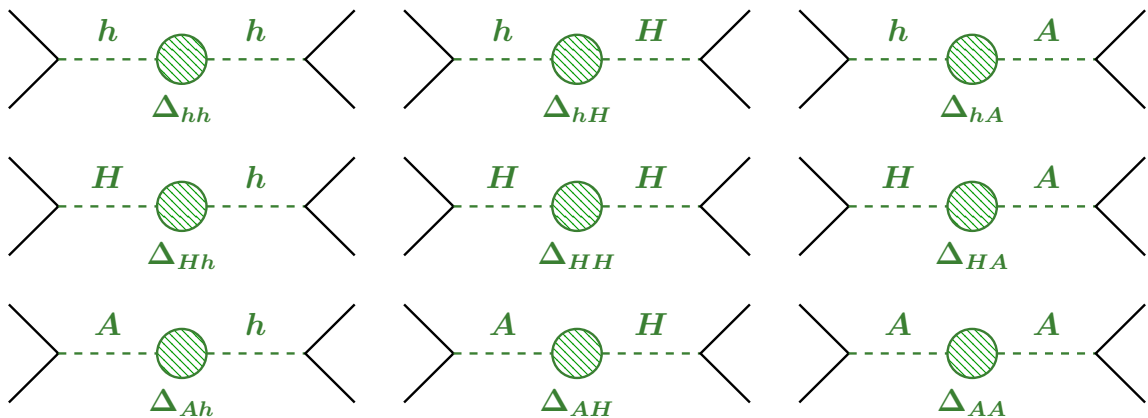

In a physical process where neutral Higgs bosons can appear as intermediate particles, all of them need to be included in the prediction, see Fig.\,\ref{fig:FullPropAmpl} and Ref.\,\cite{Fowler:2010eba}. The Higgs part of the amplitude then contains a sum over the irreducible vertex functions $\gh i^{X}$ (for a coupling of Higgs $i$ at the first vertex $X$) and $\gh j^{Y}$ (for a coupling of Higgs $j$ at the second vertex $Y$) times the fully momentum-dependent mixing propagators,
\begin{align}
 \mathcal{A} = \sum_{i,j=h,H,A} \gh i^{X} \,\Delta_{ij}\ps\, \gh j^{Y} \label{eq:FullSum}.
\end{align}
Applying Eq.\,(\ref{eq:ijBWsum}), the amplitude in Eq.\,(\ref{eq:FullSum}) can be approximated by the sum over Breit--Wigner propagators multiplied by on-shell $Z$-factors, in agreement with Ref.\,\cite{Fowler:2010eba},
\begin{align}
 \mathcal{A} &\simeq \sum_{i,j=h,H,A} \gh i^{X} \,\left[\sum_{a=1}^{3}\Zb_{ai}\,\BW_a\ps\,\Zb_{aj}\right]\, \gh j^{Y} \label{eq:BWsum}\\
 &= \sum_{a=1}^{3} \left(\sum_{i=h,H,A}\Zb_{ai}\gh i^{X}\right)\,\BW_a\ps\,\left(\sum_{j=h,H,A}\Zb_{aj}\gh j^{Y}\right)\label{eq:BWsums}\\
 &=\sum_{a=1}^{3} \gh{h_a}^{X}\,\BW_a\ps\,\gh{h_a}^{Y}\label{eq:VerthaBW}.
\end{align}
The first bracket in Eq.\,(\ref{eq:BWsums}) represents $\gh {h_a}^{X}$,
i.e.,  the vertex $X$ connected to the mass eigenstate $h_a$ as for an
external Higgs boson in Eq.\,(\ref{eq:Vertexha}). Subsequently, the second
bracket is equal to the coupling of $h_a$ at vertex $Y$, $\gh {h_a}^{Y}$. As
opposed to Sect.\,\ref{sect:ExtH}, the $h_a$ is not on-shell here, but a
propagator with momentum $\psq$ between the vertices $X$ and $Y$,
represented by the Breit--Wigner propagator $\BW_a\ps$. So the $\Zb$-factors
are not only useful for the on-shell properties of external Higgs bosons,
but they can also be used as an on-shell approximation of the mixing between
Higgs propagators. This will be investigated numerically in
Sect.\,\ref{sect:NumBWfull}.

Concerning the propagators of unstable particles, the long-standing problems
related to the treatment of unstable particles in quantum field theory 
should be kept in
mind. While the introduction of a non-zero width into the propagator of an
unstable particle is indispensable for treating the resonance region where
the unstable particle is close to its mass shell, such a modification of the
propagator in general mixes orders of perturbation theory. This can give
rise to a violation of unitarity and / or to gauge-dependent results, see
e.g.\ Ref.\,\cite{Denner:2014zga} for a recent discussion of this issue 
and references
therein. The approximation of the full off-shell propagators, where
the effective self-energy appears in the denominator of the diagonal
propagator according to 
Eq.\,(\ref{eq:Diieff}), in terms of on-shell 
$\Zb$-factors and Breit--Wigner propagators is not meant to provide a
solution to this long-standing problem. It should be noted, however, that
the expression of the off-shell propagator in terms of on-shell 
$\Zb$-factors and simple Breit--Wigner factors significantly remedies the
gauge dependence of the off-shell propagator. 
This is due to the fact that the approximation of 
Eq.~(\ref{eq:DeltaiiBWa}) and
Eq.~(\ref{eq:ijBWsum}) is on the one hand numerically close to the full
propagator (in the considered kinematic region and applying the Feynman gauge) 
but on the other hand does not contain gauge-dependent contributions that
depend on the squared momentum $p^2$ of the propagator.
While those
$p^2$-dependent contributions of the gauge part of a Dyson-resummed
self-energy in the denominator of a propagator often lead to violations of
unitarity for high values of $p^2$ (see e.g.\ Ref.\,\cite{Denner:1996gb}), the 
Breit--Wigner factors multiplied with 
$\Zb$-factors that are evaluated on-shell are much better behaved in the
limit where $p^2$ gets large. We defer a more detailed
discussion of this issue to future work.

\subsection{Calculation of interference terms in the Breit--Wigner 
formulation}\label{sect:IntBW}
In Eq.\,(\ref{eq:ijBWsum}), the Breit--Wigner propagators are combined such that they approximate a given full propagator. Conversely, we will now separate the $h_a$ part from the contribution of the other mass eigenstates in the amplitude with Higgs exchange between the vertices $X$ and $Y$: 
\begin{align}
 \mathcal{A}_{h_a} \equiv \gh {h_a}^{X}\, \BW_a\ps\, \gh {h_a}^{Y}
 = \sum_{i,j=h,H,A} \gh i^{X}\, \Zb_{ai}\, \BW_a\ps\, \Zb_{aj}\, \gh j^{Y} \label{eq:amplha},
\end{align}
i.e.\ the exchange of the state $h_a$ coupling with the mixed vertices $\gh {h_a}$ from Eq.\,(\ref{eq:Vertexha}) as for an external Higgs.
\begin{figure}[ht!]
\begin{center}
 \begin{tikzpicture}[line width=1 pt, scale=0.55]
	\node at (0.3,0) {$\hat\Gamma_{{\color{blue} h_a}}^{X}$};
	\draw (0.4,1)--(1.4,0.1);
	\draw (0.4,-1)--(1.4,-0.1);
	\fill[black!30!white] (1.7,0) circle (0.3);
	\draw[color=blue,dashed] (2,0)--(4,0);	
	\node[color=blue] at (3,0.6) {$\boldsymbol{h_a}$};
	\fill[black!30!white] (4.3,0) circle (0.3);
	\draw (5.6,1)--(4.6,0.1);
	\draw (5.6,-1)--(4.6,-0.1);
	\node at (5.5,0) {$\hat\Gamma_{{\color{blue} h_a}}^{Y}$};
	\node at (-0.5,-3) {=};
\begin{scope}[shift={(1.5,-3)}]
 	\node at (0.3,0) {$\hat\Gamma_{{\color{OliveGreen} h}}^{X}$};
	\draw (0.4,1)--(1.4,0)--(0.4,-1);
	\draw[color=OliveGreen,dashed] (1.4,0)--(2.6,0);
	\fill[black!30!white] (2.9,0) circle (0.3);
	\node[color=black] at (2.9,-0.8) {$\Zb_{{\color{blue}a}{\color{OliveGreen}h } }$};
	\draw[color=blue,dashed] (3.2,0)--(4.7,0);
	\fill[black!30!white] (5,0) circle (0.3);
	\node[color=black] at (5,-0.8) {$\Zb_{{\color{blue}a}{\color{OliveGreen}h } }$};
	\draw[color=OliveGreen,dashed] (5.3,0)--(6.5,0);	
	\node[color=blue] at (3.95,0.6) {$\boldsymbol{h_a}$};
	\node[color=OliveGreen] at (2,0.6) {$\boldsymbol{h}$};
	\node[color=OliveGreen] at (5.9,0.6) {$\boldsymbol{h}$};
	\draw (7.5,1)--(6.5,0)--(7.5,-1);
	\node at (7.4,0) {$\hat\Gamma_{{\color{OliveGreen} h}}^{Y}$};
	\node at (8.4,0) {+};	
  \begin{scope}[shift={(9,0)}]
	\draw (0.4,1)--(1.4,0)--(0.4,-1);
	\draw[color=OliveGreen,dashed] (1.4,0)--(2.6,0);
	\fill[black!30!white] (2.9,0) circle (0.3);
	\draw[color=blue,dashed] (3.2,0)--(4.7,0);
	\fill[black!30!white] (5,0) circle (0.3);
	\draw[color=OliveGreen,dashed] (5.3,0)--(6.5,0);
	\draw (7.5,1)--(6.5,0)--(7.5,-1);	
	\node at (8.4,0) {+};	 
	\node[color=blue] at (3.95,0.6) {$\boldsymbol{h_a}$};
	\node[color=black] at (2.9,-0.8) {$\Zb_{{\color{blue}a}{\color{OliveGreen}h } }$};
	\node[color=black] at (5,-0.8) {$\Zb_{{\color{blue}a}{\color{OliveGreen}H } }$};
	\node[color=OliveGreen] at (2,0.6) {$\boldsymbol{h}$};
	\node[color=OliveGreen] at (5.9,0.6) {$\boldsymbol{H}$};
	\node at (0.3,0) {$\hat\Gamma_{{\color{OliveGreen} h}}^{X}$};
	\node at (7.4,0) {$\hat\Gamma_{{\color{OliveGreen} H}}^{Y}$};
  \end{scope}
  \begin{scope}[shift={(18,0)}]
	\draw (0.4,1)--(1.4,0)--(0.4,-1);
	\draw[color=OliveGreen,dashed] (1.4,0)--(2.6,0);
	\fill[black!30!white] (2.9,0) circle (0.3);
	\draw[color=blue,dashed] (3.2,0)--(4.7,0);
	\fill[black!30!white] (5,0) circle (0.3);
	\draw[color=OliveGreen,dashed] (5.3,0)--(6.5,0);	
	\draw (7.5,1)--(6.5,0)--(7.5,-1);
	\node[color=blue] at (3.95,0.6) {$\boldsymbol{h_a}$};
	\node[color=black] at (2.9,-0.8) {$\Zb_{{\color{blue}a}{\color{OliveGreen}h } }$};
	\node[color=black] at (5,-0.8) {$\Zb_{{\color{blue}a}{\color{OliveGreen}A } }$};
	\node[color=OliveGreen] at (2,0.6) {$\boldsymbol{h}$};
	\node[color=OliveGreen] at (5.9,0.6) {$\boldsymbol{A}$};
	\node at (0.3,0) {$\hat\Gamma_{{\color{OliveGreen} h}}^{X}$};
	\node at (7.4,0) {$\hat\Gamma_{{\color{OliveGreen} A}}^{Y}$};
  \end{scope}
\end{scope}
\begin{scope}[shift={(1.5,-6)}]
	\node at (-0.6,0) {+};	
	\draw (0.4,1)--(1.4,0)--(0.4,-1);
	\draw[color=OliveGreen,dashed] (1.4,0)--(2.6,0);
	\fill[black!30!white] (2.9,0) circle (0.3);
	\draw[color=blue,dashed] (3.2,0)--(4.7,0);
	\fill[black!30!white] (5,0) circle (0.3);
	\draw[color=OliveGreen,dashed] (5.3,0)--(6.5,0);	
	\draw (7.5,1)--(6.5,0)--(7.5,-1);
	\node at (8.4,0) {+};	
	\node[color=blue] at (3.95,0.6) {$\boldsymbol{h_a}$};
	\node[color=black] at (2.9,-0.8) {$\Zb_{{\color{blue}a}{\color{OliveGreen}H } }$};
	\node[color=black] at (5,-0.8) {$\Zb_{{\color{blue}a}{\color{OliveGreen}h } }$};
	\node[color=OliveGreen] at (2,0.6) {$\boldsymbol{H}$};
	\node[color=OliveGreen] at (5.9,0.6) {$\boldsymbol{h}$};
	\node at (0.3,0) {$\hat\Gamma_{{\color{OliveGreen} H}}^{X}$};
	\node at (7.4,0) {$\hat\Gamma_{{\color{OliveGreen} h}}^{Y}$};
  \begin{scope}[shift={(9,0)}]
	\draw (0.4,1)--(1.4,0)--(0.4,-1);
	\draw[color=OliveGreen,dashed] (1.4,0)--(2.6,0);
	\fill[black!30!white] (2.9,0) circle (0.3);
	\draw[color=blue,dashed] (3.2,0)--(4.7,0);
	\fill[black!30!white] (5,0) circle (0.3);
	\draw[color=OliveGreen,dashed] (5.3,0)--(6.5,0);
	\draw (7.5,1)--(6.5,0)--(7.5,-1);	
	\node at (8.4,0) {+};	 
	\node[color=blue] at (3.95,0.6) {$\boldsymbol{h_a}$};
	\node[color=black] at (2.9,-0.8) {$\Zb_{{\color{blue}a}{\color{OliveGreen}H } }$};
	\node[color=black] at (5,-0.8) {$\Zb_{{\color{blue}a}{\color{OliveGreen}H } }$};
	\node[color=OliveGreen] at (2,0.6) {$\boldsymbol{H}$};
	\node[color=OliveGreen] at (5.9,0.6) {$\boldsymbol{H}$};
	\node at (0.3,0) {$\hat\Gamma_{{\color{OliveGreen} H}}^{X}$};
	\node at (7.4,0) {$\hat\Gamma_{{\color{OliveGreen} H}}^{Y}$};
  \end{scope}
  \begin{scope}[shift={(18,0)}]
	\draw (0.4,1)--(1.4,0)--(0.4,-1);
	\draw[color=OliveGreen,dashed] (1.4,0)--(2.6,0);
	\fill[black!30!white] (2.9,0) circle (0.3);
	\draw[color=blue,dashed] (3.2,0)--(4.7,0);
	\fill[black!30!white] (5,0) circle (0.3);
	\draw[color=OliveGreen,dashed] (5.3,0)--(6.5,0);	
	\draw (7.5,1)--(6.5,0)--(7.5,-1);
	\node[color=blue] at (3.95,0.6) {$\boldsymbol{h_a}$};
	\node[color=black] at (2.9,-0.8) {$\Zb_{{\color{blue}a}{\color{OliveGreen}H } }$};
	\node[color=black] at (5,-0.8) {$\Zb_{{\color{blue}a}{\color{OliveGreen}A } }$};
	\node[color=OliveGreen] at (2,0.6) {$\boldsymbol{H}$};
	\node[color=OliveGreen] at (5.9,0.6) {$\boldsymbol{A}$};
	\node at (0.3,0) {$\hat\Gamma_{{\color{OliveGreen} H}}^{X}$};
	\node at (7.4,0) {$\hat\Gamma_{{\color{OliveGreen} A}}^{Y}$};
  \end{scope}
\end{scope}
\begin{scope}[shift={(1.5,-9)}]
	\node at (-0.6,0) {+};	
	\draw (0.4,1)--(1.4,0)--(0.4,-1);
	\draw[color=OliveGreen,dashed] (1.4,0)--(2.6,0);
	\fill[black!30!white] (2.9,0) circle (0.3);
	\draw[color=blue,dashed] (3.2,0)--(4.7,0);
	\fill[black!30!white] (5,0) circle (0.3);
	\draw[color=OliveGreen,dashed] (5.3,0)--(6.5,0);	
	\draw (7.5,1)--(6.5,0)--(7.5,-1);
	\node at (8.4,0) {+};	
	\node[color=blue] at (3.95,0.6) {$\boldsymbol{h_a}$};
	\node[color=black] at (2.9,-0.8) {$\Zb_{{\color{blue}a}{\color{OliveGreen}A } }$};
	\node[color=black] at (5,-0.8) {$\Zb_{{\color{blue}a}{\color{OliveGreen}h } }$};
	\node[color=OliveGreen] at (2,0.6) {$\boldsymbol{A}$};
	\node[color=OliveGreen] at (5.9,0.6) {$\boldsymbol{h}$};
	\node at (0.3,0) {$\hat\Gamma_{{\color{OliveGreen} A}}^{X}$};
	\node at (7.4,0) {$\hat\Gamma_{{\color{OliveGreen} h}}^{Y}$};
  \begin{scope}[shift={(9,0)}]
	\draw (0.4,1)--(1.4,0)--(0.4,-1);
	\draw[color=OliveGreen,dashed] (1.4,0)--(2.6,0);
	\fill[black!30!white] (2.9,0) circle (0.3);
	\draw[color=blue,dashed] (3.2,0)--(4.7,0);
	\fill[black!30!white] (5,0) circle (0.3);
	\draw[color=OliveGreen,dashed] (5.3,0)--(6.5,0);
	\draw (7.5,1)--(6.5,0)--(7.5,-1);	
	\node at (8.4,0) {+};	 
	\node[color=blue] at (3.95,0.6) {$\boldsymbol{h_a}$};
	\node[color=black] at (2.9,-0.8) {$\Zb_{{\color{blue}a}{\color{OliveGreen}A } }$};
	\node[color=black] at (5,-0.8) {$\Zb_{{\color{blue}a}{\color{OliveGreen}H } }$};
	\node[color=OliveGreen] at (2,0.6) {$\boldsymbol{A}$};
	\node[color=OliveGreen] at (5.9,0.6) {$\boldsymbol{H}$};
	\node at (0.3,0) {$\hat\Gamma_{{\color{OliveGreen} A}}^{X}$};
	\node at (7.4,0) {$\hat\Gamma_{{\color{OliveGreen} H}}^{Y}$};
  \end{scope}
  \begin{scope}[shift={(18,0)}]
	\draw (0.4,1)--(1.4,0)--(0.4,-1);
	\draw[color=OliveGreen,dashed] (1.4,0)--(2.6,0);
	\fill[black!30!white] (2.9,0) circle (0.3);
	\draw[color=blue,dashed] (3.2,0)--(4.7,0);
	\fill[black!30!white] (5,0) circle (0.3);
	\draw[color=OliveGreen,dashed] (5.3,0)--(6.5,0);	
	\draw (7.5,1)--(6.5,0)--(7.5,-1);
	\node[color=blue] at (3.95,0.6) {$\boldsymbol{h_a}$};
	\node[color=black] at (2.9,-0.8) {$\Zb_{{\color{blue}a}{\color{OliveGreen}A } }$};
	\node[color=black] at (5,-0.8) {$\Zb_{{\color{blue}a}{\color{OliveGreen}A } }$};
	\node[color=OliveGreen] at (2,0.6) {$\boldsymbol{A}$};
	\node[color=OliveGreen] at (5.9,0.6) {$\boldsymbol{A}$};
	\node at (0.3,0) {$\hat\Gamma_{{\color{OliveGreen} A}}^{X}$};
	\node at (7.4,0) {$\hat\Gamma_{{\color{OliveGreen} A}}^{Y}$};
  \end{scope}
\end{scope}
\begin{scope}[shift={(3,-13)}]
	\node at (-2,0) {$=\displaystyle{\sum\limits_{{\color{OliveGreen}\boldsymbol{i,j=h,H,A}}}}$};
 	\node at (0.3,0) {$\hat\Gamma_{{\color{OliveGreen} \boldsymbol{i} }}^{X}$};
	\draw (0.4,1)--(1.4,0)--(0.4,-1);
	\draw[color=OliveGreen,dashed] (1.4,0)--(2.6,0);
	\fill[black!30!white] (2.9,0) circle (0.3);
	\node[color=black] at (2.9,-0.8) {$\Zb_{{\color{blue}a}{\color{OliveGreen}i } }$};
	\draw[color=blue,dashed] (3.2,0)--(4.7,0);
	\fill[black!30!white] (5,0) circle (0.3);
	\node[color=black] at (5,-0.8) {$\Zb_{{\color{blue}a}{\color{OliveGreen}j } }$};
	\draw[color=OliveGreen,dashed] (5.3,0)--(6.5,0);	
	\node[color=blue] at (3.95,0.6) {$\boldsymbol{h_a}$};
	\node[color=OliveGreen] at (2,0.6) {$\boldsymbol{i}$};
	\node[color=OliveGreen] at (5.9,0.6) {$\boldsymbol{j}$};
	\draw (7.5,1)--(6.5,0)--(7.5,-1);
	\node at (7.4,0) {$\hat\Gamma_{{\color{OliveGreen} \boldsymbol{j}}}^{Y}$};
\end{scope}
   \end{tikzpicture}
 \end{center}
\caption{Diagrammatic representation of the contribution $\mathcal{A}_{h_a}$ from Eq.\,(\ref{eq:amplha}) of $h_a$ ($a=1,2,3$) to the amplitude $\mathcal{A}$. The blue lines labelled by $h_a$ denote the Breit--Wigner propagator $\BW_a\ps$, and the green lines labelled by $i,j=h,H,A$ denote lowest order propagators of $h,H,A$.}
\end{figure}
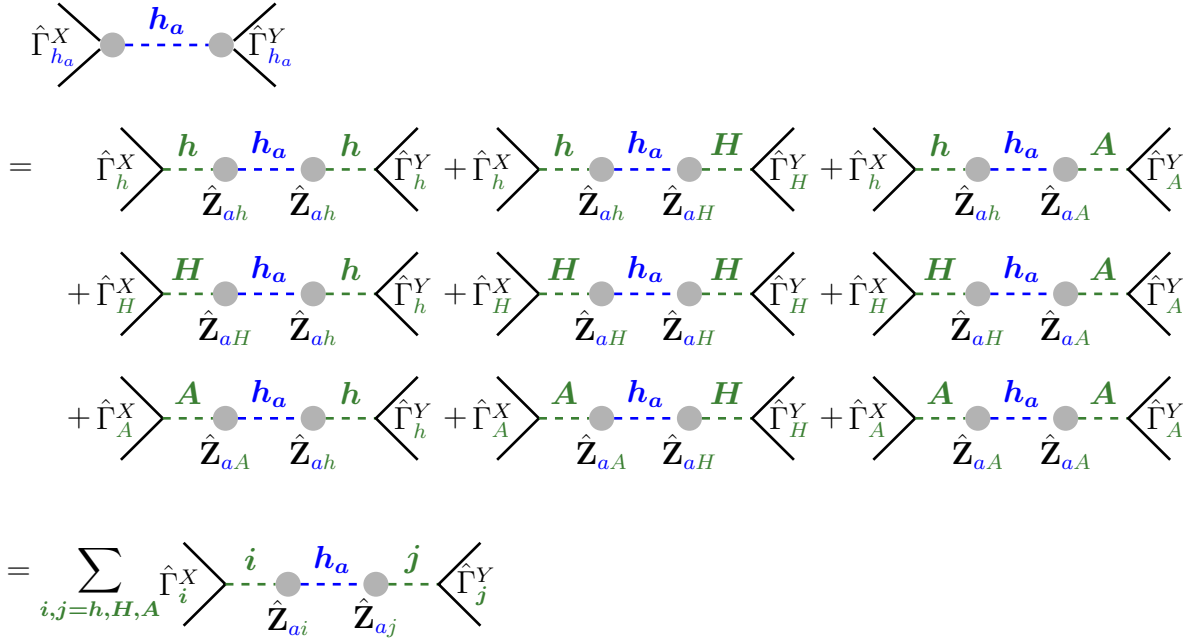

In order to calculate the squared amplitude as a \textit{coherent} sum, all contributions of $h_1, h_2, h_3$ are summed up first before taking the absolute square,
\begin{align}
 |\mathcal{A}|^{2}_{\textrm{coh}} =  \bigg\lvert \sum_{a=1}^{3} \mathcal{A}_{h_a} \bigg\rvert^{2} \label{eq:amplcoh}\,.
\end{align}
On the contrary, the \textit{incoherent} sum is the sum of the squared individual amplitudes, which misses the interference contribution,
\begin{align}
 |\mathcal{A}|^{2}_{\textrm{incoh}} =   \sum_{a=1}^{3} \bigg\lvert\mathcal{A}_{h_a} \bigg\rvert^{2} \label{eq:amplincoh}\,.
\end{align}
Thus, an advantage of the Breit--Wigner propagators is also the possibility to conveniently discern the interference of several resonances from their individual contributions in a squared amplitude
\begin{align}
 |\mathcal{A}|^{2}_{\textrm{int}} = |\mathcal{A}|^{2}_{\textrm{coh}} -|\mathcal{A}|^{2}_{\textrm{inccoh}}
  = \sum_{a<b}2\,\textrm{Re}\left[\mathcal{A}_{h_a} \mathcal{A}_{h_b}^{*}\right].
\end{align}
In contrast, the squared amplitude based on the full propagators
\begin{align}
 \lvert \mathcal{A}_{\textrm{full}}\rvert^{2} &= \Big\lvert\sum_{i,j=h,H,A} \gh i^{X} \,\Delta_{ij}\ps\, \gh j^{Y} \Big\rvert^{2} \label{eq:fullA}
\end{align}
does not allow for a straightforward determination of the pure interference term.

%%%%%%%%%%%%%%%%%%%%%%%%%%%%%%%%%%%%%%%%%%%%%%%%%%%%%%%%%%%%%%%%%%%%%%%%%%%%%
\section{Numerical comparison in the MSSM with complex parameters}
\label{sect:NumBWfull}
%%%%%%%%%%%%%%%%%%%%%%%%%%%%%%%%%%%%%%%%%%%%%%%%%%%%%%%%%%%%%%%%%%%%%%%%%%%%%
After the analytical considerations so far, we will now compare the full 
propagators with their approximation as a combination of Breit--Wigner 
propagators and $\Zb$-factors numerically.
For definiteness, we will evaluate the propagators in the MSSM with 
complex parameters. We use \texttt{FeynHiggs-2.10.2}
\footnote{The additional contributions contained in more recent versions 
are not essential for the numerical comparison carried out here.}\,\cite{Heinemeyer:1998np, 
Heinemeyer:1998yj, Degrassi:2002fi, Heinemeyer:2007aq} for the numerical 
evaluation of Higgs masses, widths and mixing properties. 
In order to investigate the applicability of the expansion of the full 
propagators in one or all three resonance regions, we will first use 
as input a complex 
squared momentum around the three complex poles. 
For the later application to 
physical processes where the squared momentum equals the centre-of-mass energy 
$s$, we will also evaluate the propagators at $\psq=s$ near the real parts of 
the complex poles. In Sect.\,\ref{sect:prop3light} we choose a scenario where 
all three Higgs bosons are relatively light so that we can study their mutual 
overlap. As a test of the $\Zb$-factor approximation, we work in a scenario with 
large mixing between $H$ and $A$ in Sect.\,\ref{sect:propMixextreme}.

\subsection{The MSSM with complex parameters at tree level}\label{chap:MSSM}
In this section we fix the notation for the different particle sectors of 
the MSSM with complex parameters, following Ref.\,\cite{Frank:2006yh}.

\paragraph{Sfermion sector}\label{sect:sfermion}
The mixing of sfermions $\tilde{f}_L, \tilde{f}_R$ within one generation into mass eigenstates $\tilde{f}_1, \tilde{f}_2$ is parametrised  by the matrix
\begin{align}
 M_{\tilde{f}}^{2}&=\left(
 \begin{matrix}
      M_{\tilde{f}_L}^{2}+m_f^{2}+M_Z^{2}\cos 2 \beta (I_f^{3}-Q_f\sw^{2}) & m_f X_f^{*}\\
      m_f X_f & M_{\tilde{f}_R}^{2}+m_{f}^{2}+M_Z^{2}\cos2\beta Q_f\sw^{2}
                        \end{matrix} \right),\label{eq:Msf}\\
 X_f &:= A_f - \mu^{*}\cdot \left\{ \begin{matrix}
                               \cot \beta,~f = \text{up-type}~~~~\\\tan \beta,~f = \text{down-type}.
                               \end{matrix} \right.
\end{align}
The trilinear couplings $A_f=|A_f|e^{i\phi_{A_f}}$, as well as $\mu=|\mu|e^{i\phi_{\mu}}$, can be complex. These phases enter the Higgs sector via sfermion loops starting at one-loop order.
Diagonalising $M_{\tilde{f}}^{2}$ for all $\tilde{f}$ separately, one obtains the sfermion masses $m_{\tilde{f}_1}\leq m_{\tilde{f}_2}$.

\paragraph{Gluino sector}
The gluino $\tilde{g}^{a},~a=1,2,3,$ has a mass of 
$ m_{\tilde{g}}=|M_3|$,
where $M_3=|M_3|\,e^{i\phi_{M_3}}$ is the possibly complex gluino mass parameter. 
Since the gluino does not directly couple to Higgs bosons, the phase $\phi_{M_3}$ enters the Higgs sector only at the two-loop level, but has an impact for example on the bottom Yukawa coupling already at one-loop order. 

\paragraph{Neutralino and chargino sector}
\label{sect:neucha}
At tree-level, mixing in the chargino sector is governed by the higgsino and wino mass parameters $\mu$ and $M_2$, respectively. In the neutralino sector it additionally depends on the bino mass parameter $M_1$.
The charginos $\widetilde{\chi}_i^{\pm},~i=1,2$, as mass eigenstates are superpositions of the charged winos $\widetilde{W}^{\pm}$ and higgsinos $\widetilde{H}^{\pm}$, with the mass matrix,
\begin{equation}
  X =  \bpm M_2 & \sqrt{2}M_W\sib\\ \sqrt{2}M_W \cb & \mu \epm.
  \label{eq:X}
\end{equation}
Likewise in the neutralino sector, the neutral electroweak gauginos $\widetilde{B},\,\widetilde{W}^{3}$ and the neutral Higgsinos $\tilde{h}_d^0,\, \tilde{h}_u^0$ mix into the mass eigenstates $\ci,\,i=1,...,4$. The mixing is encoded in the gaugino mass matrix $Y$,
\begin{equation}
Y = \left(\begin{matrix}
	M_1 & 0 & -M_Z \cb s_W & M_Z \sinb s_W\\
	0 & M_2 & M_Z \cb c_W & -M_Z \sinb c_W \\
	-M_Z \cb s_W & M_Z \cb c_W & 0 &-\mu\\
	M_Z \sinb s_W & -M_Z \sinb c_W &  -\mu &0\\
	     \end{matrix} \right) .
\label{eq:neutmix}
\end{equation}
The gaugino mass parameters $M_1$ and $M_2$ as well as the higgsino mass 
parameter can in principle be complex. However, only two of 
the phases are independent, and a frequently used convention is to set $\phi_{M_2}=0$.

\paragraph{Higgs sector}\label{sect:Higgstree}
The MSSM requires two complex scalar Higgs doublets with opposite hypercharge $Y_{\mathcal{H}_{1,2}} = \pm 1$,
\begin{align}
 \mathcal{H}_1 &= \begin{pmatrix}h_d^{0}\\h_d^{-}\end{pmatrix}
                  =\bpm v_d+\frac{1}{\sqrt{2}}(\phi_1^{0}-i\chi_1^{0})\\ -\phi_1^{-}\epm \label{eq:H1doublet}\\
 \mathcal{H}_2 &= \begin{pmatrix}h_u^{+}\\h_u^{0}\end{pmatrix}
                  = \bpm \phi_2^{+}\\v_u+\frac{1}{\sqrt{2}}(\phi_2^{0}+i\chi_2^{0})\epm. \label{eq:H2doublet}
\end{align}
A relative phase between both Higgs doublets
vanishes at the minimum of the 
Higgs potential, and the possible phase of the coefficient of the bilinear term 
in the Higgs potential can be rotated away.  
Hence, the Higgs sector conserves $\CP$ at lowest order, and the $4\times 4$ 
mass matrix of the neutral states $\mathbf{M}_{\phi\phi \chi\chi}$ becomes 
block-diagonal. The neutral tree level mass eigenstates are obtained by a 
diagonalisation of $\mathbf{M}_{\phi\phi \chi\chi}$:
\begin{align}
 \bpm h\\H\\A\\G \epm = \bpm -\sa& \ca & 0 & 0\\ \ca &\sa&0&0\\ 0&0& -\sibn &\cbn \\ 0&0 & \cbn& \sibn \epm \bpm \phi_1^{0}\\ \phi_2^{0}\\ \chi_1^{0}\\ \chi_2^{0}\epm,\hspace*{1cm}
 \bpm H^{\pm}\\G^{\pm}\epm = \bpm  -\sibc& \cbc\\ \cbc &\sibc\epm 
 \bpm \phi_1^{\pm}\\ \phi_2^{\pm}\epm\label{eq:ncHmix},
\end{align}
where we introduced the short-hand notation $s_x\equiv\sin x,~c_x\equiv\cos x$. 
For later use we define $\tb\equiv \tan\beta$. The mixing angle $\alpha$ is 
applied for the $\mathcal{CP}$-even Higgs bosons $h,H$; $\beta_n$ for the 
neutral $\mathcal{CP}$-odd Higgs $A$ and Goldstone boson $G$, and $\beta_c$ for 
the charged Higgs $H^{\pm}$ and the charged Goldstone boson $G^{\pm}$. The 
minimum conditions for $V_H$ lead to $\beta = \beta_n = \beta_c$ at tree level. 
At higher orders, however, $\tan\beta$ is renormalised whereas the mixing 
angles $\alpha, \beta_n$ and $\beta_c$ are not renormalised.
The masses of the $\mathcal{CP}$-odd and the charged Higgs bosons are at tree level related by
\begin{eqnarray}
 m_{H^{\pm}}^{2} &=& m_A^{2} + M_W^{2}\label{eq:mH+}.
\end{eqnarray}
At lowest order, the Higgs sector is fully determined by the two SUSY 
input parameters (in addition to SM masses and gauge couplings) $\tan\beta$ and 
$m_{H^{\pm}}$ (or, for conserved $\mathcal{CP}$, equivalently $m_A$).

\paragraph{Higher-order corrections}
Higher-order corrections in the MSSM Higgs sector are very relevant.
Particles from other sectors contribute via loop diagrams to Higgs observables, 
in particular the trilinear couplings $A_f$, the stop and sbottom masses, and in 
the sub-leading terms the higgsino mass parameter $\mu$. 
Thus, beyond the lowest order, the Higgs sector is influenced by more parameters 
than only $M_{H^{\pm}}$ (or $M_A$) and $\tan\beta$. 
We adopt the hybrid on-shell and $\overline{\rm{DR}}$-renormalisation scheme 
defined in Ref.\,\cite{Frank:2006yh}.

\subsection{Scenario with three light Higgs bosons}\label{sect:prop3light}
For the numerical evaluation of the propagators, we work in the $\mhmod$ scenario\,\cite{Carena:2013qia}.
In this example, we fix the variable parameters
\begin{align}
\mu &= 200\gev,\nonumber\\
 \mhp &= 160\gev,\nonumber\\
 \tan\beta &= 50,\label{eq:choiceTbMHp}
\end{align}
and choose for the complex phase $\pat=\pi/4$ to allow for $\CP$-violating mixing. 
This parameter choice is not meant to be the experimentally viable; in fact it has been experimentally ruled out by searches for $H,A\rightarrow \tau\tau$. The purpose is rather to provide a setting for illustration with nearby, but resolvable resonances so that the test of the Breit--Wigner approximation is not limited to well separated poles.
These parameter values result in the following complex poles:
\begin{align}
 \Mm_1 &=(15791-70i)\gev^{2},~~~\Mm_2=(16202-525i)\gev^{2},~~~\Mm_3=(17388-385i)\gev^{2}\label{eq:polevalues}.
 \end{align}
All of the loop-corrected masses obtained from the real parts of the complex poles listed above are relatively light:
 \begin{align}
 M_{h_1}&=125.7\gev,~~~M_{h_2}=127.3\gev,~~~M_{h_3}=131.9\gev\label{eq:massvalues}
 \end{align}
so that the mass differences are of the order of -- but not smaller than -- the total widths from the imaginary parts of the complex poles, $\Gamma_{h_1}=0.6\gev,~\Gamma_{h_2}=4.1\gev,~\Gamma_{h_3}=2.9\gev$. 
The phase of $A_t$ induces $\CP$-violating mixing, and the on-shell mixing properties are reflected by the $\Zb$-matrix obtained with \texttt{FeynHiggs}, 
 \begin{align}
 \Zb &= \bpm	 0.95-0.04i 	& 0.34+0.09i 	&-0.05-0.05i\\
		 0.05-0.05i	& 0.02+0.03i	& 0.99-0.02i\\
		-0.35-0.09i	& 0.94-0.05i	&-0.006-0.003i	
 \epm\label{eq:Zvalues},
\end{align}
which indicates that $h_1$ couples mostly $h$-like, $h_2$ mostly $A$-like and $h_3$ mostly $H$-like. 
Thus, the contribution of $h_a$ to $\Delta_{ij}$, i.e.
\begin{equation}
 \Delta_{ij}\Big\rvert_{h_a}\ps=\Zb_{ai}\,\BW_a\ps\,\Zb_{aj}\label{eq:hacontrib},
\end{equation}
is only significant if the product $\Zb_{ai}\Zb_{aj}$ is not suppressed. In this case, we can already estimate that, for example, $h_3$ hardly contributes to $\Delta_{AA}$.

We have analysed all propagators $\Delta_{ij}$ around $\Mm_1,\,\Mm_2$ and $\Mm_3$ as well as for real momenta. In the following, we show and discuss a selection of these cases. 

\subsubsection{Propagators depending on complex momenta}
\begin{figure}[ht!]
 \subfigure[]{\includegraphics[width=0.495\textwidth]{ReDhh}\label{fig:diag1Re}} 
 \subfigure[]{\includegraphics[width=0.495\textwidth]{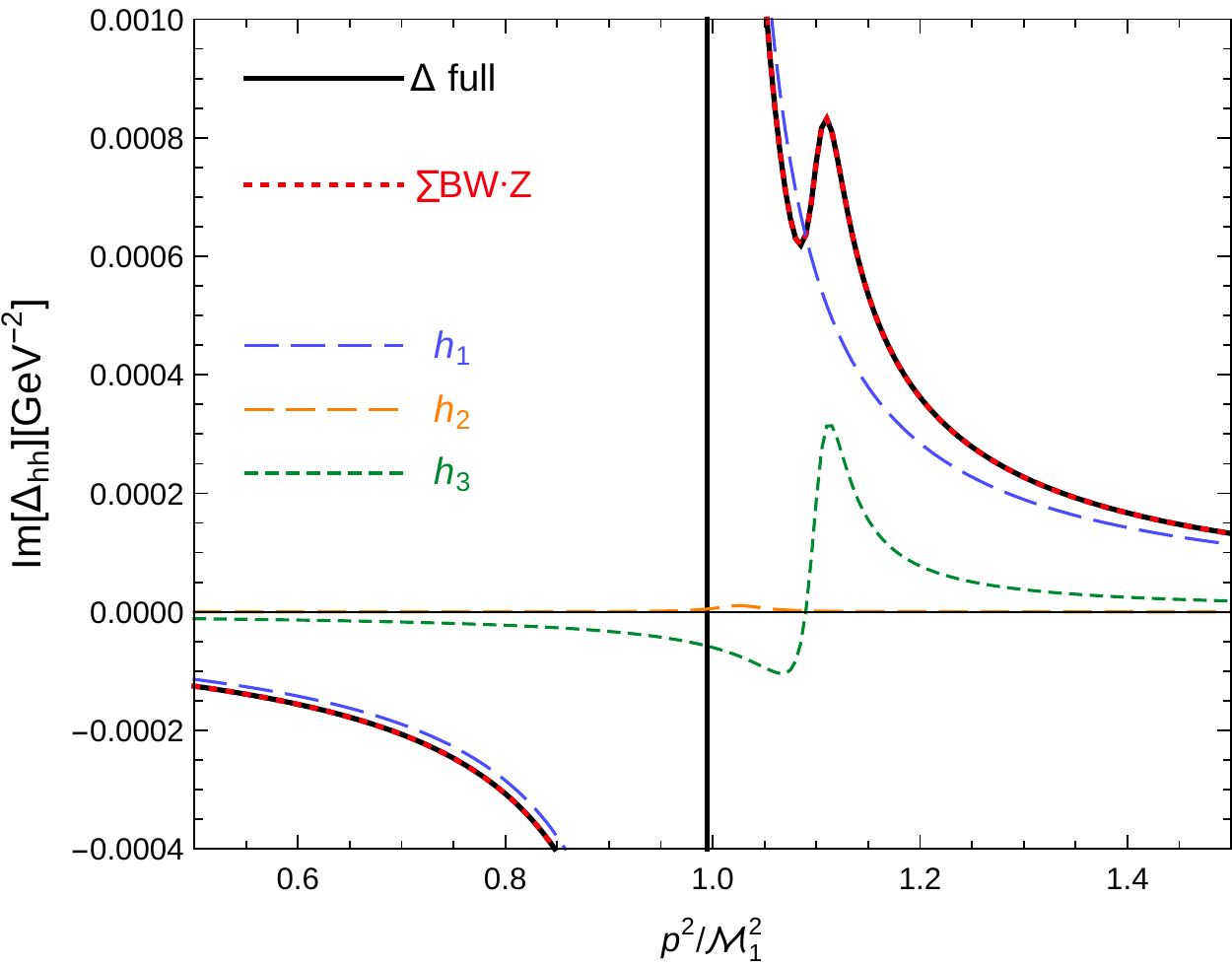}\label{fig:diag1Im}}
 \caption{Diagonal propagator $\Delta_{hh}\ps$ depending on the complex
momentum $\psq$ around $\Mm_1$ with $\psq/\Mm_1=0.5...1.5$. \textbf{(a)}
real part, \textbf{(b)} imaginary part. The full propagator $\Delta_{hh}$ (black, labelled by $\Delta$ full) is compared to the \textit{sum} of Breit--Wigner propagators weighted by $\Zb$-factors according to Eq.\,(\ref{eq:iiBWsum}) (red dotted, labelled by $\sum$BW$\cdot$Z). The individual contribution of $h_a$, i.e.\ $\Zb_{ah}^{2}\BW_a$, is shown for $h_1$ (blue, long-dashed), $h_2$ (orange, dashed) and $h_3$ (green, short-dashed).}
  \label{fig:diag1}
\end{figure}

The analytical derivation of Eq.\,(\ref{eq:ijBWsum}) builds on the expansion of 
the full propagators around the complex poles, and the on-shell condition in 
Eq.\,(\ref{eq:SolvePole}) holds exactly only at complex momentum. Therefore, we 
evaluate the self-energies and propagators around the complex poles. 
Fig.\,\ref{fig:diag1} displays $\Delta_{hh}\ps$ for $\psq = 
0.5\,\Mm_1~...~1.5\,\Mm_1$. In particular, Fig.\,\ref{fig:diag1Re} shows 
$\textrm{Re}\left[\Delta_{hh}\right]$, and Fig.\,\ref{fig:diag1Im} shows 
$\textrm{Im}\left[\Delta_{hh}\right]$ versus the ratio $x_1=\psq/\Mm_1$ such 
that $x_1=1$ corresponds to the complex pole
$\psq=\Mm_1$. The black line 
(labelled by $\Delta$ full) represents the fully momentum-dependent mixing 
propagator from Eq.\,(\ref{eq:Diieff}). Since the three poles do not have the 
same ratio between the real and imaginary parts, scaling $x_1$ does not run into 
$\Mm_2$ and $\Mm_3$. $\Delta_{hh}$ has a pole at $x=1$ and a second peak at 
$x\simeq1.1$ which is close to the real part of $\Mm_3$. This structure is very 
precisely reproduced (for a discussion of the uncertainties, see below) by the sum $\sum_{a=1}^{3}\Zb_{ah}^{2}\,\BW_a\ps$ according 
to Eq.\,(\ref{eq:iiBWsum}) -- as can be seen from the red dotted line (labelled 
as $\sum\textrm{BW}\cdot Z$), which lies directly on top of the black solid 
line.

In order to understand which of the Breit--Wigner propagators and $\Zb$-factors dominate at which momentum,
the individual contributions are shown by the dashed curves.
The blue line (labelled by $h_1$) represents the contribution of $h_1$ to $\Delta_{hh}$, i.e., $\Zb_{1h}^{2}\,\BW_1\ps$. It clearly
displays the pole at $x=1$, but strongly deviates from the full propagator at 
momenta away from $\Mm_1$. The orange line (labelled by $h_2$) represents $\Zb_{2h}^{2}\,\BW_2\ps$. Since $\Zb_{2h}$ is small in this scenario, the contribution of $h_2$ to $\Delta_{hh}$ is numerically suppressed, but a tiny share is visible.
The green line (labelled by $h_3$) stands for $\Zb_{3h}^{2}\,\BW_3\ps$ and it contributes significantly to $\Delta_{hh}$ near $\Mm_3$ because $\Zb_{3h}=-0.35-0.09i$ is sizeable in this scenario. So we notice that none of the individual Breit--Wigner propagators multiplied by the appropriate $\Zb$-factors suffices to approximate the full propagator, which has three complex poles.
On the other hand, the sum of all three Breit--Wigner propagators times $\Zb$-factors yields an accurate approach to the full mixing. This holds for the real and the imaginary part.

\graphicspath{{figures/cMhmod3lightH/CP1Plots/}}

\begin{figure}[ht!]
 \subfigure[]{\includegraphics[width=0.49\textwidth]{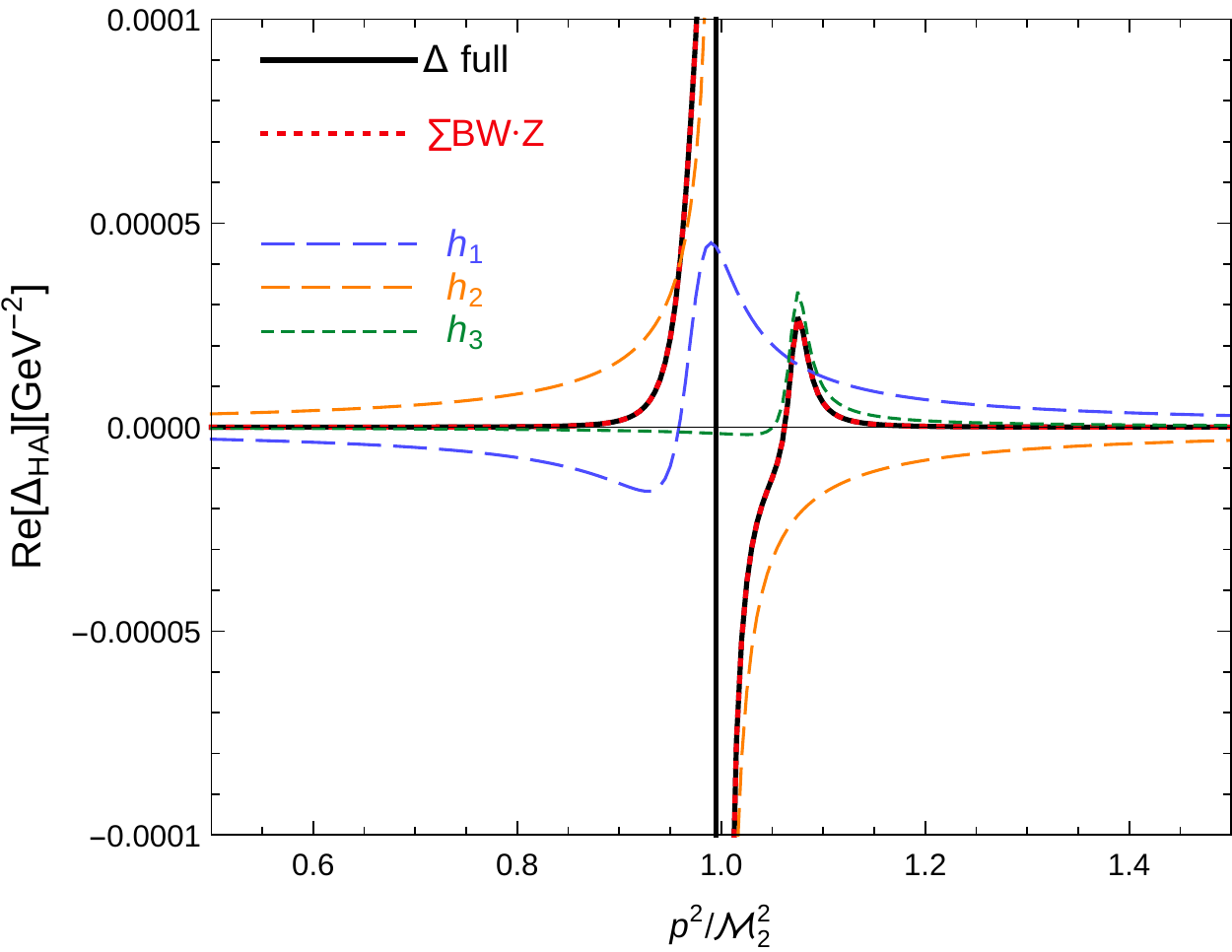}\label{fig:HARe}}
 \subfigure[]{\includegraphics[width=0.49\textwidth]{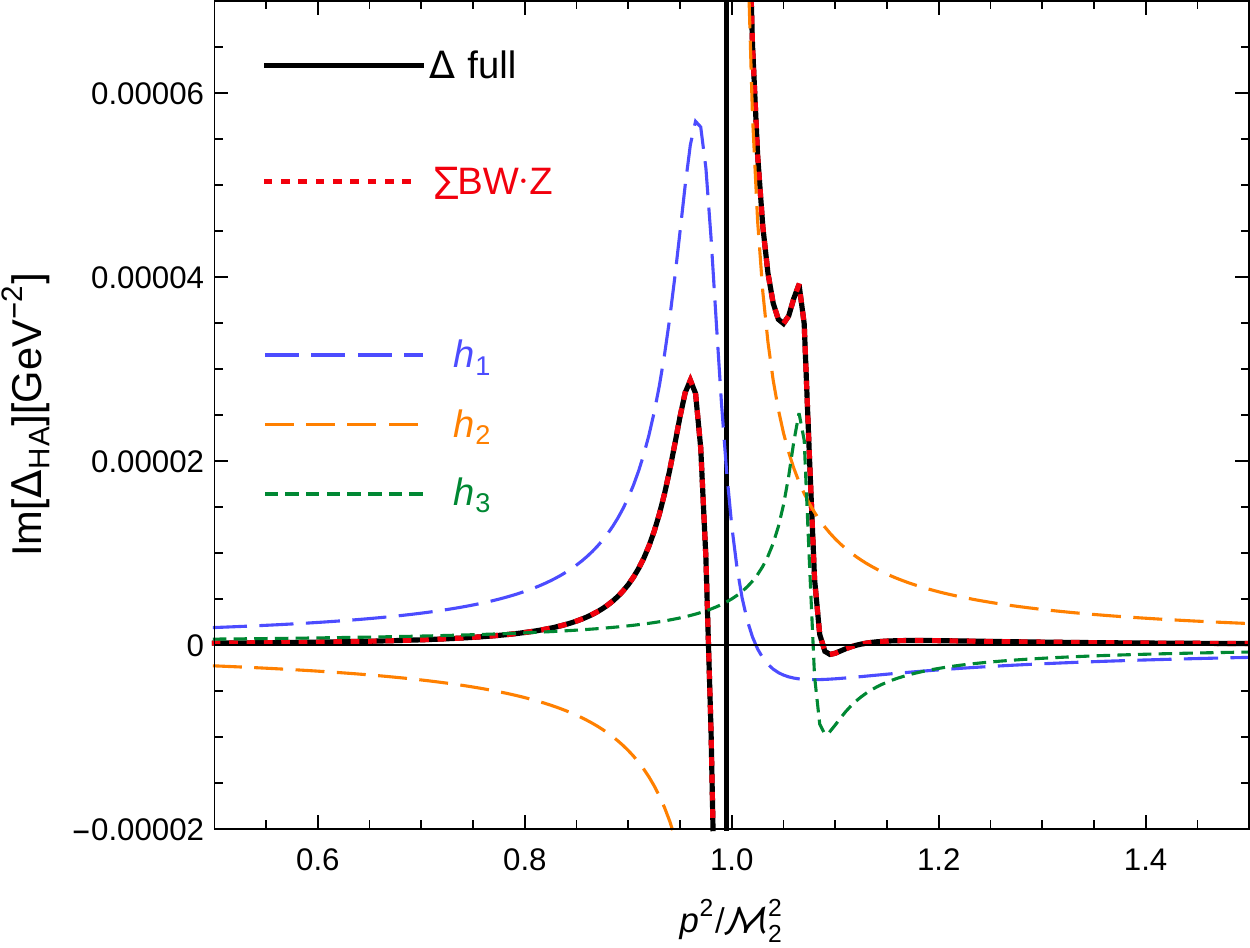}\label{fig:HAIm}}
 \caption[Off-diagonal propagator $\Delta_{HA}\ps$ depending on complex momenta.]{Off-diagonal propagator $\Delta_{HA}\ps$ depending on the complex momentum $\psq$ around $\Mm_2$ with $\psq/\Mm_2=0.5...1.5$. \textbf{(a)} real part, \textbf{(b)} imaginary part. Labelling as in Fig.\,\ref{fig:diag1}.}
  \label{fig:offdiag2}
\end{figure}

Having discussed the example of a diagonal propagator, we will now assess whether the $\Zb$-factor approximation succeeds also for off-diagonal propagators. For instance, Fig.\,\ref{fig:offdiag2} depicts $\Delta_{HA}$ versus $x_2=\psq/\Mm_2$ such that $x=1$ matches $\psq=\Mm_2$ where the propagator diverges. As above, owing to the different ratio between the real and imaginary part of each complex pole, scaling $x_2$ does not run into $\Mm_1$ and $\Mm_3$, but $\Delta_{HA}$ peaks close to their real parts. As in Fig.\,\ref{fig:diag1}, the black line representing the full propagator and the red, dotted line representing the sum of Breit--Wigner propagators according to Eq.\,(\ref{eq:ijBWsum}) agree very well. Additionally, 
one can see the individual Breit--Wigner shapes. Because the products of the relevant $\Zb$-factors, here $\Zb_{aH}\Zb_{aA}$, are non-negligible for all $a=1,2,3$, each Breit--Wigner propagator is important in the approximation of both the real part (Fig.\,\ref{fig:HARe}) and the imaginary part (Fig.\,\ref{fig:HAIm}) of $\Delta_{HA}$. The other diagonal and off-diagonal propagators in this scenario which are not displayed here have an equally good agreement between the full calculation and the approximation.

In order to quantify this agreement, we compute the relative deviation of
the approximated propagators from the full ones, for each diagonal and
off-diagonal element of the propagator matrix, depending on the momentum
around each of the complex poles.
For $\psq=x\cdot \Mm_1$, $0.5\leq x\leq 1.5$ 
as in Fig.\,\ref{fig:diag1},
the relative deviation of the approximated propagators from the full ones 
is below $4\cdot 10^{-3}$ for $\Delta_{hh}$, $\Delta_{HH}$ 
and below $9\cdot 10^{-3}$ for $\Delta_{AA}$
(only $\Delta_{hh}$ is displayed here),
apart from the regions where the
propagator itself goes through zero and changes sign.
A similar agreement holds both for the diagonal and off-diagonal
propagators such as in Fig.\,\ref{fig:offdiag2}, and for their real and imaginary parts. As expected, the relative
deviation is minimal at the complex pole ($x=1$).
The largest deviation between the full and the approximated propagators occurs (apart from the sign flips) at momenta far away from the pole 
where the full propagators approach zero and in some cases the individual propagators of the states $h_a$ contribute with opposite signs to sum.

The relative deviation 
for the displayed momenta 
should be compared to the expected uncertainty based
on the estimates in Sect.\,\ref{sec:uncertainty}. The higher powers of the
imaginary part of the momentum that are neglected in the approximate 
evaluation of the
self-energies at complex momenta give rise to only a small effect
($\epsilon^{\rm Im}_{ij}\sim \mathcal{O}(10^{-7})$. This induces an effect
on the propagator of $\mathcal{O}(10^{-9}-10^{-6})$), which does not
account for the
actual difference (as explained above, both the full and the approximated
propagators are subject to this kind of uncertainties arising 
from the evaluation at complex momenta).
In contrast, the relative effect of the expansion of the momentum-dependent
effective self-energy around the complex pole 
has a more significant effect of up to $\epsilon^p_{a,i}\lesssim 6\cdot 10^{-3}$ for the analysed momentum range, where the largest deviations occur furthest away from the poles.
As a conclusion, the
uncertainty 
arising from the approximate evaluation of the propagators at complex
momenta is found to be negligible, while the
expected uncertainty from the expansion around a complex pole
is very close to the observed deviation.

\graphicspath{{figures/cMhmod3lightH/CP2Plots/}}

\subsubsection{Propagators depending on the real momentum $\psq=s$}
The calculation of the propagators at and around the complex poles together with the evaluation of the self-energies at complex momenta according to Eq.\,(\ref{eq:SigijReIm}) was needed to fulfil the assumptions of the approximation. However, in collider processes, the Higgs propagator might appear for example in the s-channel of a $2\rightarrow 2$ scattering process where the squared momentum equals the square of the centre-of-mass energy $s$. Therefore here we will check the Breit--Wigner approximation around the real parts of the complex poles.

\begin{figure}[ht!]
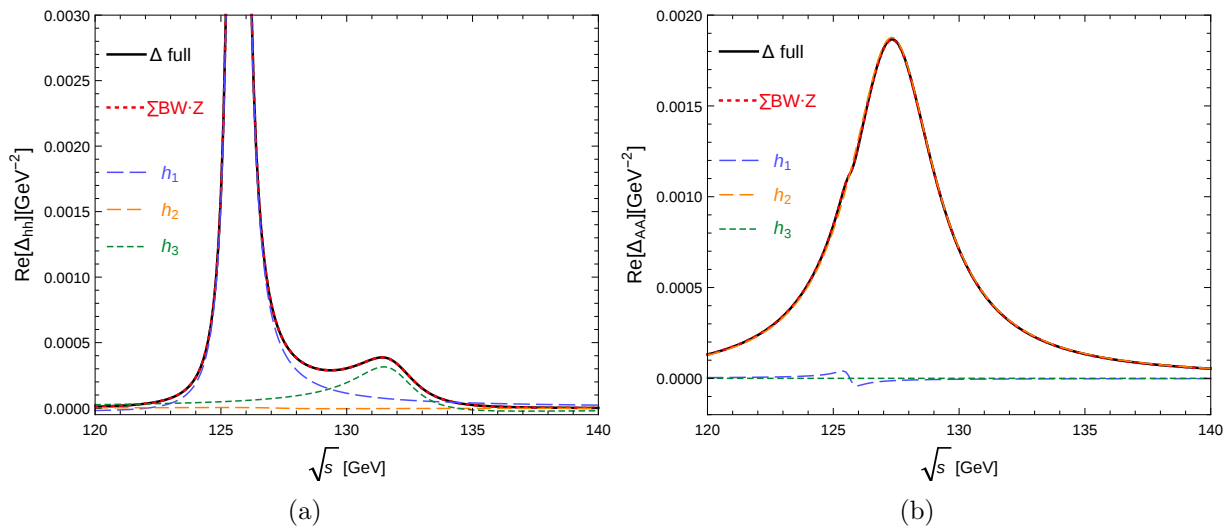

 \subfigure[]{\includegraphics[width=0.495\textwidth]{ReDhh}\label{fig:hhsRe}}
 \subfigure[]{\includegraphics[width=0.495\textwidth]{ReDAA}\label{fig:AAsRe}}
  \caption[Diagonal propagators $\Delta_{hh}\ps$ and $\Delta_{AA}\ps$ for $\psq=s$.]{Diagonal propagators $\Delta_{hh}\ps$ (\textbf{(a)}) and $\Delta_{AA}\ps$ (\textbf{(b)}) depending on the real momentum $\psq=s$ around $\sqrt{s}\simeq M_{h_1},\,M_{h_2},\,M_{h_3}$. Labelling as in Fig.\,\ref{fig:diag1}.}
  \label{fig:diagS}
\end{figure}

Fig.\,\ref{fig:hhsRe} shows $\textrm{Re}\left[\Delta_{hh}\right]$ in the range $\sqrt{\psq}\simeq M_{h_1}, M_{h_2}, M_{h_3}$ of the three loop-corrected masses given in Eq.\,(\ref{eq:massvalues}). The propagator has a pronounced peak around $M_{h_1}$ and a smaller and broader one at $M_{h_3}$. Again, the approximation (red, dotted) defined in Eq.\,(\ref{eq:iiBWsum}) meets the full propagator (black) very precisely. The contribution of $h_1$ multiplied by $\Zb_{1h}^{2}$ of $\mathcal{O}(1)$ dominates near $M_{h_1}$. At $M_{h_3}$ the Breit--Wigner shape of $h_3$ is dominant although multiplied only by $\Zb_{3h}^{2}=(-0.35-0.09i)^{2}$, but also the tail of $\BW_1$ is relevant. The resonance of $h_2$ is strongly suppressed by the small $\Zb_{2h}$.
The described behaviour is analogous in
$\textrm{Re}\left[\Delta_{HH}\right]$ (not shown here), where the strongest
peak around $M_{h_3}$ is dominated by the $h_3$-contribution and the second
peak near $M_{h_1}$ by the $h_1$-contribution, whereas the $h_2$ component
is negligible. 
Also in this case, the tails of the propagators of the relevant mass eigenstates extend also to the other resonance regions.
Fig.\,\ref{fig:AAsRe} visualises $\textrm{Re}\left[\Delta_{AA}\right]$ with a broad peak at $M_{h_2}$. The black curve of the full propagator is again directly 
underneath the red, dotted curve of the Breit--Wigner approximation, which in this case stems nearly entirely from $h_2$ because of $\Zb_{2A}\simeq 1$. Within $\Delta_{AA}$, the contribution of $h_1$ only has a minor impact, which can be seen as a small kink in Fig.\,\ref{fig:AAsRe}. 
Although $\BW_1\ps$ gets close to its pole, the resonance of $h_1$ is strongly suppressed by the small $\Zb$-factor $\Zb_{1A}=0.05 (1+i)$.
As we anticipated above from the structure of $\Zb$ in Eq.\,(\ref{eq:Zvalues}), $\BW_{3}$ is a negligible component of $\Delta_{AA}$ for this parameter point.\\

\graphicspath{{figures/cMhmod3lightH/SPlots/}}
So far we have seen that the Breit--Wigner formulation combined with on-shell $\Zb$-factors accurately reproduces (up to a relative deviation of $\order(10^{-3}- 10^{-2})$)
the momentum dependence of the full diagonal and off-diagonal propagators by adding the contributions from all resonance regions. If all resonances are sufficiently separated (not shown in this example) or if all but one contributing products of $\Zb$-factors are negligible, a single Breit--Wigner term is enough to approximate the full propagator in one of the resonance regions. In the general case, however, all Breit--Wigner terms need to be included. Even if the peaks are not located very close to each other compared to their widths, the tail of one resonance supported by a substantial product of $\Zb$-factors can leak into another resonance region.

\subsection{Scenario with large mixing}\label{sect:propMixextreme}
While the scenario in the previous section was characterised by three relatively similar masses, we now choose a setting with quasi degenerate heavy states $h_2$ and $h_3$. In Sect.\,\ref{sect:prop3light} we considered the $M_h^{\rm{mod}+}$-scenario with the standard value of $\mu= 200\gev$ in combination with the phase $\phi_{A_t}=\pi/4$, leading to a moderate mixing predominantly between $h$ and $A$. In Ref.\,\cite{Carena:2013qia} it was suggested to choose also different values, $\mu=\pm200,\pm500, \pm1000\gev$. So in addition to the choice above, we now apply the following modification of the parameters in Eq.\,(\ref{eq:choiceTbMHp}):
\begin{align}
 \mu &= 1000\gev,\nonumber\\
 \mhp &= 650\gev,\nonumber\\
 \tan\beta &= 20.\label{eq:choicemue1000}
\end{align}
This results in the complex poles
\begin{equation}
 \Mm_1 = (15797-0.2i)\gev^{2},~~\Mm_2 = (415336-1673i)\gev^{2},~~\Mm_3 = (415554-1857i)\gev^{2},
\end{equation}
and therefore in similar masses of the heavy Higgs bosons,
\begin{align}
 M_{h_1} &= 125.33\,\gev,~~~M_{h_2}= 644.47\gev,~~~M_{h_3}= 644.63\gev\,.
\end{align}
There is a large mixing between $H$ and $A$, visible in the $\Zb$-factors evaluated with \texttt{FeynHiggs},
\begin{align}
 \Zb &\simeq \bpm
 1.01 	&0 		&0\\
 0 	&1.15-0.27i	&-0.47-0.66i\\
 0	&0.49+0.65i	&1.13-0.28i
 \epm\,.
\end{align}
In the scenario in Sect.\,{\ref{sect:prop3light}}, $\Zb$ is approximately
unitary, 
\begin{align}
 \Zb\cdot\Zb^{\dagger}\Big\rvert_{\textrm{Eq.}~(\ref{eq:choiceTbMHp})} &\simeq \bpm
    1	&-0.1i	&0.2i\\
    0.1i	&1	&0\\
    -0.2i	&0	&1
 \epm
 \simeq \mathbb{1}.
\end{align}
On the contrary, in the present scenario with $\mu=1000\gev$, the product $\Zb\cdot\Zb^{\dagger}$ deviates strongly from $\mathbb{1}$: 
\begin{align}
 \Zb\cdot\Zb^{\dagger}\Big\rvert_{\textrm{Eq.}~(\ref{eq:choicemue1000})}&\simeq \bpm
 1	&0	&0\\
 0	&2	&-1.7i\\
 0	&1.7i	&2 \label{eq:Znonunitary}
 \epm.
\end{align}
Hence it is useful to examine whether the Breit--Wigner propagators with $\Zb$-factors still yield a viable approximation of the fully momentum-dependent propagators in this scenario with large mixing.

\subsubsection{Propagators depending on complex momenta}
\begin{figure}[ht!]
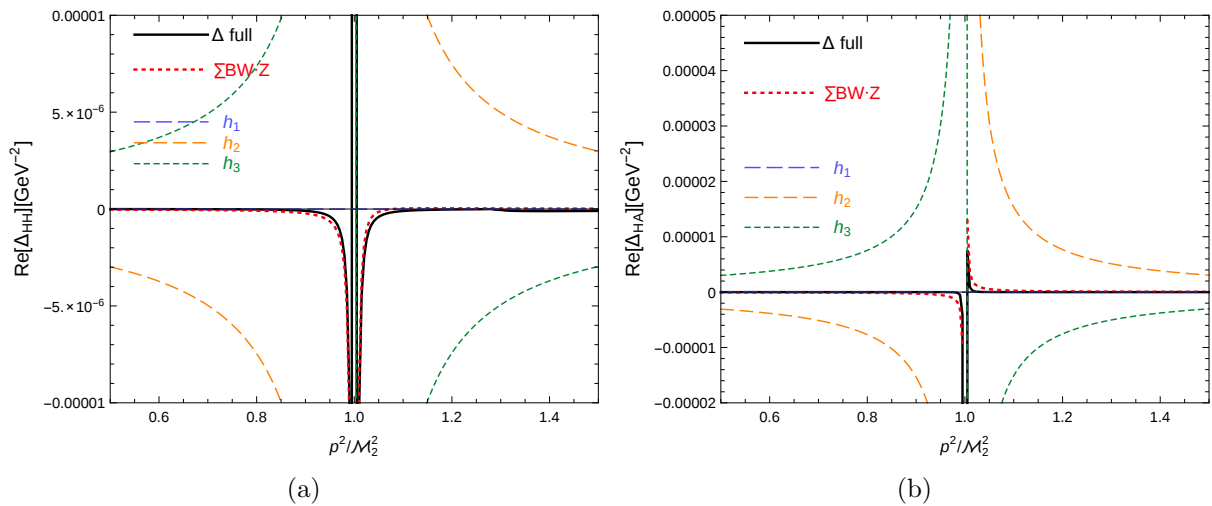

 \subfigure[]{\includegraphics[width=0.485\textwidth]{ReDHH} \label{fig:mue1000ReHH}}
 \subfigure[]{\includegraphics[width=0.485\textwidth]{ReDHA} \label{fig:mue1000ReHA}}
 \caption[Real parts of $\Delta_{HH}\ps$ and $\Delta_{HA}\ps$ depending on
complex momenta.]{Real parts of \textbf{(a)} the diagonal propagator
$\Delta_{HH}\ps$ and \textbf{(b)} the off-diagonal propagator $\Delta_{HA}\ps$ depending on the complex momentum $\psq$ around $\Mm_2$ with $\psq/\Mm_2=0.5...1.5$. Labelling as in Fig.\,\ref{fig:diag1}.}
 \label{fig:mue1000Mm2}
\end{figure}

Due to the large difference between $M_{h_1}$ and $M_{h_2}\simeq M_{h_3}$, the propagators $\Delta_{hh},\,\Delta_{hH}$ and $\Delta_{hA}$ are strongly suppressed around $\Mm_2$. Fig.\,\ref{fig:mue1000ReHH} shows $\textrm{Re}\left[\Delta_{HH}\right]$ for complex momentum around $\Mm_2$. The full calculation (black) and the Breit--Wigner approximation (red, dotted) are in 
relatively
good agreement with each other although the curves do not lie directly on top of one another as in the scenario of Sect.\,\ref{sect:prop3light}. The two heavy states $h_2$ and $h_3$ with a mass difference of less than $0.2\gev$ and total widths of $\Gamma_{h_2}=2.6\gev$ and $\Gamma_{h_3}=2.9\gev$ are too close to be resolved. $\Zb_{2H}^{2}\Delta_{h_2}$ (orange) and $\Zb_{3H}^{2}\Delta_{h_3}$ (green) contribute with similar magnitude, but opposite signs so that the result differs strongly from the single terms. Indeed, only their sum provides a reliable approximation.
A comparable situation is shown in Fig.\,\ref{fig:mue1000ReHA} for $\textrm{Re}\left[\Delta_{HA}\right]$.

In this scenario of highly admixed, almost mass-degenerate states
$h_2,\,h_3$, the relative deviation of the approximated propagators from the
full ones is larger, in particular in the region where the propagator itself
is close to zero.
The uncertainty of the evaluation of complex momenta is negligible here, as in the scenario of Sect.\,\ref{sect:prop3light}.
Instead, the reasons for the resulting deviation are the following.
The neglected terms from the expansion around the complex pole (see Eq.\,(\ref{eq:epsp})) amount to $\epsilon^p_{a,i}\sim\order(10^{-2}-10^{-1})$ 
for the states $i=H,\,A$ involved in the mixing 
around the nearby complex poles $\Mm_a,\,a=2,3$.
The relative effect of the deviation is enhanced by 
the fact that the full propagator is close to zero. The approximation
consists of the sum of two resonant contributions ($h_2,\,h_3$), which are in
this case several orders of magnitude larger than the full propagator and
occur with opposite signs. Hence, despite a cancellation of several orders
of magnitude the approximation yields 
the correct sign of the propagator and
a good description of its behaviour and size.
In contrast, a single-resonance approach would result 
in a prediction that would be off by several orders of magnitude. As a result,
although slight deviations between the
the approximated propagator and the full one are visible in this 
scenario that is characterised by large mixing effects,
the approximation based on Eq.\,(\ref{eq:ijBWsum}) still represents a very
significant improvement compared to a single-resonance treatment.

\graphicspath{{figures/cMhmodMue1000/CP2Plots/}}

\subsubsection{Propagators depending on the real momentum $\psq=s$}
\begin{figure}[ht!]
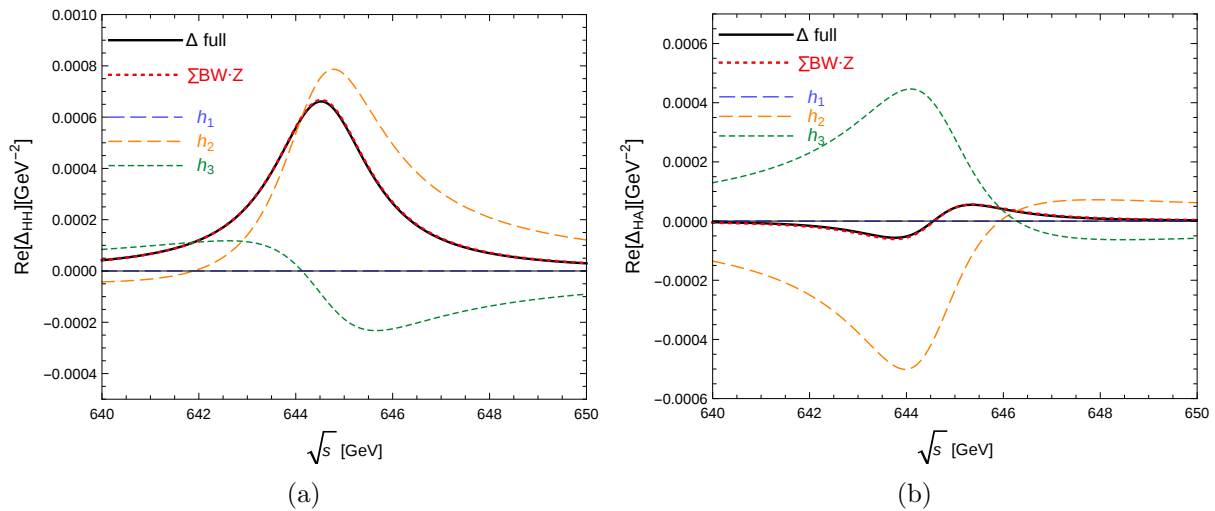

\subfigure[]{\includegraphics[width=0.485\textwidth]{ReDHH} \label{fig:mue1000sReHH}}
\subfigure[]{\includegraphics[width=0.485\textwidth]{ReDHA} \label{fig:mue1000sReHA}}
 \caption[Real parts of $\Delta_{HH}\ps$ and $\Delta_{HA}\ps$ for $\sqrt{s}\simeq M_{h_2},\,M_{h_3}$.]{Real parts of \textbf{(a)} the diagonal propagator $\Delta_{HH}\ps$ and \textbf{(b)} the off-diagonal propagator $\Delta_{HA}\ps$ depending on the real momentum $\psq=s$ around $\sqrt{s}\simeq M_{h_2},\,M_{h_3}$. Labelling as in Fig.\,\ref{fig:diag1}.}
 \label{fig:mue1000S}
\end{figure}

Fig.\,\ref{fig:mue1000S} shows the same selection of propagators as in
Fig.\,\ref{fig:mue1000Mm2}, but in this case evaluated at real momentum. The approximation from Eq.\,(\ref{eq:ijBWsum}) leads to a good agreement 
between the propagators in the full mixing calculation (black) and the Breit--Wigner formulation (red, dotted), displayed for $\textrm{Re}[\Delta_{HH}]$ in Fig.\,\ref{fig:mue1000sReHH} and for $\textrm{Re}[\Delta_{HA}]$ in Fig.\,\ref{fig:mue1000sReHA}. While the $h_1$-part (blue) is negligible due to the much lower mass $M_{h_1}$, $h_2$ (orange) and $h_3$ (green) both contribute substantially because their complex poles are very close to each other. 
For the evaluation at real momentum, the diagonal propagators $\Delta_{HH}, \Delta_{AA}$ are reproduced with an accuracy of $\order(10^{-3})$ by the approximation while the relative deviation is $\order(10^{-1})$ for $\Delta_{HA}$ where a cancellation of larger terms occurs. 
These comparisons show that the Breit--Wigner approximation is also applicable in scenarios of quasi-degenerate states and a strong resonance-enhanced mixing. 
We note, however, that the agreement between the full propagators and those with on-shell mixing factors is less 
accurate here than in the scenario with moderate, nearly unitary mixing. 

\graphicspath{{figures/cMhmodMue1000/SPlots/}}

\subsubsection{Comparison of $\Zb$-factors with effective couplings}
\label{sect:UZ}
The effective coupling approach mentioned in Sect.\,\ref{sect:Ueff} makes use of the unitary, real $\Ub$-matrix instead of the $\Zb$-matrix. 
However, $\Ub$ is not evaluated at the complex pole, but at $\psq=0$ (see 
Sect.\,\ref{sect:Ueff}) and it does not comprise the imaginary parts of the 
self-energies. We compare the effective coupling 
approach based on the matrix $\Ub$ with the $\Zb$-factor approach which does take the 
imaginary parts into account, but cannot be directly interpreted as a unitary 
transformation between the states of the different bases.

\begin{figure}[hb!]
 \subfigure[]{\includegraphics[width=0.485\textwidth]{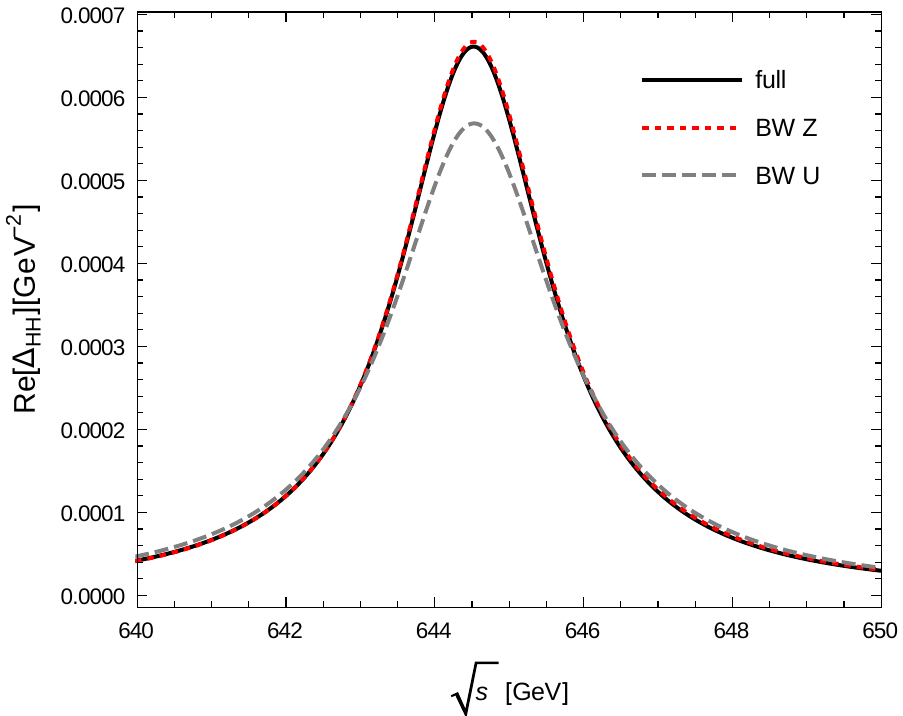} 
\label{fig:UZReHH}}
 \subfigure[]{\includegraphics[width=0.485\textwidth]{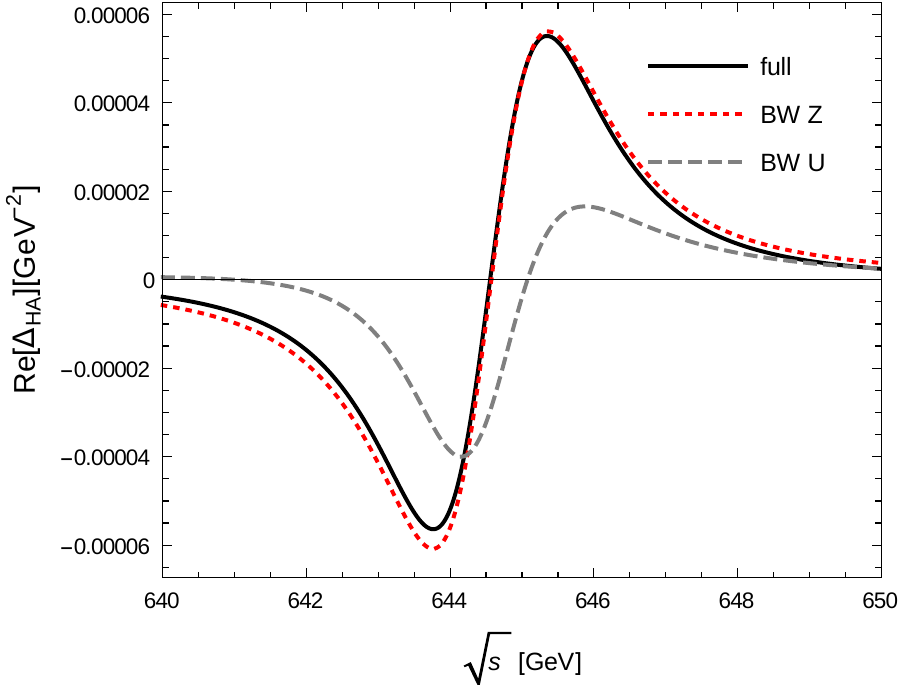} 
\label{fig:UZReHA}}
 \caption{Comparison of the 
 full propagators (black, solid) with the Breit--Wigner approximations based on 
 the $\Ub$-matrix (grey, dashed) and $\Zb$-matrix (red, dotted) 
 for real parts of the 
diagonal propagator $\Delta_{HH}\ps$  \textbf{(a)} and the off-diagonal $\Delta_{HA}\ps$ \textbf{(b)} 
depending on the real momentum $\psq=s$ around $\sqrt{s}\simeq 
M_{h_2},\,M_{h_3}$.}
 \label{fig:mue1000UZ}
\end{figure}

Based on $\Zb$- and $\Ub$-factors from \texttt{FeynHiggs}, 
Fig.\,\ref{fig:mue1000UZ} displays 
in the scenario of Eq.\,(\ref{eq:choicemue1000})
real parts of $\Delta_{HH}$ and 
$\Delta_{HA}$ at real momentum around $M_{h_2}\simeq M_{h_3}$, calculated as a 
fully momentum-dependent mixing propagator (black), using the $\Zb$-matrix 
approach (red, dotted) defined in Eq.\,(\ref{eq:ijBWsum}), i.e.\
$\Delta_{ij}^{Z}(\psq)\simeq \sum_{a=1}^{3}\Zb_{ai}\,\BW_a(\psq)\,\Zb_{aj}$, 
and the $\Ub$-matrix variant (grey, dashed) in the $\psq=0$ approximation as
\begin{equation}
 \Delta_{ij}^{U}(\psq)\simeq \sum_{a=1}^{3}\Ub_{ai}(\psq=0)\,\BW_a(\psq)\,\Ub_{aj}(\psq=0) \label{eq:UijBWsum}.
\end{equation}
\\
The black curves in Fig.\,\ref{fig:mue1000UZ} representing the full propagators are identical to those shown in Fig.\,\ref{fig:mue1000S}.
While in Fig.\,\ref{fig:UZReHH} the curve representing the approximation in terms of $\Zb$-factors 
is almost identical to the curve of the full $\Delta_{HH}$ (with a relative 
deviation at the peak of $0.8\%$), the approximation in terms of $\Ub$-factors differs from the full result 
by up to $14\%$. In Fig.\,\ref{fig:UZReHA}, not even the shape of $\Delta_{HA}$ 
is correctly approximated by the approach using the $\Ub$-matrix whereas the approach using the $\Zb$-matrix comes 
close to the full $\Delta_{HA}$ (up to the small deviations that can be seen
in the plots). 
 
This analysis indicates that the $\Zb$-factors combined with Breit--Wigner propagators are well-suited to describe the Higgs propagators including their mixing
also in scenarios with close-by resonances and strong mixing. This approach captures the leading momentum dependence and adequately accounts for the imaginary parts. In contrast, the combination of $\Ub$-factors and Breit--Wigner propagators is -- despite its unitary nature -- incomplete with respect to the mixing effects in the resonance region and regarding the significance of imaginary parts.

%%%%%%%%%%%%%%%%%%%%%%%%%%%%%%%%%%%%%%%%%%%%%%%%%%%%%%%%%%%%%%%%%%%%%%%%%%%%%%%
\subsection{Breit--Wigner and full propagators in cross sections}
\label{sect:fullBWxsec}
As an application of the derivations above, we calculate a cross section with Higgs exchange.
We study the example process $b\overline{b} \rightarrow h, H, A \rightarrow \tau^{+}\tau^{-}$, where the intermediate Higgs bosons are once represented by the full mixing propagators $\Delta_{ij}$ and once by Breit--Wigner propagators multiplied by $\Zb$-factors. In order to disentangle this investigation from other higher-order effects, we restrict the Higgs-fermion-fermion vertices to the tree-level and do not include the emission of real particles in the initial or final state, but focus on the propagator corrections.

For the implementation of the full propagator method, we extended a \texttt{FeynArts} model file. New scalars $ij$ are introduced that correspond to the full propagator $\Delta_{ij}\ps$ and couple to the first vertex as the interaction eigenstate $i$ and to the second vertex as $j$. Those propagators are used in the \texttt{FormCalc} calculation supplemented by self-energies from \texttt{FeynHiggs} incorporating corrections up to the two-loop level and 
the full momentum dependence at the one-loop level. 

Considering only mixing between $h$ and $H$ at this point, we choose as a 
$\CP$-conserving scenario the 
$M_{h}^{\textrm{max}}$-scenario\,\cite{Carena:1999xa,Carena:2002qg} with 
$\tan\beta = 50,\, \mhp = 153\,$GeV, but we modify it by setting 
$A_{f_3}=2504\,$GeV. 
As before, this scenario has been selected for illustration purposes
and it is not meant to be phenomenologically viable.
An outcome of this parameter choice  are large off-diagonal $Z$-factors 
$\hat{\textbf{Z}}_{12}\simeq 0.65+0.29i,\, \hat{\textbf{Z}}_{21}\simeq 
-0.64-0.29i$, and $\hat{\textbf{Z}}_{11}\simeq 0.85-0.22i,\, 
\hat{\textbf{Z}}_{22}\simeq 0.84-0.23i$. The masses of the $\mathcal{CP}$-even 
Higgs bosons are very close to each other, $M_{h_1}=126.20\,$GeV and $M_{h_2}=127.55$\,GeV, 
while the widths obtained from the imaginary part of the complex poles are 
$\Gamma_{h_1}=0.94\,$GeV and $\Gamma_{h_2}=1.21\,$GeV. Despite its large width 
of $\Gamma_A=3.58\,$GeV, the third neutral Higgs boson does not overlap 
significantly with the other two resonances due to the mass of 
$M_{h_3}=119.91\,$GeV, and no mixing with the other two states occurs because
we are considering here the $\cp$-conserving case of real parameters.

\begin{figure}[h!]
 \begin{center}
 \begin{tikzpicture}
  \node at (0,0) 
{\includegraphics[width=0.8\textwidth]{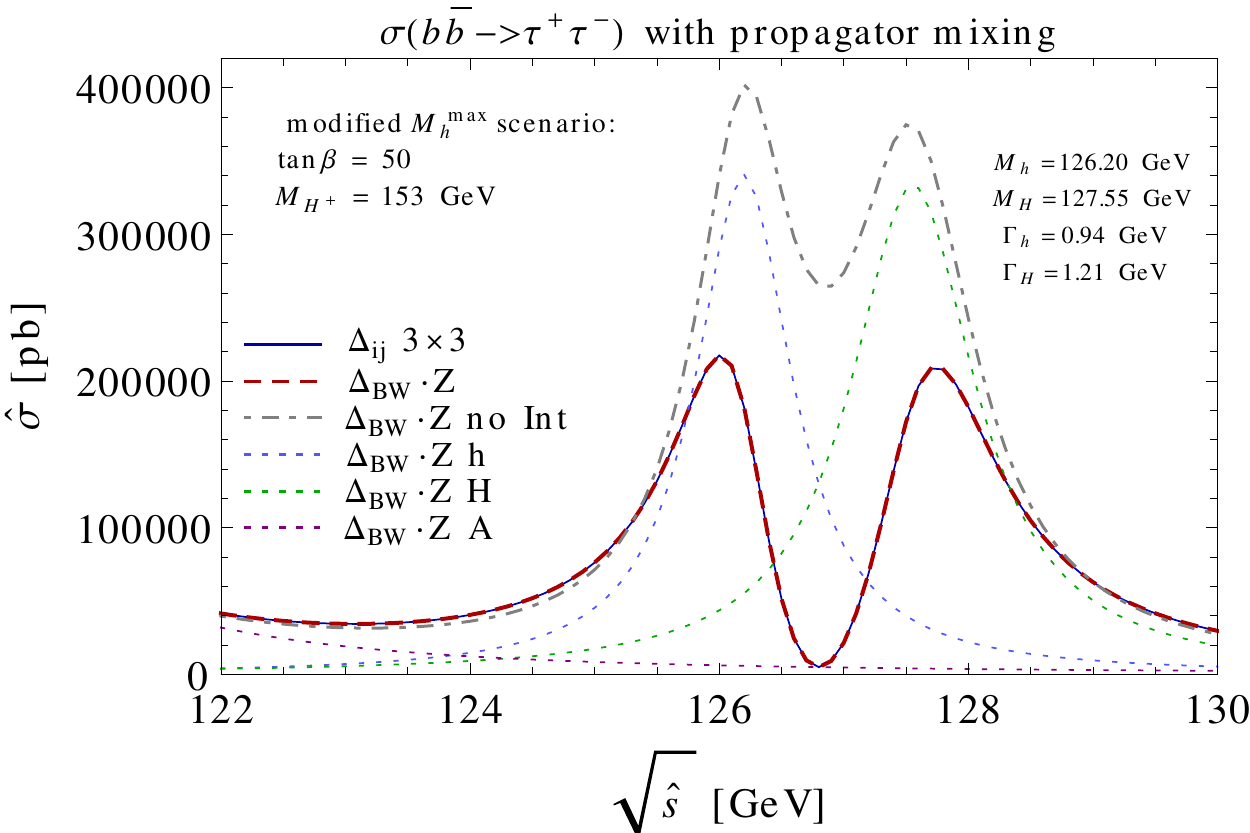}};
  \node at (-67pt,112pt) {\textbf{\textasciicircum}};
\end{tikzpicture}
\caption[Higgs exchange in $b\overline{b}\rightarrow \tau^{+}\tau^{-}$ in a 
modified $M_{h}^{\textrm{max}}$-scenario with full propagators compared to the 
coherent and the incoherent sum of Breit--Wigner propagators and 
$\Zb$-factors.]{The partonic cross section $\hat\sigma(b\overline{b}\rightarrow 
\tau^{+}\tau^{-})$ in a modified $M_{h}^{\textrm{max}}$-scenario with 
$\tan\beta 
= 50$ and $\mhp = 153\,$GeV. The cross section is calculated with the full 
mixing propagators (blue, solid), approximated by the coherent sum of 
Breit--Wigner propagators times $\Zb$-factors with the interference term (red, 
dashed) and the incoherent sum without the interference term (grey, 
dot-dashed). 
The individual contributions mediated by $h_1$ (light blue), $h_2$ (green) and 
$h_3$ (purple) are shown as dotted lines.}
\label{fig:DeltaBWMhmax}
 \end{center}
\end{figure}

Fig.\,\ref{fig:DeltaBWMhmax} shows the partonic cross section
$\hat\sigma(b\overline{b} \rightarrow h, H, A \rightarrow \tau^{+}\tau^{-})$
as a function of the centre-of-mass energy $\sqrt{\hat s}$, where $\hat s =
(p_b+p_{\overline{b}})^{2}$ is the squared sum of the momenta of the $b$-
and $\overline{b}$-quarks in the initial state. The calculation based on the
full propagators (represented by the blue, solid line) is in very good
agreement with the cross section based on the coherent sum of the $h_1, h_2,
h_3$ contributions of the Breit--Wigner propagators multiplied with $\Zb$-factors
(red, dashed) according to Eq.\,(\ref{eq:amplcoh}). Both
curves lie on top of each other and contain two peaks originating from $h_1$
(light blue, dotted) and $h_2$ (green, dotted). The resonances of $h_1$ and
$h_2$ partly overlap 
as the mass difference is of the order of the total widths, but the two
peaks can still be distinguished. The $h_3$ contribution peaks at a lower mass in
this scenario, but for completeness it is also shown (purple, dotted). The
incoherent sum $|h_1|^{2}+|h_2|^{2}+|h_3|^{2}$ (grey, dash-dotted) from
Eq.\,(\ref{eq:amplincoh}) clearly overestimates the full cross section 
as a consequence of the missing interference term that turns out to be destructive in this case. 
It is taken into account in the full calculation and in the coherent sum of Breit--Wigner propagators with $\Zb$-factors. For the efficient calculation of interference terms of quasi degenerate resonances, see e.g. Refs.\,\cite{Fuchs:2014ola,Fuchs:2014zra}.
 
\begin{figure}[ht!]
 \begin{center}
\includegraphics[width=
0.6\textwidth]{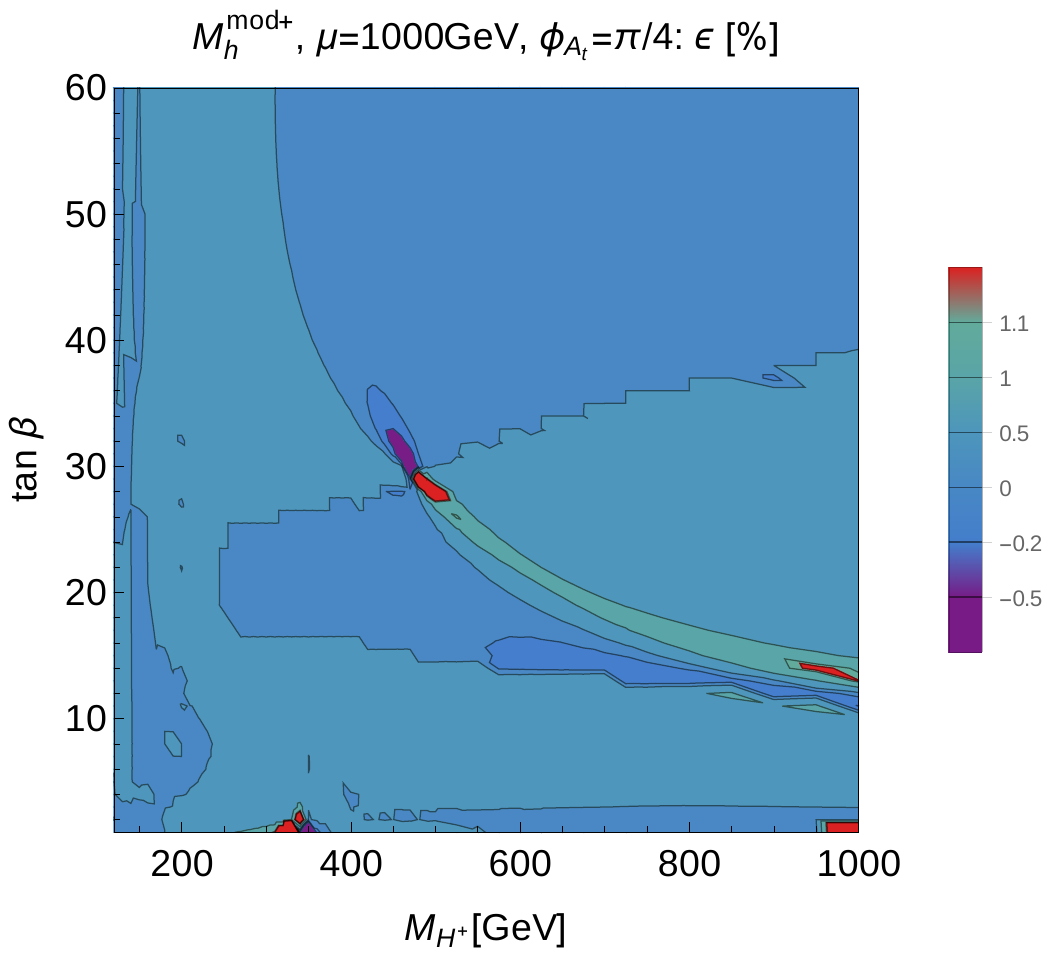}
  \caption{Relative difference 
$\epsilon$ in \% between cross sections based on the full propagator mixing and the 
Breit--Wigner approximation with $\Zb$-factors using the total width 
$\Gamma^{\rm{Im}}$ from the imaginary part of the complex pole: the partonic 
cross section 
$\hat\sigma(b\bar b \rightarrow 
\tau^{+}\tau^{-})$ via neutral Higgs bosons in the $\mhmod$ 
scenario with 
$\mu=1000\gev$, $\phi_{A_t}=\pi/4$.}
  \label{fig:epsilon}
  \end{center}
\end{figure}

While the comparison in Fig.\,\ref{fig:DeltaBWMhmax} is restricted to the case 
of $2\times 2$-mixing due to a scenario with real parameters, the agreement of 
the cross section 
$\hat\sigma(b\overline{b} \rightarrow h, H, A \rightarrow \tau^{+}\tau^{-})$
calculated with the full or Breit--Wigner propagators can be seen also in a 
scenario with complex parameters in Fig.\,\ref{fig:epsilon}. For the numerical 
evaluation, we choose 
a scenario with large mixing defined in 
Eq.\,(\ref{eq:choicemue1000}), i.e.\ the $\mhmod$ scenario with 
$\mu=1000\gev$ and the phase $\pat=\pi/4$ where we vary $\mhp$ and 
$\tb$.

We investigate the relative deviation $\epsilon$ between the cross 
section $\sigma_{\rm{full}}$ based on the full propagators and the cross section 
$\sigma_{\rm{coh}}^{\rm{BW}\Zb}$ based on the coherent sum of Breit--Wigner 
propagators with $\Zb$-factors, where the total widths are obtained from the 
imaginary parts of the complex poles, $\Gamma^{\textrm{Im}}$,
defined in Eq.\,(\ref{eq:GammaIm}),
\begin{align}
 \epsilon &= \frac{\sigma_{\rm{coh}}^{\rm{BW}\Zb}(\pat) 
}{\sigma_{\rm{full}}(\pat) }-1\label{eq:epsilon}.
\end{align}
Fig.\,\ref{fig:epsilon} reveals that both methods agree very well, with a 
maximum deviation of $\pm2\%$ around $\mhp=500\gev,~\tan\beta=28$ and of about 
$0.8\%$ along the green band
from this parameter point to larger values of $\mhp$ and lower values of $\tb$.
Otherwise the two calculations lead to the same 
results within $0.1\%$. Hence the use of Breit--Wigner propagators is suitable 
also for phenomenological applications such as the calculation of cross 
sections even in a scenario with complex parameters and large, $\cp$-violating 
mixing. A phenomenological investigation of such a scenario 
will be addressed in a
forthcoming publication\,\cite{Fuchs:2017poi,Fuchs:2017wkq,IntCalc:InProgress}.

\subsection{Impact of the total width}
\label{sect:Gtot}
This section addresses the impact of the precise value of the total width. So far, we have obtained the Higgs widths from the imaginary part of the complex poles as in Eq.\,(\ref{eq:MGamma}) in order to consistently compare with the full propagator mixing. If the self-energies in $\seff$ are calculated at the one-loop level, the total width extracted from a complex pole of $\Delta_{ii}$ is then a tree-level width. Correspondingly, partial two-loop contributions to the imaginary parts of the self-energies give rise to partial one-loop corrections of the decay width. However, two-loop self-energies evaluated at $\psq=0$, as they are approximated in \texttt{FeynHiggs}, do not contribute to the imaginary part of the pole so that the width determined from the imaginary part of the complex pole remains at its tree-level value.
Corrections to Higgs boson decays in the MSSM at and beyond the one-loop level are known and have been found to be important, see e.g. Refs.\,\cite{Dabelstein:1995js,Djouadi:1995gt,Williams:2011bu,Heinemeyer:2015pfa}.
Thus, the sum of the partial decay widths into any final state $X$ of a Higgs boson $h_a$,
\begin{equation}
\Gamma_{h_a}^{\textrm{tot}} = \sum\limits_X \Gamma(h_a \rightarrow X),\label{eq:GammaTot}
\end{equation}
leads to a more accurate result for its total width than from the imaginary part of the corresponding complex pole,
\begin{equation}
 \Gamma_{h_a}^{\textrm{Im}}=-\textrm{Im}[\Mm_a]/M_{h_a}, \label{eq:GammaIm}
\end{equation}
even if $\Mm_a$ is based on self-energies at the same order as used for the calculation of $\Gamma_{h_a}^{\textrm{tot}} $ as in Eq.\,(\ref{eq:GammaTot}). 
\texttt{FeynHiggs} contains the partial Higgs decay widths and their sum at the leading two-loop order. Having checked the compelling agreement between the full propagators and the Breit--Wigner propagators with the width from the imaginary part of the complex pole in the previous sections, now  we implement the total width from \texttt{FeynHiggs} into the Breit--Wigner propagators in order to obtain the most precise phenomenological prediction.

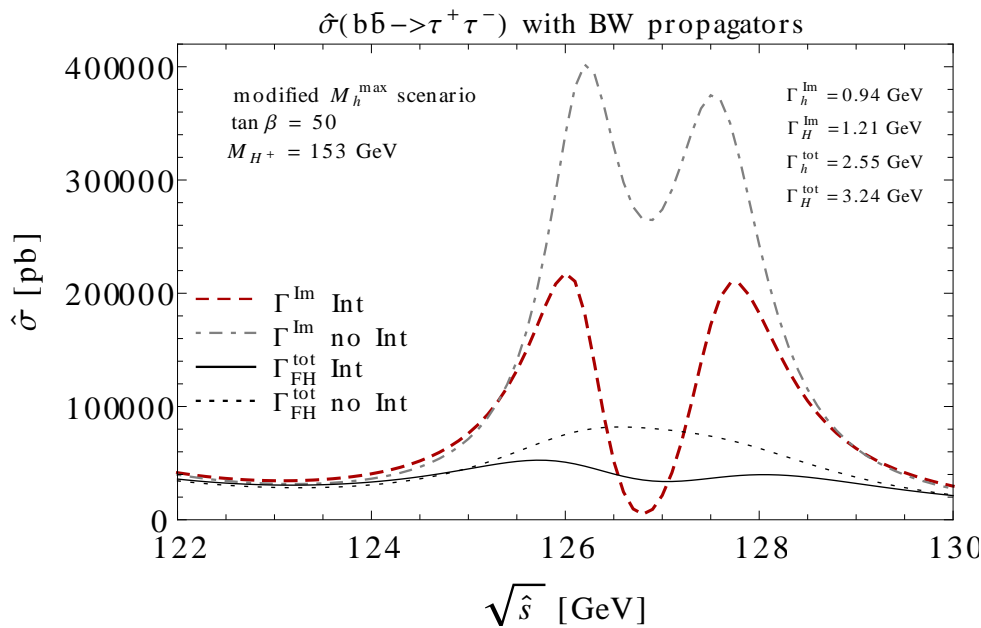
\begin{figure}[t!]
\begin{center}
\begin{tikzpicture}
 \node at (0,0) {\includegraphics[width=0.8\textwidth]{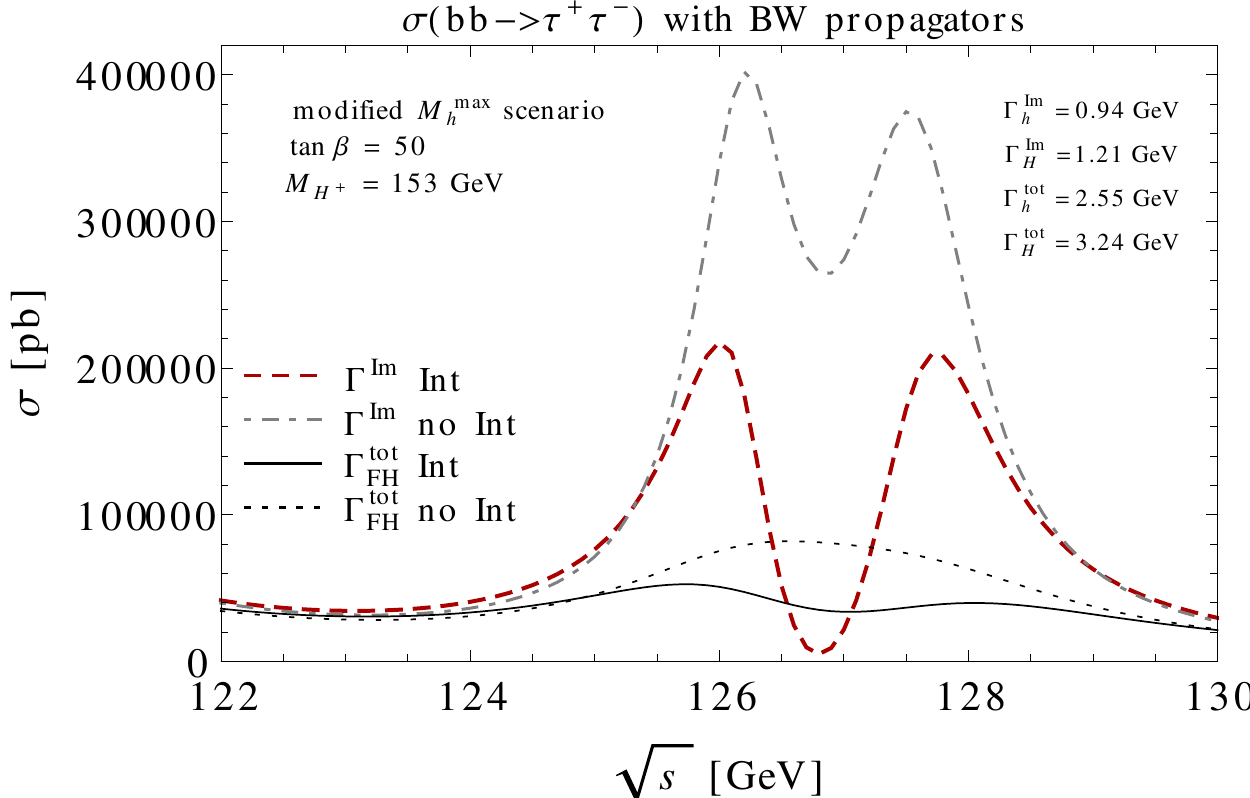}};
  \node at (-60.8pt,110.5pt) {\textbf{\textasciicircum}};
  \node at (-42pt,114.5pt) {-};
  \node[rotate=90] at (-176pt,-1pt) {\textbf{\textasciicircum}};
  \node at (13pt,-109pt) {\textbf{\textasciicircum}};
\end{tikzpicture}
\caption{Effect of the total width as an input for Breit--Wigner propagators: 
The partonic cross section 
$\hat\sigma(b\overline{b}\rightarrow \tau^{+}\tau^{-})$ in the same modified $M_{h}^{\textrm{max}}$-scenario as in Fig.\,\ref{fig:DeltaBWMhmax} with $\tan\beta = 50$ and $M_{H^{+}} = 153\,$GeV. The Breit--Wigner propagators with the total widths from the imaginary part of the complex pole including (red, dashed) and excluding (grey, dash-dotted) the interference term (as in Fig.\,\ref{fig:DeltaBWMhmax}) are compared with the Breit--Wigner propagators where the total widths are obtained from \texttt{FeynHiggs}. The corresponding results are shown including (black, solid) and excluding (black, dotted) the interference term. 
}
\label{fig:GammaImTotMhmax}
\end{center}
\end{figure}
In the modified $\mhmax$ scenario, the higher-order corrections have a significant impact on the Higgs decay widths so that $\Gamma_{h_1}^{\textrm{tot}}=2.55\,$GeV and $\Gamma_{h_2}^{\textrm{tot}}=3.24\,$GeV are much larger than the widths obtained from the imaginary part of the complex pole.
This affects the order of magnitude of the cross-section 
$\hat\sigma\left(b\overline{b}\rightarrow \tau^{+}\tau^{-}\right)$ and the 
structure of the resonances, as can be seen in Fig.\,\ref{fig:GammaImTotMhmax}. 
The coherent sum of Breit--Wigner propagators including the interference term 
(red, dashed) and the incoherent sum without the interference term (grey, 
dash-dotted) using $\Gamma^{\textrm{Im}}$ from the imaginary parts of the 
complex poles are the same as in Fig.\,\ref{fig:DeltaBWMhmax}. In contrast, the
total widths $\Gamma^{\textrm{tot}}_{\textrm{FH}}$ obtained from 
\texttt{FeynHiggs} as the sum of higher-order partial widths are implemented 
into the Breit--Wigner propagators in the cross section based on the coherent 
sum of all $h_a$-contributions (black, solid) and the incoherent sum (black, 
dotted).

The large increase in the widths from 
$\Gamma_{h_a}^{\textrm{Im}}$ to
$\Gamma_{h_a}^{\textrm{tot}}$
has a very significant effect in Fig.\,\ref{fig:GammaImTotMhmax}, since the resonant behaviour is smeared out by the larger widths.
As a consequence, the separate resonances are less pronounced, and 
the cross section is suppressed. 
Here, the incoherent sum without the interference term again overestimates the cross-sections. In addition, it lacks the two-peak structure. This observation emphasizes the importance of including the total width at the highest available precision and to take the interference term into account. One can also see that two resonances might be too close to be resolved if they are smeared by large widths.

%%%%%%%%%%%%%%%%%%%%%%%%%%%%%%%%%%%%%%%%%%%%%%%%%%%%%%%%%%%%%%%%%%%%%%%%%%%%%%%
\section{Conclusions}
\label{sect:conclusions}
%%%%%%%%%%%%%%%%%%%%%%%%%%%%%%%%%%%%%%%%%%%%%%%%%%%%%%%%%%%%%%%%%%%%%%%%%%%%%%%

We have shown that the momentum-dependent propagator matrices of systems of
unstable particles that mix with each other can be accurately approximated by a 
combination of Breit--Wigner propagators and wave function normalisation
factors, where the latter are evaluated at the complex poles of the
propagator matrix.
For illustration, we have applied this approach to the example of the
neutral Higgs bosons of the MSSM with complex parameters, where the
$\cp$-violating interactions give rise to a 
$3\times 3$ mixing between the lowest-order mass
eigenstates $h,H,A$ and the loop-corrected mass eigenstates $h_1,h_2,h_3$
(further mixing contributions with unphysical Goldstone
bosons and vector bosons can be incorporated separately up to the considered
order in perturbation theory). In the special case of $\cp$-conservation 
the mixing between the physical Higgs bosons is reduced to a 
$2\times 2$ mixing between the states $h$ and $H$.

Analysing the pole structure of propagator matrices, we have shown that for
the case of $3\times 3$ mixing each entry of the propagator matrix has three
complex poles, while for the $2\times 2$ mixing case each entry of the propagator 
matrix has two complex poles. Consequently, a single-pole
approximation is not sufficient to approximate the momentum dependence of
the full propagators. In particular for close-by states with sizeable
widths, the different resonance regions may overlap.

We have derived in a process-independent way
how the full propagators can be expanded 
around all of their complex poles. This approximation results in the sum of 
Breit--Wigner propagators of the corresponding resonances, weighted by 
wave function normalisation factors which encompass 
the relation between the lowest-order mass eigenstates and the
loop-corrected mass eigenstates.
As a key result, we have demonstrated that wave function
normalisation factors that have been derived to ensure the correct on-shell
properties of external particles in physical processes at their (in general
complex) poles are a useful tool also for describing off-shell propagators, 
i.e.\ including
momentum dependence. It has been shown that the momentum dependence of the
full propagators can be accurately approximated by simple Breit--Wigner
propagators. The complex wave function normalisation factors 
properly incorporate the 
imaginary parts of the self-energies arising from absorptive parts of loop
integrals, which in contrast are neglected in effective coupling approaches
where the contributing self-energies are evaluated at
vanishing external momentum. 

In our numerical analysis 
we have found very good agreement between the approximation presented in
this paper and the 
full propagators.
Besides the numerical comparison of the full and approximated propagators
around the complex poles or depending on real momenta, we have provided a
detailed discussion of the uncertainties involved in the approximation. In
particular, the omitted higher-order terms from the expansion of
self-energies around the real part of complex momenta, the expansion of the
momentum-dependent effective self-energies around the complex poles and
numerical effects in the on-shell condition 
have been quantified.
The uncertainty estimates of these sources are
in general well reflected 
in our numerical results. We have also discussed the validity
of the approximation 
far away from the poles.

The formalism of Breit--Wigner propagators and on-shell wave function
normalisation factors has 
several appealing advantages in describing the propagators of unstable
particles that mix with each other.
It avoids the momentum-dependent evaluation of self-energies and thereby 
significantly simplifies and accelerates the calculation 
of higher-order contributions.
Besides, it enables the separation of the individual resonant contributions 
of the mass eigenstates $h_a$ 
within a given process and the 
straightforward calculation of their interference term. 
Moreover, the Breit--Wigner propagator turns into a $\delta$-distribution in the 
limit of a vanishing width, thus facilitating the separate calculation of the 
production and decay of an intermediate unstable
particle by means of the 
narrow-width approximation
and its generalisation to the case of overlapping and interfering resonances. 
In addition, the Breit--Wigner formulation 
allows the implementation of a more precise total width 
by incorporating important higher-order effects from the partial widths 
that are not included in the imaginary part of the complex pole 
with self-energies evaluated at the same perturbative order as for
the partial widths. 
This feature is very useful for phenomenological predictions of processes 
involving the exchange of unstable particles,
benefitting from the use of quantities computed at the highest available
order in perturbation theory. 

The explicit calculations presented in this paper 
have been performed in the MSSM, 
but the introduced method itself can be easily extended to the cases of
particles with non-zero spin and mixing among more than three particles,
such as,
for example,
in different models with a non-minimal scalar sector or new vector resonances.

%%%%%%%%%%%%%%%%%%%%%%%%%%%%%%%%%%%%%%%%%%%%%%%%%%%%%%%%%%%%%%%%
\section*{Acknowledgements}
%%%%%%%%%%%%%%%%%%%%%%%%%%%%%%%%%%%%%%%%%%%%%%%%%%%%%%%%%%%%%%%%
We thank Alison Fowler for her contributions at an early stage of this work. 
E.F.\ thanks the DESY theory group where most of this work was carried out.
The work of E.F.\ was partially supported by the 
German National Academic Foundation. The work of G.W.\ is supported in part by 
the Collaborative Research Centre SFB~676 of the DFG, ``Particles, Strings and 
the Early Universe'', and by the European Commission through the ``HiggsTools'' 
Initial Training Network PITN-GA-2012-316704.
%%%%%%%%%%%%%%%%%%%%%%%%%%%%%%%%%%%%%%%%%%%%%%%%%%%%%%%%%%%%%%%%%%%%%%%%%%%%%

\bibliographystyle{utphys}
\bibliography{mthesis_literature,gNWALoop,PhDthesis,bbH,propagator}
\markboth{}{}

\end{document}